\documentclass[11pt]{article}
\usepackage{amssymb,amsmath,xcolor,graphicx,xspace}
\usepackage{amsfonts}
\usepackage{amsthm}
\usepackage{mathtools}
\usepackage{caption}
\usepackage[mag=1000,portrait]{geometry}
\usepackage{indentfirst}
\usepackage[onehalfspacing]{setspace}
\usepackage{subcaption}
\usepackage{booktabs}
\usepackage{natbib}
\usepackage[hidelinks]{hyperref}
\usepackage{enumitem}
\usepackage{threeparttable}
\DeclareGraphicsExtensions{.pdf,.eps,.ps,.png,.jpg,.jpeg}
\newtheorem{theorem}{Theorem}
\newtheorem{assumption}{Assumption}

\newtheorem{lemma}{Lemma}
\newtheorem{proposition}{Proposition}
\newtheorem{remark}{Remark}
\newtheorem{definition}{Definition}
\newtheorem{example}{Example}
\DeclareMathOperator*{\argmin}{arg\,min}

\DeclareMathOperator{\Var}{Var}

\newcommand{\R}{\mathbb{R}}

\newcommand{\EP}{\mathrm{E}_P}
\newcommand{\vecop}{\operatorname{vec}}
\newcommand{\vech}{\operatorname{vech}}

\newcommand{\dto}{\overset{d}{\rightarrow}}
\newcommand{\os}{\mathrm{os}}

\newcommand{\itr}{\mathrm{it}}

\begin{document}
\title{Orthogonal Scores and Double/Debiased Machine Learning for Possibly Misspecified GMM}
\author{ Byunghoon Kang\thanks{Email: b.kang1@lancaster.ac.uk.}\\
  Lancaster University}
\date{November 12, 2026}
\maketitle

\begin{abstract}
We develop Neyman orthogonal scores for GMM models that remain valid when the moment conditions may fail to hold. Existing score constructions typically require the moment conditions to hold exactly, which may fail in over-identified models, where the number of moment functions exceeds the number of parameters. We construct \emph{misspecification-robust} orthogonal score  that has a closed form; it is the conventional orthogonal score plus adjustment terms, for the estimation of the Jacobian and of the weighting matrix, which are proportional to the population moment and vanish under correct specification. The moment functions may depend on nonparametric or machine learning first steps, and we show that double/debiased machine learning (DML) combined with the proposed scores and cross-fitting provides uniformly valid inference for a low-dimensional GMM estimand, while the conventional DML standard errors are inconsistent when the moment conditions fail.  We illustrate the results in the partially linear instrumental variables (PLIV) model and in the local average treatment effects (LATE) model with multiple instruments.
\medskip

\noindent \textbf{Keywords:} Neyman orthogonal moments, double/debiased machine learning, GMM,  misspecification,   semiparametric  two-step estimation, cross-fitting.

\noindent \textbf{JEL Classification:} C13, C14, C26, C36.
\end{abstract}

\section{Introduction}\label{sec:intro}
Many parameters of interest in economics are defined by moment conditions and estimated by the generalized method of moments \citep[GMM,][]{hansen1982}. The moment conditions often depend on nuisance parameters, which are unknown functions or high-dimensional vectors estimated in a first step by machine learning or nonparametric methods. Plugging first-step estimates into the moment function generally produces a first-order bias, because the regularization or smoothing bias of the first step affects the estimator of the parameter of interest. A large literature \citep{neyman1959,newey1994,chs2015,ccddhnr2018,ceinr2022} shows that this bias can be removed by a Neyman orthogonal, or locally robust, moment function which is insensitive to the nuisance parameters. Double/debiased machine learning (DML) \citep{chs2015,ccddhnr2018} is now widely used in economics to estimate a low-dimensional structural parameter in such settings. Combined with cross-fitting that removes own-observation bias and avoids Donsker conditions on the first-step estimator, the resulting DML or debiased GMM estimators are $\sqrt N$-consistent and asymptotically normal under the rate conditions that many machine learning and nonparametric methods satisfy.

However, the orthogonal scores used in this literature are derived under correct specification: there exists a parameter value at which the population moments are exactly zero. In over-identified models, with more moment conditions than parameters, it is an assumption about the data generating process.\footnote{In just-identified models, where the number of moment conditions equals the number of parameters, the estimand can be defined as the unique solution to the population moment conditions. \citet{ceinr2022} note that their standard errors are robust to misspecification in the just-identified case.}  For instance, in the partially linear instrumental variables (PLIV) model, the moment conditions may fail when some of the instruments are invalid, or when the effect of the endogenous regressor is heterogeneous so that different instruments identify different local average treatment effects (LATE) \citep{imbens1994}.

When the moment conditions fail, the GMM estimator still converges to a well-defined limit, the GMM estimand or pseudo-true value, which solves the population first-order condition and depends on the weighting matrix. This estimand is often of interest in its own right; with heterogeneous treatment effects, for example, it is a weighted average of instrument-specific LATEs \citep{angristimbens1995}. However, the orthogonal score of \citet{chs2015} and \citet{ccddhnr2018} for GMM, which combines the moment functions using the Jacobian and the weighting matrix, is no longer orthogonal when the population moments are not zero, and then the conventional DML variance estimator is inconsistent.

This paper constructs Neyman orthogonal/locally robust  scores for over-identified GMM that remain valid when the moment conditions do not hold, which we label \emph{misspecification-robust} orthogonal scores. We stack the GMM first-order condition with the moment equations that define the Jacobian and the weighting matrix, and write the over-identified GMM estimator as a just-identified estimator with an augmented parameter vector, as in \citet{imbens1997}. The stacked moment conditions hold at the GMM estimand by construction, whether or not the original moment conditions hold.  However, this stacked moment condition is \emph{not} orthogonal to the Jacobian and the weighting matrix that are treated as additional finite-dimensional nuisance parameters. Then, applying the standard orthogonal score construction to the stacked system gives a score that is orthogonal with or without correct specification.

We show that the resulting score has a closed form. It equals the conventional orthogonal score plus two adjustment terms, one for the Jacobian and one for the weighting matrix, both proportional to the population moment at the GMM estimand. Under correct specification, the two terms vanish and the score reduces to the conventional one. With finite-dimensional nuisance parameters, we verify that these additional terms are those that \citet{hallinoue2003} and \citet{hansenlee2021} obtain by direct expansion for the asymptotic variance of over-identified GMM under misspecification (see also \citet{imbens1997} and \citet{hkl2022}). Although the construction does not give a new variance formula for GMM with finite-dimensional parameters, our result makes explicit the connection between the Neyman orthogonal score and the robust variance formula in misspecified GMM. Furthermore, the corrections appear here as terms of a Neyman orthogonal score, which, unlike a variance formula, can be made insensitive to a first-step estimate.

We extend our results to semiparametric two-step GMM, where the parameter of interest is finite-dimensional and the moment functions depend on unknown functions estimated in a first step, as in \citet{newey1994}, \citet{ackerberg2014} and \citet{ceinr2022}. Starting from an orthogonal/locally robust moment function (e.g., obtained by adding the first-step influence function to the identifying moment function as in \citet{ceinr2022}), 
 we derive a misspecification-robust score that is orthogonal jointly to the first-step functions, the Jacobian and the weighting matrix.

We then study the DML estimator with cross-fitting based on the misspecification-robust score. For scores that are linear in the parameter of interest with a one-step weighting matrix, we show that the estimator is asymptotically normal  and the confidence intervals are uniformly valid for the GMM estimand over a class of data generating processes. The same confidence interval is valid whether or not the moment conditions hold. With linear moment functions and one-step weighting, the point estimate is unchanged and it coincides with the DML estimator of \citet{ccddhnr2018} and the debiased GMM estimator of \citet{ceinr2022}. So, a researcher who already reports DML estimates keeps the same estimate and replaces only the standard error with the one based on the misspecification-robust score. The rate conditions on the first step are similar to those of \citet{ccddhnr2018}, with the additional requirement that the first-step biases of the Jacobian and of the weighting matrix are small when the model is misspecified.

For clarity, our main asymptotic results focus on moment functions that are linear in the parameter of interest and a one-step weighting matrix that does not depend on it. Our general results on the Neyman orthogonal score constructions allow both one-step GMM (Theorem~\ref{thm:ortho}) and efficient GMM weighting (Theorem~\ref{thm:general}(i)). Efficient weighting requires an additional correction term, because the weighting matrix is then not orthogonal to the first step, and it makes the score non-linear in the parameter of interest even when the moment functions are linear. Non-linear moment functions and general weighting, including iterated GMM, are treated in Appendix~\ref{app:general} (Theorem~\ref{thm:general}; see also Remark~\ref{rem:nocorrection}).

We consider two illustrative examples. First, the over-identified PLIV model is a leading example. With an orthogonal moment function in the style of \citet{robinson1988}, the nuisance parameters are the conditional expectations of the outcome, the endogenous regressor and the instruments given the controls. Then, the DML estimator is cross-fitted two-stage least squares on the residualized data, and the robust score adds correction terms that involve the same conditional expectations and sample moments of the residuals. Second, we consider the LATE model with multiple binary instruments, where over-identification arises from treatment effect heterogeneity rather than from invalid instruments. With a diagonal weighting matrix the GMM estimand is a convex combination of the instrument-specific LATEs, and we provide valid inference for it. In a Monte Carlo study of the PLIV model, the robust standard error under misspecification is about $15$--$20\%$ larger than the conventional one and close to the standard deviation of the estimator, and the conventional DML confidence interval undercovers at every sample size while the robust one stays close to the nominal level. Under correct specification, the two standard errors and confidence intervals are almost the same, so little is lost by using the robust score even when the moment conditions hold.

This paper is related to two strands of the literature. The first is the large literature on Neyman orthogonal or locally robust moment functions and semiparametric two-step estimation \citep{neyman1959,newey1994,chs2015,ccddhnr2018,ichimuranewey2022,ceinr2022,cns2022}. \citet{ackerberg2014} study the efficiency of semiparametric two-step GMM. 
These papers assume that the moment conditions hold, or allow misspecification only in the just-identified case. We extend \citet{chs2015} and \citet{ccddhnr2018} by providing the Neyman orthogonal score  for over-identified GMM that is robust to misspecification of the moment conditions. \citet{aichen2007} study sieve minimum distance estimation of possibly misspecified semiparametric conditional moment restriction models with a known weighting matrix; in their setting, the finite-dimensional parameter and the unknown functions are estimated jointly \citep{aichen2003,aichen2012}.  Our paper considers the semiparametric two-step GMM based on the unconditional moment conditions, and the first step is a separate object, such as a conditional expectation, whose limit is well defined whether or not the moment conditions hold.  This two-step structure is what allows the first step to be estimated by any machine learning method with cross-fitting. \citet{chensantos2018} study over-identification in a general class of regular models.

The second is the growing literature on model misspecification in general contexts \citep{kitamura2013,armstrongkolesar2021,bonhommeweidner2022,abgrs2025, act2025}, which goes back to \citet{white1982} for maximum likelihood. For GMM with finite-dimensional parameters, \citet{hallinoue2003} derive the limit distribution of the estimator under misspecification, \citet{hansenlee2021} consider the iterated GMM estimator, and \citet{lee2018} considers two-stage least squares when instruments identify different LATEs; see also \citet{imbens1997} and \citet{schennach2007}. We extend the misspecification-robust variance of this literature to models with nonparametric or machine learning first steps. Under correct specification, the adjustment terms in our score are of the order of the sample moment, and \citet{hkl2022} show that the misspecification-robust variance, which takes these terms into account, acts as a finite-sample or higher-order correction in linear GMM. We consider first-order orthogonality; recent papers by \citet{bjnw2026} and \citet{cst2026} construct moment functions that are Neyman orthogonal to higher order in order to reduce higher-order bias.

The rest of the paper is organized as follows. Section~\ref{sec:setup} reviews the Neyman orthogonal score for GMM with finite-dimensional nuisance parameters, explains why it fails under misspecification, and constructs the misspecification-robust score. Section~\ref{sec:theory} extends the construction to functional nuisance parameters, defines the DML estimator with cross-fitting and establishes its asymptotic properties. Section~\ref{sec:pliv} applies the results to the over-identified PLIV model, and Section~\ref{sec:late} to local average treatment effects identified by multiple instruments. Section~\ref{sec:sim} reports a Monte Carlo study, and Section~\ref{sec:conc} concludes. Appendix~\ref{app:proofs} collects the proofs, and Appendix~\ref{app:general} extends the results to non-linear scores and general weighting, including iterated GMM.

We use the following generic notation throughout the paper. $\Pr_P$ and $\EP[\cdot]$ denote probability and expectation under the probability law $P$, with the subscript suppressed when no ambiguity arises. We use capital letters $W$ to denote random elements and the corresponding lowercase $w$ to denote fixed values $W$ can take, and $(W_i)_{i=1}^N$ denotes a random sample of $N$ independent copies of $W$. We equip the parameter spaces with their Borel $\sigma$-fields, and assume throughout that the moment functions and each component of the score functions are measurable. For $q\ge1$ we write $\|f\|_{P,q}:=\|f(W)\|_{P,q}:=\big(\int |f(w)|^q\,dP(w)\big)^{1/q}$ for the $L^q(P)$ norm. For a column vector $x$, $x'$ is its transpose. For a differentiable map $x\mapsto f(x)$ from $\R^d$ to $\R^k$, we denote $\partial_{x'}f$ as the matrix of partial derivatives $(\partial/\partial x')f$, and write $\partial_{x'}f(x_0)$ for $\partial_{x'}f(x)|_{x=x_0}$.  $\vecop(\cdot)$ denotes column-wise vectorization, and $\otimes$ denotes the Kronecker product. 

\section{Neyman orthogonal score/moment functions for GMM}\label{sec:setup}

In this section, we formally introduce the model and provide the definitions of Neyman orthogonality for a self-contained treatment.  We closely follow the notation of \cite{ccddhnr2018}.  Let  $\theta\in\Theta\subset\R^{d_\theta}$ denote a finite-dimensional parameter vector of interest. We assume that the true value $\theta_0$ satisfies the moment conditions
\begin{equation}
\label{eq:momcond}
\EP\big[\psi(W,\theta_0,\eta_0)\big]=0,
\end{equation}
where  $\psi=(\psi_1,\dots,\psi_{d_\theta})'$ is a known score function, $W$ is a random element  in a measurable space $(\mathcal W,\mathcal A_{\mathcal W})$ with law determined by a probability measure $P\in\mathcal P_N$, where $\mathcal P_N$ is a collection of probability measures, and a nuisance parameter $\eta$ taking values in a convex subset $T$ of a normed vector space, with true value $\eta_0\in T$. To introduce the Neyman orthogonality condition, we define the pathwise (Gateaux) derivative map for $\eta\in T$, 
\[
D_r[\eta-\eta_0]:=\partial_r\Big\{\EP\big[\psi\big(W,\theta_0,\eta_0+r(\eta-\eta_0)\big)\big]\Big\},
\qquad r\in[0,1),
\]
whenever it exists, and write
$\partial_\eta\EP[\psi(W,\theta_0,\eta_0)][\eta-\eta_0]:=D_0[\eta-\eta_0]$. Let $T_N\subset T$ be a nuisance realization set such that the estimators $\widehat{\eta}_0$ of $\eta_0$ specified below take values in this set with high probability. 

\begin{definition}[Neyman orthogonality; Definition 2.1 in \cite{ccddhnr2018}]
\label{def:northo}
The score $\psi$ obeys the \emph{orthogonality condition} at $(\theta_0,\eta_0)$ with
respect to the nuisance realization set $T_N\subset T$ if \eqref{eq:momcond} holds and
the pathwise derivative map $D_r[\eta-\eta_0]$ exists for all $r\in[0,1)$ and $\eta\in
T_N$ and vanishes at $r=0$:
\begin{equation}
\label{eq:northo}
\partial_\eta\EP\big[\psi(W,\theta_0,\eta_0)\big][\eta-\eta_0]=0
\qquad\text{for all }\eta\in T_N.
\end{equation}
\end{definition}

When $\eta$ is a finite-dimensional vector, condition \eqref{eq:northo} holds with $T_N=T$ if the vector of partial derivatives $\partial_\eta\EP[\psi_j(W,\theta_0,\eta)]|_{\eta=\eta_0} = 0 $ for each $j=1,\dots,d_\theta$.

\subsection{ Construction of Neyman orthogonal scores under correct specification}
\label{ssec:nogmm}

Now, we provide the construction of the Neyman orthogonal scores for the GMM model. We first consider the GMM model with finite-dimensional nuisance parameters by closely following the same model in \citet{chs2015} and \cite{ccddhnr2018} to describe why they do not apply directly to the  misspecified GMM models, and then we consider the semiparametric GMM model with infinite-dimensional or functional nuisance parameters in Section \ref{sec:theory}.

\citet{chs2015} and \citet[Section~2.2.2]{ccddhnr2018} consider the following  correctly specified GMM model, i.e., assume that the parameters $\theta$ and $\beta$ have true values $\theta_0$ and $\beta_0$ solving the moment conditions exactly
\begin{equation}
\label{eq:ccddhnr217}
\EP\big[m(W,\theta_0,\beta_0)\big]=0,
\end{equation}
where  $m:\mathcal W\times\Theta\times \mathcal B\to\R^{d_m}$  is a known, vector-valued moment function, $\theta\in\Theta\subset\R^{d_\theta}$ is the parameter of interest and $\beta\in\mathcal B\subset\R^{d_\beta}$ is a finite-dimensional nuisance parameter with $d_m\ge d_\theta+d_\beta$.  We assume throughout that $m$ is twice continuously differentiable in $(\theta, \beta)$ with dominated derivatives, so that differentiation and expectation can be interchanged.  

Then, a  Neyman orthogonal score function for GMM is 
\begin{equation}
\label{eq:ccddhnr218}
\psi(W,\theta,\eta)=\mu\, m(W,\theta,\beta),
\qquad
\eta=\big(\beta',\vecop(\mu)'\big)'\in T=\mathcal B\times\R^{d_\theta d_m},
\end{equation}
where $\mu\in\R^{d_\theta\times d_m}$ is the orthogonalization matrix considered as an additional nuisance parameter, and 
whose true value is
\begin{equation}
\label{eq:mu0generic}
\mu_0:=A'\Omega^{-1}-A'\Omega^{-1}G_{\beta,0}\big(G_{\beta,0}'\Omega^{-1}G_{\beta,0}\big)^{-1}G_{\beta,0}'\Omega^{-1}
\ \in\R^{d_\theta\times d_m},
\end{equation}
where $A$ is a $d_m\times d_\theta$ moment selection matrix, $\Omega$ is a $d_m\times d_m$ positive definite weighting matrix, and $G(\gamma)$ is the Jacobian matrix  for $\gamma=(\theta',\beta')'\in\Theta\times\mathcal B\subset\R^{d_\gamma}, d_\gamma=d_\theta+d_\beta$,
\[
G(\gamma):=\partial_{\gamma'}\,\EP[m(W,\gamma)]
=\big[\,G_\theta(\gamma)\ \ G_\beta(\gamma)\,\big]\in\R^{d_m\times d_\gamma}, 
\]
and $G_{0}=G(\gamma_0)=[\,G_{\theta,0}\ \ G_{\beta, 0}\,]$ for  $\gamma_0=(\theta_0',\beta_0')'$. Both $A$ and $\Omega$ may be chosen arbitrarily and 
\begin{equation}
\label{eq:effchoice}
A=G_{\theta,0},
\qquad
\Omega=\Var_P\big(m(W,\theta_0,\beta_0)\big)=\EP\big[m(W,\theta_0,\beta_0)m(W,\theta_0,\beta_0)'\big]
\end{equation}
yields the efficient score, in the sense of providing the estimator of $\theta_0$ with the smallest variance in the class of GMM estimators \citep{hansen1982}, and the semiparametrically efficient score \citep{chamberlain1987}.

The following Lemma shows that the score in \eqref{eq:ccddhnr218} satisfies the Neyman orthogonality condition. Let $\eta_0=(\beta_0',\vecop(\mu_0)')'$ denote the true value of $\eta$.

\begin{lemma}[Neyman orthogonal scores for GMM setup, Lemma 2.3 of \citet{ccddhnr2018}]
\label{lem:ccddhnr}
If \eqref{eq:ccddhnr217} holds, $G_0$ exists and $\Omega$, $G_{\beta,0}'\Omega^{-1}G_{\beta,0}$ are invertible, then the
score $\psi$ in \eqref{eq:ccddhnr218} is Neyman orthogonal at $(\theta_0,\eta_0)$ with
respect to the nuisance realization set $T_N=T$.
\end{lemma}

It is straightforward to show that the Neyman orthogonality condition in Lemma \ref{lem:ccddhnr} requires the following:
\begin{equation}
\label{eq:lem1proof}
\partial_\eta\EP\big[\psi(W,\theta_0,\eta_0)\big]
=\Big[\ \mu_0 G_{\beta,0}\ ,\ \ \EP\big[m(W,\theta_0,\beta_0)\big]'\otimes I_{d_\theta}\ \Big]=0. 
\end{equation}
Orthogonality with respect to $\beta$ in the first block holds because $\mu_0G_{\beta,0}=0$ by construction in \eqref{eq:mu0generic}. However, the second block (partial derivatives with respect to $\mu$) vanishes only when the moment condition $\EP[m(W,\theta_0,\beta_0)]=0$ holds at $(\theta_0, \beta_0)$. When the moment condition fails to hold, i.e., $\EP[m(W,\theta,\beta)] \neq 0$ for all values of  $(\theta, \beta)$, the score $\psi$ does not satisfy the Neyman orthogonality condition as 
$\EP[m(W,\theta_0,\beta_0)] '\otimes I_{d_\theta}\neq0$  for any choice of $A$ and $\Omega$.

\subsection{Neyman orthogonal score for possibly misspecified GMM}
\label{ssec:nogomm_mis}

Now, we construct Neyman orthogonal scores/moment functions which are valid even when the moment condition \eqref{eq:ccddhnr217} fails to hold, i.e., $\EP[m(W,\theta,\beta)] \neq 0$ for all $\gamma \in \Theta\times\mathcal B$. Define the population moment as
\[
\bar{m}(\gamma):= \EP\big[m(W,\gamma)\big]\in\R^{d_m}, \qquad \gamma=(\theta',\beta')'.
\]
The model is \emph{correctly specified} if $\bar{m}(\gamma)=0$ for some $\gamma$ in the interior of $\Theta\times\mathcal B$. When the model is just-identified, $d_m =d_\gamma$, we define $\gamma_0$ as a solution to $\bar{m}(\gamma_0)=0$.  When the model is over-identified, $d_m>d_\gamma$, generically no $\gamma\in\Theta\times\mathcal B$ solves all $d_m$ equations. When $d_m>d_\gamma$, what a GMM estimator estimates is a GMM estimand (or pseudo-true value), $\gamma_0=(\theta_0',\beta_0')'$, defined as the solution of the population first-order condition
\begin{equation} 
\label{eq:foc}
Q(\gamma_0)=0,
\qquad
Q(\gamma):=G(\gamma)'\,\Omega(\gamma)^{-1}\,\bar{m}(\gamma)\in\R^{d_\gamma},
\end{equation}
where $\Omega(\gamma)$ is the symmetric, positive definite weight. We assume that a unique $\gamma_0$ exists and lies in the interior of $\Theta\times\mathcal B$.  We define the weighting matrix $\Omega(\gamma):=\EP[\zeta(W,\gamma)]$, where $\zeta(w,\gamma)=\zeta(w)$ for the one-step GMM, and $\zeta(w,\gamma)=m(w,\gamma)\,m(w,\gamma)'$ for the iterated GMM. For the one-step GMM,  the weight $\Omega(\gamma)\equiv\Omega$ does not vary with $\gamma$, $\gamma_0$ is a minimizer of the population criterion $\bar{m}(\gamma)'\Omega^{-1}\bar{m}(\gamma)$, and \eqref{eq:foc} is the first-order condition. For the iterated GMM, \eqref{eq:foc} is instead a fixed point, and $\gamma_0$ is a minimizer of $\bar{m}(\gamma)'\Omega(\gamma_0)^{-1}\bar{m}(\gamma)$. Under correct specification, \eqref{eq:foc} holds for any weight, so $\gamma_0$ coincides with the true value $(\theta_0',\beta_0')'$ of Section~\ref{ssec:nogmm}. Under misspecification, $\gamma_0$ depends on the weighting matrix, and we suppress the dependence in the notation for simplicity.

To construct the Neyman orthogonal scores that are valid even when the moment condition fails to hold, we stack the GMM first-order condition together with moment equations that define the Jacobian and the weight matrix as additional nuisance parameters. To be specific, define the stacked moment function as follows:
\begin{equation}
\label{eq:h}
h(w, \theta,\lambda)=
\begin{pmatrix}
\vecop\!\big(\Gamma-\partial_{\gamma'}m(w,\gamma)\big)\\[4pt]
\vecop\!\big(\Omega-\zeta(w,\gamma)\big)\\[4pt]
\Gamma'\,\Omega^{-1}\,m(w,\gamma)
\end{pmatrix}
\in\R^{d_h},
\quad \lambda:=\big(\beta',\ \vecop(\Gamma)',\ \vecop(\Omega)'\big)'\in\Lambda\subset\R^{d_\lambda}
\end{equation}
where $\lambda$ is the nuisance parameter with $d_\lambda=d_\beta+d_m d_\gamma+d_m^2$, $\Gamma\in\R^{d_m\times d_\gamma}$ is a Jacobian parameter, and $\Omega\in\R^{d_m\times d_m} $ is a weight parameter (symmetric positive definite). It is straightforward to check that the number of equations for the stacked moment function $h(w,\theta,\lambda)$ equals the number of parameters $(\theta,\lambda)$, so the system \eqref{eq:h} is just-identified. \citet{imbens1997} considers the just-identified GMM representation with an augmented parameter vector for the standard over-identified GMM model \citep[see also][]{neweymcfadden1994,chamberlainimbens1995}.\footnote{\citet{imbens1997} considers $\vech(\Omega)$ instead of $\vecop(\Omega)$. Because $\Omega$ is symmetric, the number of parameters is the same as the one considered in \citet{imbens1997} after eliminating duplicated $d_m(d_m-1)/2$ elements. To keep the matrix algebra transparent, we use $\vecop()$, and all discussions remain the same by replacing $\vecop()$ with $\vech()$.}  In one-step GMM, when $\zeta(W,\gamma)=\zeta(W)$ does not depend on $\gamma$, \eqref{eq:h} has the same three components, with the outer product replaced by $\zeta$, and for the iterated GMM, \eqref{eq:h} is exactly the moment function in equation~(4) of \citet{imbens1997}.\footnote{The two-step GMM estimator also admits a just-identified representation of the form \eqref{eq:h}, with augmented parameter vectors of the first-step estimator, its own Jacobian and the first-order condition with the first-step weighting matrix. Thus, the orthogonal score for the two-step GMM requires an additional term relative to the one-step case. The main text focuses on the one-step and the iterated GMM estimators, and we do not treat the continuously updated estimator (CUE) of \citet{hhy1996}, in which the first-order condition contains an additional term from the derivative of the weighting matrix, which is  different from \eqref{eq:foc}. Under correct specification, the two-step, iterated and CUE estimators are first-order equivalent and asymptotically efficient. Under misspecification, they estimate possibly different estimands, and the textbook variance formula $(G_0'\Omega_0^{-1}G_0)^{-1}$ is invalid.}

We can define the true value of the nuisance $\lambda$ in \eqref{eq:h}, with $\gamma_0$ defined in \eqref{eq:foc}, as
\[
\lambda_0:=\big(\beta_0',\ \vecop(G_0)',\ \vecop(\Omega_0)'\big)',
\qquad
G_0=\partial_{\gamma'}\EP\big[m(W,\gamma)\big]\big|_{\gamma=\gamma_0},
\qquad
\Omega_0=\EP\big[\zeta(W,\gamma_0)\big].
\]

\noindent Then, the following moment condition holds by construction,
\begin{equation}
\label{eq:hmoment}
\ \EP\big[h(W,\theta_0,\lambda_0)\big]=0\ 
\end{equation}
without assuming the original moment conditions $\EP\big[m(W,\theta_0,\beta_0)\big]=0$. 

The following result  extends the results of \citet{chs2015} and \citet{ccddhnr2018} by providing  the Neyman orthogonal construction of the score $\psi^{*}(w, \theta, \eta)$ for GMM that is robust to misspecification of the moment conditions \eqref{eq:ccddhnr217}. 

\begin{proposition}[Misspecification-robust Neyman orthogonal scores for GMM]
\label{prop:ortho}
Suppose that the moment condition \eqref{eq:hmoment} holds and $h(w, \theta, \lambda)$ is differentiable in
$(\theta,\lambda)$ at $(\theta_0,\lambda_0)$. We consider the following score 
\begin{equation}
\label{eq:psi}
\psi^{*}(w,\theta,\eta):=\mu\, h(w,\theta,\lambda),
\qquad
\eta=\big(\lambda',\vecop(\mu)'\big)'\in T=\Lambda\times\R^{d_\theta d_h},
\end{equation}
where $\Lambda$ is convex, $\mu\in\R^{d_\theta\times d_h}$ is the orthogonalization matrix with $d_h=d_\theta+d_\lambda$, and whose true value is
\begin{equation}
\label{eq:mu0stack}
\mu_0:=A_h'\Omega_h^{-1}-A_h'\Omega_h^{-1}J_\lambda\big(J_\lambda'\Omega_h^{-1}J_\lambda\big)^{-1}J_\lambda'\Omega_h^{-1}
\in\R^{d_\theta\times d_h},
\qquad
\eta_0:=\big(\lambda_0',\vecop(\mu_0)'\big)' , 
\end{equation}
where  $A_h\in\R^{d_h\times d_\theta}$ is a moment
selection matrix, $\Omega_h$ is a $d_h\times d_h$ positive definite matrix, and $J$ is the Jacobian of the stacked system at the pseudo-true value, 
\[
J:=\EP\big[\partial_{(\theta',\lambda')}h(W,\theta_0,\lambda_0)\big]
=\big[\,J_\theta\ \ J_\lambda\,\big]\in\R^{d_h\times d_h},
\qquad J_\theta\in\R^{d_h\times d_\theta},\quad J_\lambda\in\R^{d_h\times d_\lambda},
\]
and $J_\lambda$ has full column rank $d_\lambda$. 

Then $\EP[\psi^{*}(W,\theta_0,\eta_0)]=0$, and the score $\psi^{*}(w, \theta, \eta)$ of \eqref{eq:psi} is Neyman orthogonal at $(\theta_0,\eta_0)$  with respect to the unrestricted nuisance realization set $T_N=T$.
\end{proposition}

The Neyman orthogonality condition in Proposition~\ref{prop:ortho} is robust against moment misspecification compared to the results in \citet[Lemma~2.3]{ccddhnr2018}, in the sense that $\psi^*(w, \theta, \eta)$ is orthogonal at $(\theta_0, \eta_0)$ whether or not the original moment condition $\bar{m}(\gamma) = \bar{m}(\theta, \beta) = 0$ holds.  Analogous to  Lemma~\ref{lem:ccddhnr}, the proof of Proposition~\ref{prop:ortho} requires the following to hold for the stacked system
\begin{equation}
\label{eq:prop1proof}
\partial_\eta\EP\big[\psi^{*}(W,\theta_0,\eta_0)\big]
=\Big[\ \mu_0J_\lambda\ ,\ \ \EP\big[h(W,\theta_0,\lambda_0)\big]'\otimes I_{d_\theta}\ \Big]=0 ,
\end{equation}
where the first block holds by definition of $\mu_0$ in \eqref{eq:mu0stack}, and  the second condition holds because of \eqref{eq:hmoment}, which holds by construction.  For the original scores $\psi$ in Lemma~\ref{lem:ccddhnr}, it  is equal to $\bar{m}(\gamma_0)'\otimes I_{d_\theta}$, which is possibly non-zero under misspecification for every choice of $A$ and $\Omega$.\footnote{Similar to \citet[Definition~2.2]{ccddhnr2018}, we can also use an approximate Neyman orthogonality condition. For example, when $J_\lambda'\Omega_h^{-1}J_\lambda$ is close to singular, the exact construction \eqref{eq:mu0stack} can be replaced by  those of \citet[Lemma~2.4]{ccddhnr2018}.  The resulting score is then Neyman $\lambda_N$ near-orthogonal, i.e., \eqref{eq:northo} holds only up to a bound $\|\partial_\eta\EP[\psi^{*}(W,\theta_0,\eta_0)][\eta-\eta_0]\|\le\lambda_N$ for all $\eta\in T_N$, for a sequence of positive constants $\lambda_N=o(N^{-1/2})$.} 

\begin{remark}
\label{rem:sample}
It is important to note that the stacked moment function $h(w, \theta,\lambda)$ changes the representation, not the estimator. Write the sample moment and
sample weight as
\[
\bar m_N(\gamma):=\frac1N\sum_{i=1}^N m(W_i,\gamma),
\qquad
\widehat\Omega_N(\gamma):=\frac1N\sum_{i=1}^N \zeta(W_i,\gamma).
\]
Solving the sample analogue of \eqref{eq:h}, $N^{-1}\sum_i h(W_i,\widehat\theta,\widehat\lambda)=0$  provides the sample Jacobian $\widehat\Gamma=N^{-1}\sum_i\partial_{\gamma'}m(W_i,\widehat\gamma)$, the sample weight $\widehat\Omega=\widehat\Omega_N(\widehat\gamma)$, and the original GMM first-order condition $\widehat\Gamma'\widehat\Omega^{-1}\bar m_N(\widehat\gamma)=0$. When  $\zeta(w,\gamma) = \zeta(w)$ does not vary with $\gamma$, the M-estimator with the moment function $h(w, \theta,\lambda)$ is the same as the one-step GMM estimator
\[
\widehat\gamma^{\os}:=\argmin_{\gamma\in\Theta\times\mathcal B}\ \bar m_N(\gamma)'\,\widehat\Omega_N^{-1}\,\bar m_N(\gamma), 
\]
and when  $\zeta(w,\gamma)=m(w,\gamma)\,m(w,\gamma)'$, the estimator is the iterated GMM estimator, defined as the fixed point
\[
\widehat\gamma^{\itr}=\argmin_{\gamma\in\Theta\times\mathcal B}\ \bar m_N(\gamma)'\,\widehat\Omega_N(\widehat\gamma^{\itr})^{-1}\,\bar m_N(\gamma).
\]
The same equivalence holds for the debiased GMM estimator \citep[Section~2.2]{ceinr2022}, where $\beta_0$ is estimated in a first step and enters the moment functions. We will discuss the construction of the misspecification-robust score with functional nuisances $\beta_0$  in Section~\ref{sec:theory}.
\end{remark}

\subsection{The robust orthogonal score in a closed form}
\label{ssec:closedform}

Proposition~\ref{prop:ortho} defines the robust orthogonal score function $\psi^{*}(w, \theta, \eta)$ implicitly, but it can be made explicit. The next Proposition shows that the robust score $\psi^{*}(w, \theta, \eta)$ is related to the terms used to construct the so-called misspecification-robust variance of  \citet{hallinoue2003} and \citet{hansenlee2021} (see also \citet{act2025}), exhibiting the extra terms, relative to the textbook formula, that arise from the sampling variation of the estimated Jacobian and weight matrix.

The orthogonalization matrix $\mu_0$ in \eqref{eq:mu0stack} depends on the choice of $(A_h,\Omega_h)$, and so does the score $\psi^{*}$ below. However, the influence function of the GMM estimator $\widehat\theta$ in Proposition~\ref{prop:closed} does not depend on $(A_h, \Omega_h)$.

\begin{proposition}[The robust orthogonal score in closed form]
\label{prop:closed}
Suppose the conditions of Proposition~\ref{prop:ortho} hold, the Jacobian of the score
$\mu_0J_\theta=\partial_{\theta'}\EP\big[\psi^{*}(W,\theta,\eta_0)\big]\big|_{\theta=\theta_0}$, $H$ and $H_{\beta\beta}$ are nonsingular, where
\[
H:=\partial_{\gamma'}Q(\gamma)\big|_{\gamma=\gamma_0}
=\begin{pmatrix}H_{\theta\theta}&H_{\theta\beta}\\ H_{\beta\theta}&H_{\beta\beta}\end{pmatrix}
\in\R^{d_\gamma\times d_\gamma}, H_{\theta\theta}\in\R^{d_\theta\times d_\theta}, H_{\beta\beta}\in\R^{d_\beta\times d_\beta}, 
\]
and 
\[
J_0:=H_{\theta\theta}-H_{\theta\beta}H_{\beta\beta}^{-1}H_{\beta\theta}
\]
is also nonsingular. Then the score \eqref{eq:psi} at $(\theta_0, \eta_0)$ satisfies
\begin{equation}
\label{eq:psiexplicit}
\psi^{*}(w,\theta_0,\eta_0)=\mu_0J_\theta\,J_0^{-1}\big(s_\theta(w)-H_{\theta\beta}H_{\beta\beta}^{-1}s_\beta(w)\big),
\end{equation}
where  $s(w) = (s_\theta(w)',s_\beta(w)')'$ with $s_\theta(w) \in\R^{d_\theta}, s_\beta(w) \in\R^{d_\beta}$, 
\begin{equation}
\label{eq:s}
s(w)  :=
G_0'\Omega_0^{-1}m(w,\gamma_0)
+\big(\partial_{\gamma'}m(w,\gamma_0)-G_0\big)'\Omega_0^{-1} \bar m(\gamma_0)
-G_0'\Omega_0^{-1}\big(\zeta(w,\gamma_0)-\Omega_0\big)\Omega_0^{-1}\bar m(\gamma_0).
\end{equation}
Moreover, the GMM estimator $\widehat\theta$ of Remark~\ref{rem:sample} satisfies, under standard regularity
conditions,
\begin{equation}
\label{eq:if}
\begin{aligned}
\sqrt N\big(\widehat\theta-\theta_0\big)
&=-(\mu_0J_\theta)^{-1}\,\frac{1}{\sqrt N}\sum_{i=1}^N\psi^{*}(W_i,\theta_0,\eta_0)+o_P(1)\\
&=-J_0^{-1}\,\frac{1}{\sqrt N}\sum_{i=1}^N\big(s_\theta(W_i)-H_{\theta\beta}H_{\beta\beta}^{-1}s_\beta(W_i)\big)+o_P(1)
\ \dto\ \mathcal N\big(0,\ J_0^{-1}\Sigma_0J_0^{-1\prime}\big),
\end{aligned}
\end{equation}
where $\Sigma_0=\Var_P\big(s_\theta(W)-H_{\theta\beta}H_{\beta\beta}^{-1}s_\beta(W)\big)$.
\end{proposition}

Misspecification is what activates the orthogonalization. When $\bar m(\gamma_0) = 0 $, Proposition~\ref{prop:closed} reduces to Lemma~\ref{lem:ccddhnr}, i.e., the score in \eqref{eq:psiexplicit} becomes (up to the factor $\mu_0J_\theta J_0^{-1}$)
\[
s_\theta(w)-H_{\theta\beta}H_{\beta\beta}^{-1}s_\beta(w)
=\Big(G_{\theta,0}'\Omega_0^{-1}-G_{\theta,0}'\Omega_0^{-1}G_{\beta,0}\big(G_{\beta,0}'\Omega_0^{-1}G_{\beta,0}\big)^{-1}G_{\beta,0}'\Omega_0^{-1}\Big)m(w,\gamma_0),
\]
which is exactly the score \eqref{eq:ccddhnr218} of Lemma~\ref{lem:ccddhnr}, with $\mu_0$ from
\eqref{eq:mu0generic} evaluated at $A=G_{\theta,0}$ and $\Omega=\Omega_0$, because the Jacobian and weight matrix channels of \eqref{eq:s} vanish, $s(w)=G_0'\Omega_0^{-1}m(w,\gamma_0)$, and
$H$ reduces to $G_0'\Omega_0^{-1}G_0$. When $\Omega_0=\EP[m(W,\gamma_0)m(W,\gamma_0)']$, this is the efficient choice \eqref{eq:effchoice}.

Note that $(\Gamma,\Omega)$ are not nuisance parameters of the original GMM problem, but when $\bar m(\gamma_0)\neq0$, the first-order condition in the stacked moment function $h(w, \theta, \lambda)$ is \emph{not} orthogonal to $(\Gamma,\Omega)$, which is precisely why Jacobian and weight estimation then contribute to the orthogonal score, and thus the influence function and asymptotic variance of the GMM estimator under misspecification. The orthogonalization makes these contributions explicit: they are the Jacobian and weight matrix channels of \eqref{eq:s}, which are both proportional to $\bar m(\gamma_0)$. 

Therefore, with the finite-dimensional nuisance parameters, it is not surprising that the influence function formula in Proposition~\ref{prop:closed} coincides with those of misspecified GMM literature. \citet{hallinoue2003} and \citet{hansenlee2021}) obtain the formula by direct expansion of the first-order condition with finite-dimensional nuisance parameters;  the three terms in \eqref{eq:s} correspond to those in \citet{hallinoue2003}, and \eqref{eq:if} reproduces the misspecification-robust asymptotic variance of   \citet{hallinoue2003} for the one-step GMM, and that of \citet{hansenlee2021} for the iterated GMM. In our construction, our corrections appear as terms of a Neyman orthogonal score rather than as corrections to a variance formula. This distinction matters when a score  \eqref{eq:psi} depends on the infinite-dimensional nuisance parameters $\beta_0$, because our robust-score construction can be made insensitive to a first-step estimate of $\beta_0$.  Section~\ref{sec:theory} exploits this direction by considering semiparametric GMM setup, replacing the finite-dimensional $\beta_0$ by infinite-dimensional nuisance parameters $\beta_0$ estimated nonparametrically or by a machine learning first step.

%==================================================================
\section{Neyman orthogonal scores for GMM with infinite-dimensional nuisance parameters}
\label{sec:theory}

Section~\ref{sec:setup} considers the moment condition model, where the nuisance parameter $\beta$ is finite-dimensional. In this section, we consider the general semiparametric moment condition model, where the parameter of interest $\theta$ is still finite-dimensional, but the moment function depends on an unknown functional nuisance parameter $\beta$, estimated in a first step nonparametrically or by machine learning.

We closely follow the setup of \citet{ceinr2022} and allow the moment conditions to be misspecified. Let $\theta\in\Theta\subset\R^{d_\theta}$ be the finite-dimensional parameter of interest and let $\beta$ be an unknown nonparametric or high-dimensional first-step function, such as a conditional expectation, taking values in a convex subset $\mathcal B$ of a normed vector space with norm $\|\cdot\|_{\mathcal B}$. Let $\beta_0:=\beta(P)$ denote the probability limit of a first-step estimator $\widehat\beta$ when $P$ is the true probability measure of the data $W$. This limit is defined under general misspecification, as in \citet{newey1994}; for instance, if $\widehat\beta(x)$ is a nonparametric estimator of the conditional expectation, then $\beta(P)(x)=\EP[Y|X=x]$, where $P$ is unrestricted, except for regularity conditions.

We assume that there is a vector of known moment functions $m:\mathcal W\times\Theta\times\mathcal B\to\R^{d_m}$ with $d_m\ge d_\theta$. Under correct specification, \citet{ceinr2022} assume that $\theta_0$ is identified by the moment conditions
\begin{equation}
\label{eq:ceinrmoment}
\EP[m(W,\theta_0,\beta_0)]=0,
\end{equation}
that is, $\theta_0$ is the unique solution to $\EP[m(W,\theta,\beta_0)]=0$ over $\theta\in\Theta$.

Plugging an estimator of $\beta_0$ into $m(w, \theta, \beta)$ generally leads to a first-order bias from the first-step estimation. Thus, we consider an orthogonal, or locally robust, moment function $\psi(w,\theta,\beta,\alpha)$, where $\alpha$ is an additional unknown function with true value $\alpha_0 := \alpha(P)$, such that $\psi$ is locally insensitive to the first-step estimation of $(\beta,\alpha)$. For example, \citet{ceinr2022} consider the first-step influence function (FSIF), $\phi(w,\theta,\beta,\alpha)$, which may depend on  $\alpha$ and satisfies
\[
\frac{d}{d\tau}\EP\big[m\big(W,\theta,\beta(P_\tau)\big)\big]\Big|_{\tau=0}
=\int\phi(w,\theta,\beta_0,\alpha_0)\,\tilde P(dw),
\qquad
\EP\big[\phi(W,\theta,\beta_0,\alpha_0)\big]=0,
\]
for all $\theta\in\Theta$ and all alternative distributions $\tilde P$, where $P_\tau=(1-\tau)P+\tau\tilde P$ for $\tau\in[0,1]$ and $\tilde P$ is unrestricted, except for regularity conditions such as the existence of $\beta(P_\tau)$. Then, \citet{ceinr2022} define the orthogonal moment function as
\[
\psi(w,\theta,\beta,\alpha):=m(w,\theta,\beta)+\phi(w,\theta,\beta,\alpha)\in\R^{d_m},
\]
and show that $\psi$ is locally insensitive to $\beta$ and that $\EP[\phi(W,\theta,\beta_0,\alpha)]=0$ for all $\alpha \in \mathcal A$,  the set of possible values of $\alpha_0$ under regularity conditions. Different approaches to the construction of Neyman orthogonal moment functions $\psi$, such as those obtained by concentrating out or by projection, may be used; see \citet{newey1994}, \citet{chs2015} and \citet{ccddhnr2018}, for example.

\begin{example}[Partially linear IV model]
\label{ex:pliv}
Let $W=(Y,D,X,Z)$ with $D\in\R^{d_\theta}$ and $Z\in\R^{d_z}$, $d_z\ge d_\theta$, and consider
\begin{equation}
\label{eq:plivmodel}
Y=D'\theta+g(X)+U,\qquad \EP[U\mid X,Z]=0,
\end{equation}
where $Z$ is the vector of instrumental variables. With $\ell_0(X):=\EP[Y|X]$ and $r_0(X):=\EP[D|X]$, the model implies $g(X)=\ell_0(X)-r_0(X)'\theta$, and a natural identifying moment function is $m(w,\theta,\beta)=Z\big(Y-\ell(X)-(D-r(X))'\theta\big)$ with $\beta=(\ell,r)$. Its FSIF is $\phi(w,\theta,\beta,\alpha)=-z(X)\big(Y-\ell(X)-(D-r(X))'\theta\big)$ with the additional nuisance parameter $\alpha=z$, whose true value is $z_0(X):=\EP[Z|X]$ \citep{newey1994,ceinr2022},  so that
\begin{equation}
\label{eq:plivmoment}
\psi(w,\theta,\beta,\alpha)=\big(Z-z(X)\big)\big(Y-\ell(X)-(D-r(X))'\theta\big),
\end{equation}
which is the Robinson form of the orthogonal moment \citep{robinson1988,ccddhnr2018}. Let $\tilde Z:=Z-z_0(X)$, $\tilde Y:=Y-\ell_0(X)$ and $\tilde D:=D-r_0(X)$. Then, we have $\EP\big[m(W,\theta,\beta_0)\big]=\EP\big[\psi(W,\theta,\beta_0,\alpha_0)\big]=\EP[Z\tilde Y]-\EP[Z\tilde D']\theta=\EP[\tilde Z\tilde Y]-\EP[\tilde Z\tilde D']\theta$.
\end{example}

Throughout the paper, we define the population moment functions as follows:
\begin{equation}
\label{eq:estimandinvariance}
\bar m(\theta):=\EP\big[m(W,\theta,\beta_0)\big]=\EP\big[\psi(W,\theta,\beta_0,\alpha_0)\big]
\qquad\text{for every }\theta\in\Theta.
\end{equation}
When we use the approach of \citet{ceinr2022}, for example,  \eqref{eq:estimandinvariance} holds by Theorem~2 of \citet{ceinr2022}, where the mean-zero property $\EP\big[\phi(W,\theta,\beta_0,\alpha)\big]=0$ holds for all $\theta\in\Theta$ and $\alpha\in\mathcal A$.\footnote{
Condition \eqref{eq:estimandinvariance} holds for any construction that adds to $m$ with a correction
that has mean zero at $(\beta_0,\alpha_0)$ for every $\theta$, such as the FSIF, but it can fail when the
first-step limit depends on $\theta_0$. Theorems~\ref{thm:ortho}--\ref{thm:late} and the discussion below are unchanged with $\bar m(\theta)$ defined as $\EP[\psi(W,\theta,\beta_0,\alpha_0)]$ by taking the orthogonal moment $\psi$ as the primitive object.}

We say that the model is correctly specified if $\bar m(\theta)=0$ for some $\theta\in\Theta$, and misspecified otherwise.  As in Section~\ref{ssec:nogomm_mis}, the moment condition may fail, that is, $\bar m(\theta)\neq0$ for all $\theta\in\Theta$. In Example~\ref{ex:pliv}, the moment condition $\bar m(\theta)=\EP[ZU(\theta)]=\EP[\tilde Z\,\EP[U(\theta)\mid X,Z]]$, $U(\theta):=\tilde Y-\tilde D'\theta$,  may fail when the conditional moment restriction in \eqref{eq:plivmodel} fails to hold, for instance because some instruments are invalid or the effect of $D$ is heterogeneous.  In the just-identified case $d_z=d_\theta$ with $\EP[\tilde Z\tilde D']$ nonsingular, $\bar m(\theta)=0$ has the unique solution $\theta_0=\EP[\tilde Z\tilde D']^{-1}\EP[\tilde Z\tilde Y]$ whether or not \eqref{eq:plivmodel} holds. In the over-identified case $d_z>d_\theta$, it has a solution only when $\EP[\tilde Z\tilde Y]$ lies in the column space of $\EP[\tilde Z\tilde D']$, which does not hold in general. \citet{ccddhnr2018} consider the just-identified setup, i.e., scalar $Z$ and $D$, and thus the associated standard errors are robust to misspecification when the number of moment functions is the same as the number of parameters of interest as noted in \citet{ceinr2022}.

Under misspecification, we define the GMM estimand $\theta_0$ as the solution to the following population first-order condition
\begin{equation}
\label{eq:focfun}
Q(\theta_0)=0,
\qquad
Q(\theta):=G(\theta)'\,\Omega(\theta)^{-1}\,\bar m(\theta)\in\R^{d_\theta},
\end{equation}
where
\begin{equation}
\label{eq:nuGfun}
G(\theta):=\partial_{\theta'}\bar m(\theta)\in\R^{d_m\times d_\theta},
\qquad
\Omega(\theta):=\EP\big[\zeta(W,\theta,\beta_0,\alpha_0)\big]\in\R^{d_m\times d_m}, 
\end{equation}
with $\zeta:\mathcal W\times\Theta\times\mathcal B\times\mathcal A\to\R^{d_m\times d_m}$ a known symmetric weight function. We use $\zeta(w,\theta,\beta,\alpha)=\zeta(w,\beta,\alpha)$ for one-step GMM and $\zeta(w,\theta,\beta,\alpha)=\psi(w,\theta,\beta,\alpha)\psi(w,\theta,\beta,\alpha)'$ for iterated GMM, and define
 $\bar m_0:=\bar m(\theta_0)$, $G_0:=G(\theta_0)$ and $\Omega_0:=\Omega(\theta_0)$. 
 
 Under regularity conditions, $\theta_0$ is the probability limit of a version of the debiased GMM estimator of \citet[equations~(2.6)--(2.7)]{ceinr2022}
\begin{equation}
\label{eq:dgmm}
\widehat\theta^{\mathrm{DGMM}}:=\argmin_{\theta\in\Theta}\ \widehat\psi_N(\theta)'\,\widehat\Omega^{\mathrm{cf}}_N(\widetilde\theta)^{-1}\,\widehat\psi_N(\theta),
\end{equation}
where we take a $K$-fold partition $(I_k)_{k=1}^K$ of $\{1,\dots,N\}$, let $(\widehat\beta_k,\widehat\alpha_k)$ be estimators of $(\beta_0,\alpha_0)$ constructed from the observations not in $I_k$, and
\[
\widehat\psi_N(\theta):=\frac1N\sum_{k=1}^{K}\sum_{i\in I_k}\psi(W_i,\theta,\widehat\beta_k,\widehat\alpha_k),
\qquad
\widehat\Omega_N^{\mathrm{cf}}(\theta):=\frac1N\sum_{k=1}^{K}\sum_{i\in I_k}\zeta(W_i,\theta,\widehat\beta_k,\widehat\alpha_k)
\]
are the cross-fitted sample moment and weight. In the one-step case $\widehat\Omega^{\mathrm{cf}}_N$ is free of $\theta$, and in the iterated case $\widetilde\theta=\widehat\theta^{\mathrm{DGMM}}$, so that \eqref{eq:dgmm} is defined as a fixed point.\footnote{There is a minor difference between the estimator \eqref{eq:dgmm} and those of \citet{ceinr2022}. \citet{ceinr2022} evaluate the FSIF part of the cross-fitted moment at a preliminary estimator $\widetilde{\theta}_k$ of $\theta_0$, $m(W_i,\theta,\widehat\beta_k)+\phi(W_i,\widetilde\theta_k,\widehat\beta_k,\widehat\alpha_k)$ in place of $\psi(W_i,\theta,\widehat\beta_k,\widehat\alpha_k)$. }  Cross-fitting helps to remove the own-observation bias from the infinite-dimensional first step, and avoids the Donsker conditions for $(\widehat\beta_k,\widehat\alpha_k)$; see \citet{neweyrobins2017}, \citet{ccddhnr2018} and \citet{ceinr2022}, for example.

\subsection{The misspecification-robust score}
\label{ssec:funcscore}

The debiased GMM estimator \eqref{eq:dgmm} reduces the $d_m$ orthogonal moments to $d_\theta$ dimensions through $\Gamma'\Omega^{-1}\psi$, with true values $\Gamma=G_0$ and $\Omega=\Omega_0$, and the associated conventional variance
\begin{equation}
\label{eq:psidml}
V=\big(G_0'\Omega_0^{-1}G_0\big)^{-1}G_0'\Omega_0^{-1}\,
\EP\big[\psi(W,\theta_0,\beta_0,\alpha_0)\psi(W,\theta_0,\beta_0,\alpha_0)'\big]\,
\Omega_0^{-1}G_0\big(G_0'\Omega_0^{-1}G_0\big)^{-1}
\end{equation}
is that of \citet[Theorem~9]{ceinr2022}. Under correct specification the reduction $G_0'\Omega_0^{-1}\psi$ is Neyman orthogonal and \eqref{eq:psidml} is valid. Under misspecification neither is true, because $\Gamma'\Omega^{-1}\psi$ is not orthogonal to the estimated $\Gamma$ and $\Omega$. As in Section~\ref{ssec:nogomm_mis}, we therefore stack the moments together with $\Gamma$ and $\Omega$, taking the analogue of \eqref{eq:h} built on $\psi$ with the parameter of interest $\theta$ alone,
\begin{equation}
\label{eq:hfun}
h(w,\theta,\lambda,\beta,\alpha):=
\begin{pmatrix}
\vecop\!\big(\Gamma-\partial_{\theta'}\psi(w,\theta,\beta,\alpha)\big)\\[2pt]
\vecop\!\big(\Omega-\zeta(w,\theta,\beta,\alpha)\big)\\[2pt]
\Gamma'\,\Omega^{-1}\,\psi(w,\theta,\beta,\alpha)
\end{pmatrix}\in\R^{d_h},
\end{equation}
where $\lambda:=(\vecop(\Gamma)',\vecop(\Omega)')'\in\Lambda\subset\R^{d_\lambda}$ with $\Gamma\in\R^{d_m\times d_\theta}$, $\Omega\in\R^{d_m\times d_m}$, $d_\lambda=d_md_\theta+d_m^2$ and $d_h=d_\theta+d_\lambda$. With $\lambda_0:=(\vecop(G_0)',\vecop(\Omega_0)')'$, the first two blocks have mean zero by \eqref{eq:nuGfun}, and the third by \eqref{eq:estimandinvariance} and \eqref{eq:focfun}, so
\begin{equation}
\label{eq:hmomentfun}
\EP\big[h(W,\theta_0,\lambda_0,\beta_0,\alpha_0)\big]=0
\end{equation}
by construction whether or not $\bar m_0=0$.

As in Remark~\ref{rem:sample}, the stacking changes the representation, not the estimator. Solving the cross-fitted sample analogue of \eqref{eq:hmomentfun},
$N^{-1}\sum_{k=1}^{K}\sum_{i\in I_k}h(W_i,\widehat\theta,\widehat\lambda,\widehat\beta_k,\widehat\alpha_k)=0$, the first two blocks give $\widehat\Gamma=\partial_{\theta'}\widehat\psi_N(\widehat\theta)$ and $\widehat\Omega=\widehat\Omega^{\mathrm{cf}}_N(\widehat\theta)$, and the third block gives
\[
\partial_{\theta'}\widehat\psi_N(\widehat\theta)'\,\widehat\Omega^{\mathrm{cf}}_N(\widehat\theta)^{-1}\,\widehat\psi_N(\widehat\theta)=0, 
\]
which is the first-order condition of the debiased GMM estimator in \eqref{eq:dgmm}. The stacked cross-fitted system therefore reproduces the debiased GMM estimator \eqref{eq:dgmm} exactly in finite samples, not only asymptotically.

We impose the following assumptions for the Neyman orthogonality based on the stacked moment functions. 

\begin{assumption}[Model and orthogonal moment function]
\label{ass:model}
Let $\mathcal A$ be a convex subset of a normed vector space with norm
$\|\cdot\|_{\mathcal A}$ containing $\alpha_0$, let $\mathcal B_N\subset\mathcal B$ and
$\mathcal A_N\subset\mathcal A$ be nuisance realization sets containing $\beta_0$ and $\alpha_0$. For all $P\in\mathcal P_N$:
\begin{enumerate}[label=(\alph*),leftmargin=2.4em]
\item $\theta_0$ in \eqref{eq:focfun} exists, is unique and lies
in the interior of $\Theta$, where $\Theta$ is compact and convex;  $\bar m$ and $\Omega$ are twice continuously differentiable on  $\Theta_0$, a neighbourhood of $\theta_0$. 
$\Omega_0$ is positive definite, and $J_0:=\partial_{\theta'}Q(\theta_0)$ is
nonsingular; $\theta\mapsto\psi(w,\theta,\beta_0,\alpha_0)$ is continuously differentiable on $\Theta_0$ with $\sup_{\theta\in\Theta_0}\|\partial_{\theta'}\psi(W,\theta,\beta_0,\alpha_0)\|\le \bar d(W)$ for some $\bar d$ with $\EP[\bar d(W)]<\infty$.
\item For each $\theta\in\Theta_0$, the maps $(\beta,\alpha)\mapsto\EP[\psi(W,\theta,\beta,\alpha)]$,
$(\beta,\alpha)\mapsto\EP[\partial_{\theta'}\psi(W,\theta,\beta,\alpha)]$ and
$(\beta,\alpha)\mapsto\EP[\zeta(W,\theta,\beta,\alpha)]$ are twice continuously Gateaux dif\-fer\-en\-tiable on $\mathcal B_N\times\mathcal A_N$. Moreover, $\partial_{\theta'}$ and $\partial_{(\beta,\alpha)}$ can be interchanged at $(\theta_0,\beta_0,\alpha_0)$.

\item The following holds for $\psi(w, \theta, \beta, \alpha)$:
\begin{equation}
\label{eq:psimean}
\EP\big[\psi(W,\theta,\beta_0,\alpha_0)\big]=\EP\big[m(W,\theta,\beta_0)\big]
\qquad\text{for all }\theta\in\Theta,
\end{equation}
and
\begin{equation}
\label{eq:unifortho}
\partial_{(\beta,\alpha)}\EP\big[\psi(W,\theta,\beta_0,\alpha_0)\big][\beta-\beta_0,\alpha-\alpha_0]=0
\qquad\text{for all }\theta\in\Theta_0, (\beta,\alpha)\in\mathcal B_N\times\mathcal A_N.
\end{equation}
\item The weight function $\zeta(w, \theta, \beta, \alpha)$ satisfies the following:
\begin{equation}
\label{eq:zetaortho}
\partial_{(\beta,\alpha)}\EP\big[\zeta(W,\theta_0,\beta_0,\alpha_0)\big][\beta-\beta_0,\alpha-\alpha_0]=0
\qquad\text{for all }(\beta,\alpha)\in\mathcal B_N\times\mathcal A_N .
\end{equation}
\end{enumerate}
\end{assumption}

Assumptions~\ref{ass:model}(a) and (b) are standard regularity conditions. Assumption~\ref{ass:model}(b) is similar to \citet[Assumption~3.1(c)]{ccddhnr2018}, while Assumption~\ref{ass:model}(a) requires additional smoothness conditions for $\bar m$ and $\Omega$ in $\theta$ because the estimand $\theta_0$ is defined by the first-order 
condition.  The interchange of derivatives is imposed in Assumption~\ref{ass:model}(b), and this trivially holds when $\psi$ is linear in $\theta$. 

Assumption~\ref{ass:model}(c) imposes two conditions on the orthogonal moment function $\psi$: that it has the same population mean as the identifying moment function $m$, and that $\psi$ is Neyman orthogonal with respect to $(\beta,\alpha)$ uniformly on $\Theta_0$. Assumption~\ref{ass:model}(c) holds for $\psi=m+\phi$ constructed from the FSIF, by Theorems~1--3 of \citet{ceinr2022}.  Assumption~\ref{ass:model}(d) requires the Gateaux derivative of the weight matrix term to vanish. Assumption~\ref{ass:model}(d) holds for a weight function that does not depend on parameters, $\zeta(w,\theta,\beta,\alpha)=\zeta(w)$, but it also allows weight functions that depend on nuisance parameters. In Example~\ref{ex:pliv}, both $\zeta(w)=ZZ'$ and $\zeta(w,\alpha)=\big(Z-z(X)\big)\big(Z-z(X)\big)'$ satisfy \eqref{eq:zetaortho}. Under $\bar m_0=0$, the Jacobian and weight matrix channels vanish, and condition \eqref{eq:zetaortho} is not needed, so that orthogonality at $\theta_0$ suffices and the conditions reduce to those of \citet{ccddhnr2018}.   Assumption~\ref{ass:model}(d) fails to hold for the efficient weight function $\zeta=\psi\psi'$, whose Gateaux derivative involves the moment residual, and the orthogonal moment construction involves the additional correction terms; see Remark~\ref{rem:nocorrection} below and Appendix \ref{app:general} for general weight function.

\begin{remark}[Partially linear IV model (cont.)]
\label{rem:whichresid}
In the partially linear IV model in Example \ref{ex:pliv}, one can also consider the identifying moment function $m(w,\theta,\beta)=Z\big(Y-D'\theta-g(X)\big)$ with $\beta=g$, and the orthogonal moment function $\psi(w,\theta,\beta,\alpha)=\big(Z-z(X)\big)\big(Y-D'\theta-g(X)\big)$ with $\alpha=z$ \citep[equation~(4.7)]{ccddhnr2018}. However, its true value $g_0(X)=\EP[Y-D'\theta_0|X]$ depends on $\theta_0$ and hence, under misspecification, on the weighting matrix. Furthermore, this alternative form of $\psi$ does not satisfy Assumption~\ref{ass:model}(c); it is orthogonal at $\theta_0$ but not uniformly in $\theta$: at $\theta\neq\theta_0$, its Gateaux derivative in the direction $z-z_0$ is $-\EP\big[(z-z_0)(X)\,r_0(X)'\big](\theta_0-\theta)\neq0$ in general.  The Robinson form \eqref{eq:plivmoment} has advantages because its nuisance parameters $\beta=(\ell,r)$ and $\alpha=z$ are conditional expectations defined without reference to $\theta_0$ or to the model, so Assumption~\ref{ass:model}(c) above holds by construction.
\end{remark}

The next result provides the Neyman orthogonal moment $\psi^* (w, \theta, \eta)$ that is robust to misspecification of the moment conditions  \eqref{eq:ceinrmoment}, i.e., without imposing $\EP\big[m(W,\theta_0,\beta_0)\big] = 0 $.

\begin{theorem}[Misspecification-robust Neyman orthogonal scores with infinite-dimensional nuisance parameters]
\label{thm:ortho}
Suppose Assumption~\ref{ass:model} holds, and consider the score
\begin{equation}
\label{eq:psirdml}
\psi^{*}(w,\theta,\eta):=\mu\,h(w,\theta,\lambda,\beta,\alpha),
\qquad
\eta:=\big(\lambda',\vecop(\mu)',\beta,\alpha\big)\in T:=\Lambda\times\R^{d_\theta d_h}\times\mathcal B\times\mathcal A,
\end{equation}
where $\mu\in\R^{d_\theta\times d_h}$ has true value $\mu_0$ given by \eqref{eq:mu0stack} with
\[
J:=\EP\big[\partial_{(\theta',\lambda')}h(W,\theta_0,\lambda_0,\beta_0,\alpha_0)\big]
=\big[\,J_\theta\ \ J_\lambda\,\big]\in\R^{d_h\times d_h},
\]
and $(A_h,\Omega_h)$ chosen such that $\mu_0J_\theta=J_0$, and
$\eta_0:=(\lambda_0',\vecop(\mu_0)',\beta_0,\alpha_0)$. Then
\begin{equation}
\label{eq:rdmlmoment}
\EP\big[\psi^{*}(W,\theta_0,\eta_0)\big]=0,
\end{equation}
and $\psi^{*}$ is Neyman orthogonal at $(\theta_0,\eta_0)$ with respect to
$T_N:=\Lambda\times\R^{d_\theta d_h}\times\mathcal B_N\times\mathcal A_N$. Furthermore, for all $\theta\in\Theta$,
\begin{equation}
\label{eq:psirdmlclosed}
\begin{aligned}
\psi^{*}(w,\theta,\eta_0)
&=G_0'\Omega_0^{-1}\psi(w,\theta,\beta_0,\alpha_0)
+\big(\partial_{\theta'}\psi(w,\theta,\beta_0,\alpha_0)-G_0\big)'\Omega_0^{-1}\bar m_0\\
&\qquad-G_0'\Omega_0^{-1}\big(\zeta(w,\theta,\beta_0,\alpha_0)-\Omega_0\big)\Omega_0^{-1}\bar m_0,
\end{aligned}
\end{equation}
and $\partial_{\theta'}\EP[\psi^{*}(W,\theta,\eta_0)]|_{\theta=\theta_0}=J_0$, where
\begin{equation}
\label{eq:J0fun}
J_0=G_0'\Omega_0^{-1}G_0
+\big(I_{d_\theta}\otimes\bar m_0'\Omega_0^{-1}\big)\,\partial_{\theta'}\vecop G(\theta_0)
-\big(\bar m_0'\Omega_0^{-1}\otimes G_0'\Omega_0^{-1}\big)\,\partial_{\theta'}\vecop\Omega(\theta_0).
\end{equation}
\end{theorem}

Theorem~\ref{thm:ortho} generalizes Propositions~\ref{prop:ortho} and \ref{prop:closed}, and  also extends the misspecified GMM literature (e.g., \citet{hallinoue2003}, \citet{hansenlee2021}) to functional nuisance parameters; it shows that \eqref{eq:psirdmlclosed} is insensitive to the first-step estimation of $\beta_0$. The first term of \eqref{eq:psirdmlclosed} is the standard orthogonal score $\psi$, and the second and third are the Jacobian and weight matrix orthogonalization parts, both proportional to $\bar m_0$. Under correct specification,  $\mu_0=[\,0\ \ 0\ \ I_{d_\theta}\,]$,
$J_0=G_0'\Omega_0^{-1}G_0$ and $\psi^{*}(w,\theta,\eta_0)=G_0'\Omega_0^{-1}\psi(w,\theta,\beta_0,\alpha_0)$, so that the asymptotic variance based on $\psi^{*}$ coincides with \eqref{eq:psidml}.

\subsection{DML estimation and inference}
\label{ssec:dml}

 We now define the DML estimator of \citet[Definition~3.2 (DML2)]{ccddhnr2018} with the misspecification-robust score
$\psi^{*}$, using the cross-fitted sample moments and weights defined below, and develop
general asymptotic theory for the DML estimator. 

\begin{definition}[DML with the misspecification-robust score]
\label{def:dml2}
\begin{enumerate}[label=(\alph*),leftmargin=2.4em]
\item Take a $K$-fold random partition $(I_k)_{k=1}^K$ of $[N]=\{1,\dots,N\}$ such that the
size of each fold $I_k$ is $n=N/K$, and for each $k\in[K]$ define $I_k^c:=[N]\setminus I_k$.
\item For each $k\in[K]$, construct estimators $\widehat\beta_k$ and $\widehat\alpha_k$ of
$\beta_0$ and $\alpha_0$ from $(W_i)_{i\in I_k^c}$.
\item Let $\widetilde\theta$ be a preliminary estimator of $\theta_0$, for instance the debiased
GMM estimator $\widehat\theta^{\mathrm{DGMM}}$ of \eqref{eq:dgmm}. Construct the cross-fitted
Jacobian, weight matrix and moment,
\[
\begin{aligned}
\widehat\Gamma&=\frac1N\sum_{k=1}^{K}\sum_{i\in I_k}\partial_{\theta'}\psi(W_i,\widetilde\theta,\widehat\beta_k,\widehat\alpha_k), \quad \widehat\Omega=\frac1N\sum_{k=1}^{K}\sum_{i\in I_k}\zeta(W_i,\widetilde\theta,\widehat\beta_k,\widehat\alpha_k)=\widehat\Omega^{\mathrm{cf}}_N(\widetilde\theta),\\
\widehat{\bar m}&=\frac1N\sum_{k=1}^{K}\sum_{i\in I_k}\psi(W_i,\widetilde\theta,\widehat\beta_k,\widehat\alpha_k),
\end{aligned}
\]
and set $\widehat\lambda:=(\vecop(\widehat\Gamma)',\vecop(\widehat\Omega)')'$,
\[
\widehat\mu:=\Big[-\big(I_{d_\theta}\otimes\widehat{\bar m}{}'\widehat\Omega^{-1}\big)\ \ \
\widehat{\bar m}{}'\widehat\Omega^{-1}\otimes\widehat\Gamma'\widehat\Omega^{-1}\ \ \ I_{d_\theta}\Big],
\qquad
\widehat\eta_k:=\big(\widehat\lambda',\vecop(\widehat\mu)',\widehat\beta_k,\widehat\alpha_k\big).
\]
\item Construct $\check\theta$ as the solution of
\begin{equation}
\label{eq:dml2eq}
\frac1N\sum_{k=1}^{K}\sum_{i\in I_k}\psi^{*}(W_i,\check\theta,\widehat\eta_k)=0,
\end{equation}
where, as in Theorem \ref{thm:ortho},
\begin{equation}
\label{eq:psistarhat}
\begin{aligned}
\psi^{*}(W_i,\theta,\widehat\eta_k)=\widehat\mu\,h(W_i,\theta,\widehat\lambda,\widehat\beta_k,\widehat\alpha_k)
&=\widehat\Gamma'\widehat\Omega^{-1}\psi(W_i,\theta,\widehat\beta_k,\widehat\alpha_k)
+\big(\partial_{\theta'}\psi(W_i,\theta,\widehat\beta_k,\widehat\alpha_k)-\widehat\Gamma\big)'\widehat\Omega^{-1}\widehat{\bar m}\\
&\qquad-\widehat\Gamma'\widehat\Omega^{-1}\big(\zeta(W_i,\theta,\widehat\beta_k,\widehat\alpha_k)-\widehat\Omega\big)\widehat\Omega^{-1}\widehat{\bar m}.
\end{aligned}
\end{equation}
\end{enumerate}
\end{definition}

The DML estimator of \citet{ccddhnr2018} based on the original orthogonal moment function $\psi$ derived under correct specification, as in Lemma~\ref{lem:ccddhnr}, instead solves
\begin{equation}
\label{eq:dml2first}
\frac1N\sum_{k=1}^{K}\sum_{i\in I_k}\widehat\Gamma'\widehat\Omega^{-1}\psi(W_i,\theta,\widehat\beta_k,\widehat\alpha_k)=0,
\end{equation}
whose estimating function is the first term of \eqref{eq:psistarhat}. The two estimating
equations differ by the sample means of the Jacobian and weight matrix channels, which are
proportional to $\widehat{\bar m}$. Under correct specification $\widehat{\bar m}=O_P(N^{-1/2})$
and the two estimators are first-order equivalent. Under misspecification the reduction
$\widehat\Gamma'\widehat\Omega^{-1}\psi$ is not orthogonal to $(\Gamma,\Omega)$, and the
conventional variance \eqref{eq:psidml} attached to \eqref{eq:dml2first} is invalid.

Next, we focus on the moment condition models with linear scores same as in
\citet[Section~3.2]{ccddhnr2018}, and with a one-step weight function $\zeta$.

\begin{assumption}[Linear scores and one-step weighting]
\label{ass:linear}
For all $w\in\mathcal W$, $\theta\in\Theta$, $\beta\in\mathcal B$ and $\alpha\in\mathcal A$,
\begin{equation}
\label{eq:mlinear}
\psi(w,\theta,\beta,\alpha)=\psi^{b}(w,\beta,\alpha)+\psi^{a}(w,\beta,\alpha)\,\theta, 
\end{equation}
and $\zeta(w,\theta,\beta,\alpha)=\zeta(w,\beta,\alpha)$.
\end{assumption}

Under Assumption~\ref{ass:linear}, $\partial_{\theta'}\psi(w,\theta,\beta,\alpha)=\psi^{a}(w,\beta,\alpha)$, and substituting \eqref{eq:mlinear} into \eqref{eq:psirdmlclosed} gives $\psi^{*}(w,\theta,\eta_0)=\psi^{*a}(w,\eta_0)\,\theta+\psi^{*b}(w,\eta_0)$ with
$\psi^{*a}(w,\eta_0)=G_0'\Omega_0^{-1}\psi^{a}(w,\beta_0,\alpha_0)$. Also the estimated score \eqref{eq:psistarhat} is linear in $\theta$, 
\[
\psi^{*}(W_i,\theta,\widehat\eta_k)=\psi^{*a}(W_i,\widehat\eta_k)\,\theta+\psi^{*b}(W_i,\widehat\eta_k),
\]
where
\[
\begin{aligned}
\psi^{*a}(W_i,\widehat\eta_k)&:=\widehat\Gamma'\widehat\Omega^{-1}\psi^{a}(W_i,\widehat\beta_k,\widehat\alpha_k),\\
\psi^{*b}(W_i,\widehat\eta_k)&:=\widehat\Gamma'\widehat\Omega^{-1}\psi^{b}(W_i,\widehat\beta_k,\widehat\alpha_k)
+\big(\psi^{a}(W_i,\widehat\beta_k,\widehat\alpha_k)-\widehat\Gamma\big)'\widehat\Omega^{-1}\widehat{\bar m}\\
&\qquad-\widehat\Gamma'\widehat\Omega^{-1}\big(\zeta(W_i,\widehat\beta_k,\widehat\alpha_k)-\widehat\Omega\big)\widehat\Omega^{-1}\widehat{\bar m},
\end{aligned}
\]
and here $\widehat\Gamma=N^{-1}\sum_{k=1}^{K}\sum_{i\in I_k}\psi^{a}(W_i,\widehat\beta_k,\widehat\alpha_k)$ and $\widehat\Omega$ of Definition~\ref{def:dml2} do not depend on $\widetilde\theta$. Further, 
\[
J_0=\partial_{\theta'}\EP[\psi^{*}(W,\theta,\eta_0)]|_{\theta=\theta_0} = \EP\big[\psi^{*a}(W,\eta_0)\big]=G_0'\Omega_0^{-1}G_0,
\]
because $\EP[\psi^{a}(W,\beta_0,\alpha_0)]=G_0$.  The conventional variance \eqref{eq:psidml} and the robust variance of Theorem~\ref{thm:dml}
below share the same Jacobian $J_0$, and differ only in their middle terms,
$G_0'\Omega_0^{-1}\EP[\psi\psi']\Omega_0^{-1}G_0$ against $\EP[\psi^{*}\psi^{*\prime}]$, which
coincide when $\bar m_0=0$, because $\psi^{*}=G_0'\Omega_0^{-1}\psi$.

Under Assumption~\ref{ass:linear}, the estimating equation \eqref{eq:dml2eq} has a closed-form solution
\[
\check\theta=-\big(\frac1N\sum_{k=1}^{K}\sum_{i\in I_k}\psi^{*a}(W_i,\widehat\eta_k)\big)^{-1}
\frac1N\sum_{k=1}^{K}\sum_{i\in I_k}\psi^{*b}(W_i,\widehat\eta_k) = -\big(\widehat\Gamma'\widehat\Omega^{-1}\widehat\Gamma\big)^{-1}
\frac1N\sum_{k=1}^{K}\sum_{i\in I_k}\psi^{*b}(W_i,\widehat\eta_k).
\]
Also, under Assumption~\ref{ass:linear},
\[
\frac1N\sum_{k=1}^{K}\sum_{i\in I_k}\partial_{\theta'}\psi(W_i,\theta,\widehat\beta_k,\widehat\alpha_k)=\widehat\Gamma
\quad\text{for all }\theta,
\qquad
\frac1N\sum_{k=1}^{K}\sum_{i\in I_k}\zeta(W_i,\widehat\beta_k,\widehat\alpha_k)=\widehat\Omega,
\]
so the sample means of the last two terms in \eqref{eq:psistarhat} vanish
and \eqref{eq:dml2eq} reduces to \eqref{eq:dml2first}. Hence DML based on
$\psi^{*}$, DML based on the first term of $\psi^{*}$ ($\widehat\Gamma'\widehat\Omega^{-1}\psi$), and the debiased GMM estimator coincide identically in finite
samples. The misspecification-robust
score construction therefore leaves the estimator exactly as it is and changes only the variance
estimator; a practitioner who already reports DML/debiased GMM keeps the same point estimate and replaces the conventional standard error with the robust standard error. The conventional variance \eqref{eq:psidml} omits the Jacobian and weight matrix
parts of $\psi^{*}$ in \eqref{eq:psirdmlclosed}; these are of the same order as the first term and vanish only when $\bar m_0=0$, so the conventional variance is inconsistent when
$\bar m_0\neq0$. This is the semiparametric counterpart of the results in
\citet{hallinoue2003} and \citet{hansenlee2021}. If the score is non-linear in $\theta$, or the
weight matrix depends on $\theta$, the DML estimator $\check\theta$ differs from both the DML estimator solving \eqref{eq:dml2first} and the debiased GMM estimator.\footnote{In Definition~\ref{def:dml2}, the first-step estimators $(\widehat\beta_k,\widehat\alpha_k)$ are
cross-fitted as in \citet[Definition~3.2]{ccddhnr2018}, whereas the finite-dimensional components $(\widehat\Gamma,\widehat\Omega,\widehat{\bar m})$, and hence $\widehat\mu$, are averages
over all folds. The nuisance parameters $\widehat\eta_k$ could instead be constructed for each fold $k$ from $(W_i)_{i\in I_k^c}$ alone, as in \citet{ccddhnr2018}. We keep the above definition in the main paper for its computational simplicity and its relationship with the debiased GMM estimator \eqref{eq:dgmm}, but we consider the usual fold-specific DML definition in Appendix \ref{app:general} with non-linear scores and efficient weights.}

\begin{assumption}[Regularity conditions]
\label{ass:rates}
Let $q>2$, $0<c_0\le c_1$ and $C<\infty$ be constants, and let $\{\delta_N\}$ and $\{\Delta_N\}$ be sequences of positive constants converging to zero with $\delta_N\ge N^{-[(1-2/q)\wedge1/2]}$.
For all $N\ge3$ and $P\in\mathcal P_N$, the following conditions hold. 
\begin{enumerate}[label=(\alph*),leftmargin=2.4em]
\item With probability at least $1-\Delta_N$, $(\widehat\beta_k,\widehat\alpha_k)\in\mathcal B_N\times\mathcal A_N$
for all $k\in[K]$.
\item The following conditions hold uniformly over
$(\beta,\alpha)\in\mathcal B_N\times\mathcal A_N$, 
\[
\begin{aligned}
&\big(\EP\|\psi(W,\theta_0,\beta,\alpha)\|^{q}\big)^{1/q}
\vee\big(\EP\|\psi^{a}(W,\beta,\alpha)\|^{q}\big)^{1/q}
\vee\big(\EP\|\zeta(W,\beta,\alpha)\|^{q}\big)^{1/q}\le C,\\
&\big\|\psi(W,\theta_0,\beta,\alpha)-\psi(W,\theta_0,\beta_0,\alpha_0)\big\|_{P,2}
\vee\big\|\psi^{a}(W,\beta,\alpha)-\psi^{a}(W,\beta_0,\alpha_0)\big\|_{P,2}\\
&\qquad\qquad
\vee\big\|\zeta(W,\beta,\alpha)-\zeta(W,\beta_0,\alpha_0)\big\|_{P,2}\le\delta_N.
\end{aligned}
\]
\item For all $(\beta,\alpha)\in\mathcal B_N\times\mathcal A_N$,
\[
\begin{aligned}
&\big\|\EP\big[\psi(W,\theta_0,\beta,\alpha)\big]- \EP[\psi(W,\theta_0,\beta_0,\alpha_0)]\big\|\\
&\qquad+\|\EP[\psi(W,\theta_0,\beta_0,\alpha_0)]\|\Big(\big\|\EP\big[\psi^{a}(W,\beta,\alpha)\big]-\EP[\psi^{a}(W,\beta_0,\alpha_0)]\big\|\\
&\qquad+\big\|\EP\big[\zeta(W,\beta,\alpha)\big]-\EP[\zeta(W,\beta_0,\alpha_0)]\big\|\Big)\le\delta_NN^{-1/2},\\
&\big\|\EP\big[\psi^{a}(W,\beta,\alpha)\big]-\EP[\psi^{a}(W,\beta_0,\alpha_0)]\big\|
+\big\|\EP\big[\zeta(W,\beta,\alpha)\big]-\EP[\zeta(W,\beta_0,\alpha_0)]\big\|\le\delta_N.
\end{aligned}
\]
\item The singular values of $G_0$ and the eigenvalues of $\Omega_0$ lie in $[c_0,c_1]$, and all
eigenvalues of $\EP[\psi^{*}(W,\theta_0,\eta_0)\psi^{*}(W,\theta_0,\eta_0)']$ are bounded below
by $c_0$.
\end{enumerate}
\end{assumption}

Assumption~\ref{ass:rates} imposes primitive conditions on $(\psi,\zeta)$ and the first-step estimators $(\widehat\beta_k,\widehat\alpha_k)$. Assumption~\ref{ass:rates}(a) is Assumption~3.2(a) of \citet{ccddhnr2018}.  Assumption~\ref{ass:rates}(b) corresponds to Assumption~3.2(b) and  (c) of \citet{ccddhnr2018} on the moment and the rate conditions, extended to the Jacobian and the weight matrix parts of \eqref{eq:psirdmlclosed}, which contribute to the limit distribution.

Assumption~\ref{ass:rates}(c) is required to control biases of the Jacobian and the weight matrix with the factor $\|\bar m_0\|$. Under correct specification, only the bias of $\psi$ at $\theta_0$ is restricted, while under misspecification the first-order biases of the Jacobian and of the weight matrix must also be
$o(N^{-1/2})$. A sufficient condition for (c) is that the maps $(\beta,\alpha)\mapsto\EP[\psi(W,\theta_0,\beta,\alpha)]$, $\EP[\psi^{a}(W,\beta,\alpha)]$ and $\EP[\zeta(W,\beta,\alpha)]$  have bounded second Gateaux derivatives and that the first step $(\widehat\beta_k,\widehat\alpha_k)$ converges at the familiar rate $o(N^{-1/4})$.  Assumption~\ref{ass:rates}(d) corresponds to Assumptions~3.1(e) and 3.2(d) of \citet{ccddhnr2018}, with the non-degeneracy condition imposed on the variance of $\psi^{*}$, which includes the Jacobian and weight matrix, instead of $\psi$.

\begin{theorem}[DML inference for misspecified semiparametric GMM with linear scores]
\label{thm:dml}
Suppose Assumptions~\ref{ass:model}, \ref{ass:linear} and \ref{ass:rates} hold. Then, the DML estimator $\check\theta$ of Definition~\ref{def:dml2} concentrates in a
$1/\sqrt N$ neighbourhood of $\theta_0$ and is approximately linear and centred Gaussian,
\begin{equation}
\label{eq:dmlclt}
\sqrt N\,\sigma^{-1}\big(\check\theta-\theta_0\big)
=-\frac{1}{\sqrt N}\sum_{i=1}^{N}\sigma^{-1}J_0^{-1}\psi^{*}(W_i,\theta_0,\eta_0)+O_P(\delta_N)
\ \dto\ \mathcal N\big(0,I_{d_\theta}\big),
\end{equation}
uniformly over $P\in\mathcal P_N$, where
$\sigma^2:=J_0^{-1}\EP[\psi^{*}(W,\theta_0,\eta_0)\psi^{*}(W,\theta_0,\eta_0)']J_0^{-1\prime}$, and $J_0=G_0'\Omega_0^{-1}G_0$. Moreover, let
\begin{equation}
\label{eq:varhat}
\widehat\sigma^2:=\widehat J_0^{-1}\Big(\frac1N\sum_{k=1}^{K}\sum_{i\in I_k}
\psi^{*}(W_i,\check\theta,\widehat\eta_k)\,\psi^{*}(W_i,\check\theta,\widehat\eta_k)'\Big)\widehat J_0^{-1\prime},
\qquad
\widehat J_0:=\widehat\Gamma'\widehat\Omega^{-1}\widehat\Gamma .
\end{equation}
If $\|\widetilde\theta-\theta_0\|=O_P(\delta_N)$ uniformly over $P\in\mathcal P_N$,  then $\widehat\sigma^2=\sigma^2+O_P(\delta_N)$, and, for a nominal level $a\in(0,1)$ and any
$\ell\in\R^{d_\theta}$ with $\|\ell\|=1$, the confidence interval
\begin{equation}
\label{eq:dmlci}
\mathrm{CI}_N:=\Big[\,\ell'\check\theta\pm\Phi^{-1}(1-a/2)\sqrt{\ell'\widehat\sigma^2\ell/N}\,\Big]
\end{equation}
is uniformly valid for $\ell'\theta_0$ over $P\in\mathcal P_N$,
\[
\sup_{P\in\mathcal P_N}\Big|\Pr\nolimits_{P}\big(\ell'\theta_0\in\mathrm{CI}_N\big)-(1-a)\Big|\longrightarrow0,
\]
where $\Phi$ is the standard normal distribution function.
\end{theorem}

\begin{remark}[Efficient weighting and non-linear scores]
\label{rem:nocorrection}
Theorem~\ref{thm:dml} focuses on one-step weighting, $\zeta(w,\theta,\beta,\alpha)=\zeta(w,\beta,\alpha)$, while Theorem~\ref{thm:ortho} only requires the weight function to satisfy Assumption~\ref{ass:model}(d). Under misspecification, Assumption~\ref{ass:model}(d) fails to hold for the efficient weight function $\zeta=\psi\psi'$. However, the orthogonal moment can be constructed without Assumption~\ref{ass:model}(d), by adding a correction term $\kappa(w,\theta,\eta)$ to $\psi^{*}$, where $\eta$ is enlarged to include any further nuisances, such that $\EP[\kappa(W,\theta_0,\eta_0)]=0$, the derivatives of $\EP[\kappa]$ vanish at the true values, and
\begin{equation}
\label{eq:phicondition}
\partial_{(\beta,\alpha)}\EP\big[\kappa(W,\theta_0,\eta_0)\big][\beta-\beta_0,\alpha-\alpha_0]
=G_0'\Omega_0^{-1}\partial_{(\beta,\alpha)}\EP[\zeta(W,\theta_0,\beta_0,\alpha_0)][\beta-\beta_0,\alpha-\alpha_0]\,\Omega_0^{-1}\bar m_0.
\end{equation}
Then, the proof of Theorem~\ref{thm:ortho} applies to $\psi^{*}+\kappa$. The right-hand side of
\eqref{eq:phicondition} is also proportional to $\bar m_0$, so $\kappa=0$  under correct specification or under Assumption~\ref{ass:model}(d).

Theorem~\ref{thm:dml} not only excludes the efficient weight $\zeta=\psi\psi'$, which is quadratic in
$\theta$, but also excludes moment functions that are non-linear in $\theta$. In both cases, we can construct the Neyman orthogonal moment $\psi^{*}$ and obtain valid inference for the DML estimator that solves the cross-fitted estimating equation for $\psi^{*}$, and Appendix \ref{app:general} provides the general results  (Theorem \ref{thm:general}). 
\end{remark}

%==================================================================
\section{The partially linear IV model}
\label{sec:pliv}

We now consider our main example, the partially linear IV model in Example~\ref{ex:pliv}. To keep
the notation simple, throughout this section we consider  a scalar endogenous regressor case, 
$d_\theta=1$. Let $W=(Y,D,X,Z)$, where
$D\in\R$ is the endogenous regressor, $X$ is a vector of controls, and $Z\in\R^{d_z}$ is a vector
of instrumental variables with $d_z>1$, so that the model is over-identified. 

The correctly specified model is
\begin{equation}
\label{eq:plivmodel2}
Y=D\,\theta_0+g_0(X)+U,\qquad \EP[U\mid X,Z]=0,
\end{equation}
for some true values $\theta_0\in\Theta$ and $g_0$, which we do not impose: the conditional moment restriction in \eqref{eq:plivmodel2} may fail, for instance because some instruments are invalid or
because the effect of $D$ is heterogeneous.

The nuisance parameters are the conditional expectations
\begin{equation}
\label{eq:plivbeta}
\beta_0:=(\ell_0,r_0),\quad \alpha_0:=z_0,
\qquad
\ell_0(X):=\EP[Y|X],\quad r_0(X):=\EP[D|X],\quad z_0(X):=\EP[Z|X],
\end{equation}
with residuals $\tilde Y=Y-\ell_0(X)\in\R$, $\tilde D=D-r_0(X)\in\R$ and
$\tilde Z=Z-z_0(X)\in\R^{d_z}$. Every component of \eqref{eq:plivbeta} is defined without
reference to $\theta_0$ or to the model \eqref{eq:plivmodel2}, as described in Remark~\ref{rem:whichresid}. The identifying moment function is
$m(w,\theta,\beta)=Z\big(Y-\ell(X)-(D-r(X))\theta\big)$. The model is misspecified when there is no solution to
\[
\EP[m(W,\theta,\beta_0)]=\EP\big[Z\big(Y-\ell_0(X)-(D-r_0(X))\theta\big)\big]=0
\]
over $\theta\in\Theta$. Adding the first-step influence function to $m$ gives the
orthogonal moment function \citep[equation~(4.8)]{ccddhnr2018},
\begin{equation}
\label{eq:plivmoment2}
\psi(w,\theta,\beta,\alpha)=\big(Z-z(X)\big)\big(Y-\ell(X)-(D-r(X))\theta\big)\in\R^{d_z},
\end{equation}
which is linear in $\theta$ in the sense of \eqref{eq:mlinear} with
\begin{equation}
\label{eq:plivab}
\psi^{b}(w,\beta,\alpha)=\big(Z-z(X)\big)\big(Y-\ell(X)\big),
\qquad
\psi^{a}(w,\beta,\alpha)=-\big(Z-z(X)\big)\big(D-r(X)\big) .
\end{equation}
Let $\Pi:=\EP[\tilde Z\tilde D]\in\R^{d_z}$. Then
\[
\bar{m}(\theta)=\EP\big[\tilde Z\tilde Y\big]-\Pi\,\theta,
\qquad
G(\theta)\equiv G_0=\EP\big[\psi^{a}(W,\beta_0,\alpha_0)\big]=-\Pi. 
\]
We consider the usual 2SLS weight matrix
\begin{equation}
\label{eq:plivzeta}
\zeta(w,\alpha)=\big(Z-z(X)\big)\big(Z-z(X)\big)',
\qquad
\Omega_0=S:=\EP\big[\tilde Z\tilde Z'\big]\in\R^{d_z\times d_z}.
\end{equation}
Then, $\theta_0$ solves the population first-order condition \eqref{eq:focfun},
\begin{equation}
\label{eq:plivtheta0}
\theta_0=\big(\Pi'S^{-1}\Pi \big)^{-1} \Pi'S^{-1}\,\EP\big[\tilde Z\tilde Y\big] ,
\qquad
\bar m_0=\EP\big[\tilde Z\tilde U\big],\quad \tilde U:=\tilde Y-\tilde D\theta_0 .
\end{equation}
As noted in Example~\ref{ex:pliv}, $\bar m(\theta)=0$ for some $\theta$ if and only if $\EP[\tilde Z\tilde Y]$
is proportional to $\Pi$, which for $d_z>1$ is a genuine restriction: provided every $\Pi_j\neq0$,
it requires that the $d_z$ instrument-specific IV estimands
$\EP[\tilde Z_j\tilde Y]/\EP[\tilde Z_j\tilde D]$, $j=1,\dots,d_z$, are all the same. If \eqref{eq:plivmodel2}
holds, then $\bar m_0=\EP[\tilde Z\,\EP[U|X,Z]]=0$ and the pseudo-true value
\eqref{eq:plivtheta0} is the structural $\theta_0$ of \eqref{eq:plivmodel2}.

It is straightforward to check the conditions in Assumption~\ref{ass:model} for the PLIV model. Assumption~\ref{ass:model}(a) holds when $\Pi\neq0$, the relevance
condition, and $\theta_0$ is interior to $\Theta$. Assumption~\ref{ass:model}(b) is also immediate. Because \eqref{eq:plivmoment2} is the identifying moment function plus its FSIF, it satisfies the
uniform orthogonality condition of Assumption~\ref{ass:model}(c) by
\citet[Theorems~1--3]{ceinr2022}, at every $\theta$; see also
\citet[Section~4.2]{ccddhnr2018}. The Gateaux derivatives in the directions of
$\beta=(\ell,r)$ and $\alpha=z$ are computed in the proof of Theorem~\ref{thm:pliv}; they vanish
because $\EP[\tilde Z|X]=0$ and $\EP[\tilde Y|X]=\EP[\tilde D|X]=0$ by the law of
iterated expectations, and do not use the conditional moment restriction in
\eqref{eq:plivmodel2}. Assumption~\ref{ass:model}(d) also holds with
\eqref{eq:plivzeta}.

\subsection{Robust scores and the implementation}
\label{ssec:plivscores}

Substituting \eqref{eq:plivab}--\eqref{eq:plivzeta} into \eqref{eq:psirdmlclosed} gives the misspecification-robust score
\begin{equation}
\label{eq:plivpsirdml}
\psi^{*}(w,\theta_0,\eta_0)=-\,\Pi'S^{-1}\tilde Z\tilde U
-\big(\tilde Z\tilde D-\Pi\big)'S^{-1}\bar m_0
+\Pi'S^{-1}\big(\tilde Z\tilde Z'-S\big)S^{-1}\bar m_0,
\end{equation}
with $J_0=G_0'\Omega_0^{-1}G_0=\Pi'S^{-1}\Pi$. The standard (correct-specification) Neyman orthogonal score of $\psi$ is
\begin{equation}
\label{eq:plivpsidml}
G_0'\Omega_0^{-1}\psi(w,\theta_0,\beta_0,\alpha_0)=-\,\Pi'S^{-1}\tilde Z\tilde U.
\end{equation}
For comparison, it is useful to see \eqref{eq:plivpsirdml} and \eqref{eq:plivpsidml} explicitly with full conditional expectations:
\begin{equation}
\label{eq:plivpop}
\begin{gathered}
\Pi=\EP\big[\big(Z-z_0(X)\big)\big(D-r_0(X)\big)\big],
\qquad
S=\EP\big[\big(Z-z_0(X)\big)\big(Z-z_0(X)\big)'\big],\\
\bar m_0=\EP\Big[\big(Z-z_0(X)\big)\big(Y-\ell_0(X)-(D-r_0(X))\theta_0\big)\Big],
\end{gathered}
\end{equation}
and the two orthogonal scores,
\begin{align}
G_0'\Omega_0^{-1}\psi(w,\theta_0,\beta_0,\alpha_0)
&=-\,\Pi'S^{-1}\big(Z-z_0(X)\big)\big(Y-\ell_0(X)-(D-r_0(X))\theta_0\big),\notag\\[4pt]
\label{eq:plivpsirdmlfull}
\psi^{*}(w,\theta_0,\eta_0)
&=-\,\Pi'S^{-1}\big(Z-z_0(X)\big)\big(Y-\ell_0(X)-(D-r_0(X))\theta_0\big)\notag\\
&\qquad-\Big[\big(Z-z_0(X)\big)\big(D-r_0(X)\big)-\Pi\Big]'S^{-1}\bar m_0\notag\\
&\qquad+\Pi'S^{-1}\Big[\big(Z-z_0(X)\big)\big(Z-z_0(X)\big)'-S\Big]S^{-1}\bar m_0 .
\end{align}
Every nuisance function appearing in \eqref{eq:plivpop}--\eqref{eq:plivpsirdmlfull} is a
conditional expectation, directly estimable by machine learning whether or not the model is
correctly specified, and $\Pi$, $S$ and $\bar m_0$ are simple population moments of the residuals. Both correction terms in \eqref{eq:plivpsirdmlfull} are multiplied by $\bar m_0$, so
it reduces to the standard orthogonal score under $\bar m_0 = 0$.

Now, we define the DML estimator in the PLIV setup following Definition~\ref{def:dml2}. Let 
$\widehat\beta_k=(\widehat\ell_k,\widehat r_k)$ and $\widehat\alpha_k=\widehat z_k$ be machine learning or nonparametric estimators of the regressions of $Y$, $D$ and $Z$ on $X$, fitted on $(W_i)_{i\in I_k^c}$, and for $i\in I_k$ form the cross-fitted residuals
\[
\widehat{\tilde Y}_i:=Y_i-\widehat\ell_k(X_i),
\qquad
\widehat{\tilde D}_i:=D_i-\widehat r_k(X_i),
\qquad
\widehat{\tilde Z}_i:=Z_i-\widehat z_k(X_i),
\qquad i\in I_k.
\]
The cross-fitted Jacobian,
weight matrix and moment in Definition~\ref{def:dml2}(c) are  the fold averages
\[
\widehat\Pi:=\frac1N\sum_{k=1}^{K}\sum_{i\in I_k}\widehat{\tilde Z}_i\widehat{\tilde D}_i,
\qquad
\widehat S:=\frac1N\sum_{k=1}^{K}\sum_{i\in I_k}\widehat{\tilde Z}_i\widehat{\tilde Z}_i',
\qquad
\overline{\tilde Z\tilde Y}:=\frac1N\sum_{k=1}^{K}\sum_{i\in I_k}\widehat{\tilde Z}_i\widehat{\tilde Y}_i.
\]

The DML estimator $\check\theta$ solves \eqref{eq:dml2eq},
$N^{-1}\sum_{k}\sum_{i\in I_k}\psi^{*}(W_i,\theta,\widehat\eta_k)=0$ with $\psi^{*}$ given by
\eqref{eq:psistarhat}. Because $\widehat\Gamma$ and $\widehat\Omega$ are  the pooled
averages of $\psi^{a}$ and $\zeta$ above, the last two terms of \eqref{eq:psistarhat} vanish, and
\eqref{eq:dml2eq} reduces to \eqref{eq:dml2first},
$\widehat\Gamma'\widehat\Omega^{-1}N^{-1}\sum_{k}\sum_{i\in I_k}\psi(W_i,\theta,\widehat\beta_k,\widehat\alpha_k)=0$.
This  gives
\begin{equation}
\label{eq:plivest}
\check\theta=\big( \widehat\Pi'\widehat S^{-1}\widehat\Pi\big)^{-1} \widehat\Pi'\widehat S^{-1}\,\overline{\tilde Z\tilde Y}, 
\end{equation}
which is exactly cross-fitted two-stage least squares on the residualized data. 

 The estimated scores and Jacobian are
\begin{equation}
\label{eq:plivpsihat}
\begin{gathered}
\widehat\psi^{*}_i=-\widehat\Pi'\widehat S^{-1}\widehat{\tilde Z}_i\widehat{\tilde U}_i
-\big(\widehat{\tilde Z}_i\widehat{\tilde D}_i-\widehat\Pi\big)'\widehat S^{-1}\widehat{\bar m}
+\widehat\Pi'\widehat S^{-1}\big(\widehat{\tilde Z}_i\widehat{\tilde Z}_i'-\widehat S\big)\widehat S^{-1}\widehat{\bar m},
\qquad i\in I_k,\\
\widehat J_0=\widehat\Pi'\widehat S^{-1}\widehat\Pi,
\qquad
\widehat\sigma^2=\widehat J_0^{-1}\,\frac1N\sum_{k=1}^{K}\sum_{i\in I_k}\big(\widehat\psi^{*}_i\big)^{2} \widehat J_0^{-1 \prime} ,
\end{gathered}
\end{equation}
where $\widehat{\bar m}=\frac1N\sum_{k=1}^{K}\sum_{i\in I_k}\widehat{\tilde Z}_i\widehat{\tilde U}_i$ is  the cross-fitted moment, and $\widehat{\tilde U}_i=\widehat{\tilde Y}_i-\widehat{\tilde D}_i\check\theta$ is a residual. 

The misspecification-robust orthogonal score does not change the estimator and changes only the
asymptotic variance, $\widehat\sigma^2$. The conventional DML standard error is obtained by dropping the last two terms of $\widehat\psi^{*}_i$ in \eqref{eq:plivpsihat}; the difference between the two standard errors is
entirely due to $\widehat{\bar m}$, and is asymptotically negligible when the moment conditions hold.

\subsection{Asymptotic distribution}
\label{ssec:plivasym}

\begin{assumption}[Regularity conditions for the PLIV]
\label{ass:pliv}
Let $q>4$ and $0<c_0\le c_1$, $C<\infty$ be constants and $\{\delta_N\},\{\Delta_N\}$
sequences of positive constants converging to zero. For all $P\in\mathcal P_N$, the following conditions hold.
\begin{enumerate}[label=(\alph*),leftmargin=2.4em]
\item $\theta_0$ in \eqref{eq:plivtheta0} exists, is unique and lies in the interior of $\Theta$, where $\Theta\subset\R$ is a compact interval.
\item $\|Y\|_{P,q}+\|D\|_{P,q}+\|Z\|_{P,q}\le C$.
\item $(\beta_0,\alpha_0)$ is given by \eqref{eq:plivbeta} and the conditional second moments are
bounded: $\|\EP[\tilde Y^2|X]\|_{P,\infty}\vee\|\EP[\tilde D^2|X]\|_{P,\infty}
\vee\|\EP[\|\tilde Z\|^2|X]\|_{P,\infty}\le C$.
\item $c_0\le\|\Pi\|\le c_1$, the eigenvalues of $S$ lie in $[c_0,c_1]$, and
$\EP\big[\psi^{*}(W,\theta_0,\eta_0)^{2}\big]\ge c_0$ for $\psi^{*}$ of \eqref{eq:plivpsirdml}.

\item With $P$-probability at least $1-\Delta_N$ the
cross-fitted estimators satisfy $(\widehat\beta_k,\widehat\alpha_k)\in\mathcal B_N\times\mathcal A_N$
for all $k$, where every $\beta=(\ell,r)\in\mathcal B_N$ and $\alpha=z\in\mathcal A_N$ obey
\[
\|\ell-\ell_0\|_{P,q}\vee\|r-r_0\|_{P,q}\vee\|z-z_0\|_{P,q}\le C,
\qquad
\|\ell-\ell_0\|_{P,2}\vee\|r-r_0\|_{P,2}\vee\|z-z_0\|_{P,2}\le\delta_N,
\]
and the rate conditions
\begin{equation}
\label{eq:plivrates}
\|z-z_0\|_{P,2}\big(\|\ell-\ell_0\|_{P,2}+\|r-r_0\|_{P,2}\big)\le\delta_N N^{-1/2}
\qquad\text{and}\qquad
\|z-z_0\|_{P,2}^{2}\le\delta_N N^{-1/2} .
\end{equation}
\end{enumerate}
\end{assumption}

Assumption~\ref{ass:pliv} differs from \citet[Assumption~4.2]{ccddhnr2018} in several aspects.
First of all, we do not impose the conditional moment restriction \eqref{eq:plivmodel2}, imposed in
their Assumption~4.2(a) as their equation~(4.5). We consider the over-identified model with a vector $Z$, $d_z>d_\theta$, and $\theta_0$ is the 2SLS estimands \eqref{eq:plivtheta0}  with residualization. The moment conditions in Assumption~\ref{ass:pliv}(b) are their Assumption~4.2(b), and the bounded conditional second moments in Assumption~\ref{ass:pliv}(c) are similar to their Assumption~4.2(d). Assumption~\ref{ass:pliv}(d) is similar to their Assumption~4.2(c), with the non-degeneracy here imposed on the variance of the
misspecification-robust score \eqref{eq:plivpsirdml} instead. The two norm bounds on the first step
in Assumption~\ref{ass:pliv}(e) are their Assumption~4.2(e). The first product rate in
\eqref{eq:plivrates} is exactly \citet[Assumption~4.2(e)(ii)]{ccddhnr2018}, while the second rate condition is
new, and is required by the weight matrix channel and can be dropped when
$\bar m_0=0$.

\begin{theorem}[DML inference in the misspecified partially linear IV model]
\label{thm:pliv}
Suppose Assumption~\ref{ass:pliv} holds. Then the DML estimator $\check\theta$ in
\eqref{eq:plivest} satisfies
\[
\sqrt N\,\sigma^{-1}\big(\check\theta-\theta_0\big)\ \dto\ \mathcal N(0,1),
\qquad
\sigma^{2}=\big(\Pi'S^{-1}\Pi\big)^{-1}\EP\big[\psi^{*}(W,\theta_0,\eta_0)^{2}\big]\big(\Pi'S^{-1}\Pi\big)^{-1} ,
\]
uniformly over $P\in\mathcal P_N$. Moreover $\widehat\sigma^{2}$ of \eqref{eq:plivpsihat} satisfies
$\widehat\sigma^2=\sigma^2+o_P(1)$, and, for a nominal level $a\in(0,1)$, the confidence interval based on $\check\theta$ is uniformly asymptotically valid for $\theta_0$,
\[
\sup_{P\in\mathcal P_N}\Big|\Pr\nolimits_{P}\big(\theta_0\in\check\theta\pm\Phi^{-1}(1-a/2)\,\widehat\sigma/\sqrt N\big)-(1-a)\Big|\longrightarrow0 .
\]
\end{theorem}

Theorem \ref{thm:pliv} extends the result of \citet[Theorem~4.2]{ccddhnr2018} to be robust to misspecification of the
moment condition \eqref{eq:plivmodel2}. Under the just-identified setup or under correct specification, $\bar m_0=0$, so that
$\psi^{*}(w,\theta_0,\eta_0)=-\Pi'S^{-1}\tilde Z\tilde U$ is the standard orthogonal score
\eqref{eq:plivpsidml}, and the conclusion reduces to \citet[Theorem~4.2]{ccddhnr2018}.

%==================================================================
\section{Local average treatment effects with multiple instruments}
\label{sec:late}

In this section, we consider estimation and inference on the local average treatment effects (LATEs) with a binary treatment variable and multiple binary instruments. This section parallels \citet[Section~5.2]{ccddhnr2018}, who treat a single binary instrument. 

The data are $W=(Y,D,X,Z)$ with outcome $Y\in\R$, binary treatment $D\in\{0,1\}$, covariates $X$,
and a vector of binary instruments $Z=(Z_1,\dots,Z_{d_z})'\in\{0,1\}^{d_z}$ with $d_z>1$.  Consider the functions $\beta_0:=\big(\ell_{0j},\nu_{0j}\big)_{j\le d_z}, \alpha_0:=\big(p_{0j}\big)_{j\le d_z}$, 
\begin{equation}
\label{eq:latebeta}
\begin{gathered}
\ell_{0j}(z,X):=\EP[Y\mid Z_j=z,X],\quad
\nu_{0j}(z,X):=\EP[D\mid Z_j=z,X],\quad
p_{0j}(X):=\Pr\nolimits_P(Z_j=1\mid X), 
\end{gathered}
\end{equation}
which are equations~(5.6)--(5.8) of \citet{ccddhnr2018}. These are again conditional expectations, which exist and are well defined for every $P\in\mathcal P_N$ with finite second moments, and are estimable by machine learning regression.

Under the assumptions of \citet{imbens1994} and \citet{frolich2007},\footnote{
Let $D(z_1,\dots,z_{d_z})$ denote the potential treatment, and $Y(d)$ the potential outcome under treatment $d$, so that $D=D(Z)$ and $Y=Y(D)$, and write $D_j(z):=D(z,Z_{-j})$ for the potential treatment when $Z_j$ is set to $z$. We can consider the following LATE conditions, which are imposed for each instrument $Z_j$, $j = 1, \dots, d_z$: 1) $\big(Y(1),Y(0),D(\cdot)\big)\perp Z\mid X$, and
$Z_1,\dots,Z_{d_z}$ are mutually independent conditional on $X$; 2) $\epsilon\le p_{0j}(X)=\Pr\nolimits_P(Z_j=1\mid X)\le1-\epsilon$ almost surely, for some $\epsilon\in(0,1/2)$; 3) $D(1,z_{-j})\ge D(0,z_{-j})$ almost surely for every value $z_{-j}$ of the other instruments; 4) $\Pr\nolimits_P\big(D_j(1)>D_j(0)\big)$ is bounded
away from zero.}  each instrument $Z_j$ identifies
\[
\theta^{(j)}:=\EP\big[Y(1)-Y(0)\mid D_j(1)>D_j(0)\big]
=\frac{\EP\big[\ell_{0j}(1,X)-\ell_{0j}(0,X)\big]}{\EP\big[\nu_{0j}(1,X)-\nu_{0j}(0,X)\big]} ,
\]
the LATE for the compliers of instrument $Z_j$. Taking this as the identifying moment function, $m_j(w,\theta,\beta)=\big[\ell_j(1,X)-\ell_j(0,X)\big]-\theta\big[\nu_j(1,X)-\nu_j(0,X)\big]$ for each instrument $j=1,\dots,d_z$, and adding its first-step influence function, which is the inverse propensity score correction with
additional nuisance parameter $\alpha=(p_j)_j$, gives
\begin{equation}
\label{eq:latepsi}
\psi=(\psi_1,\dots,\psi_{d_z})'\in\R^{d_z}, \qquad \psi_j(w,\theta,\beta,\alpha)=\psi^{Y}_j(w,\beta,\alpha)-\theta\,\psi^{D}_j(w,\beta,\alpha),
\end{equation}
where, as in \citet[Section~5.2]{ccddhnr2018},
\begin{equation}
\label{eq:latescores}
\begin{aligned}
\psi^{Y}_j(w,\beta,\alpha)&=\ell_j(1,X)-\ell_j(0,X)
+\frac{Z_j\big(Y-\ell_j(1,X)\big)}{p_j(X)}-\frac{(1-Z_j)\big(Y-\ell_j(0,X)\big)}{1-p_j(X)},\\
\psi^{D}_j(w,\beta,\alpha)&=\nu_j(1,X)-\nu_j(0,X)
+\frac{Z_j\big(D-\nu_j(1,X)\big)}{p_j(X)}-\frac{(1-Z_j)\big(D-\nu_j(0,X)\big)}{1-p_j(X)} .
\end{aligned}
\end{equation}
By construction \eqref{eq:latepsi} is linear in $\theta$ in the sense of \eqref{eq:mlinear}, with
\begin{equation}
\label{eq:lateab}
\psi^{b}(w,\beta,\alpha)=\psi^{Y}(w,\beta,\alpha),
\qquad
\psi^{a}(w,\beta,\alpha)=-\psi^{D}(w,\beta,\alpha) ,
\end{equation}
and  the population moment and Jacobian are
\begin{equation}
\label{eq:latemean}
\bar m(\theta)=\tau^{Y}-\tau^{D}\theta,
\qquad
G(\theta)\equiv G_0=\EP\big[\psi^{a}(W,\beta_0,\alpha_0)\big]=-\tau^{D}, 
\end{equation}
where $\tau^{Y}:=\EP\big[\psi^{Y}(W,\beta_0,\alpha_0)\big]\in\R^{d_z}$, $\tau^{D}:=\EP\big[\psi^{D}(W,\beta_0,\alpha_0)\big]\in\R^{d_z}$. Here $\tau^{Y}_j=\EP[\ell_{0j}(1,X)-\ell_{0j}(0,X)]$ and $\tau^{D}_j = \EP[\nu_{0j}(1,X)-\nu_{0j}(0,X)]$, so $\theta^{(j)}=\tau^{Y}_j/\tau^{D}_j$.

With a single instrument, $d_z=1$, the system is just-identified, $\bar m(\theta)=0$ always has a 
solution, and this is the case of \citet[Section~5.2]{ccddhnr2018}, in which misspecification in our sense does not arise. With $d_z>1$, the over-identifying restriction is that $\bar m(\theta)=0$
for some $\theta$, i.e., $\tau^{Y}$ is proportional to $\tau^{D}$, which  imposes
\[
\theta^{(1)}=\theta^{(2)}=\dots=\theta^{(d_z)}.
\]
In general, different instruments identify different groups of compliers, and the average treatment effect need not be the same in each complier group. This is the source of misspecification here; it is not a failure of instrument validity.

The population GMM first-order condition in \eqref{eq:focfun} is
$\tau^{D\prime}\Omega_0^{-1}(\tau^{Y}-\tau^{D}\theta_0)=0$, so that $\theta_0=\frac{\tau^{D\prime}\Omega_0^{-1}\tau^{Y}}{\tau^{D\prime}\Omega_0^{-1}\tau^{D}}$ with the weighting matrix $\Omega_0$. In order for the target parameter to be a convex combination of LATEs, we use the following diagonal weight matrix that does not depend on $\theta$,
\begin{equation}
\label{eq:latezeta}
\zeta(w)=\Omega_0=\operatorname{diag}(\varpi_1,\dots,\varpi_{d_z}),
\qquad \varpi_j>0.
\end{equation}
The weights $\varpi_j$ are assumed to be known throughout, with the identity matrix $\Omega_0=I_{d_z}$ being one  example. Another natural choice is $\zeta(w,\beta,\alpha)=\operatorname{diag}\big(\psi^{D}_j(w,\beta,\alpha)\big)$, for which $\Omega_0=\operatorname{diag}\big(\EP[\psi^{D}_j(W,\beta_0,\alpha_0)]\big)=\operatorname{diag}(\tau^{D}_j)$ and each instrument is weighted by its complier share, so that an instrument with a stronger first stage receives more weight. This weight satisfies Assumption~\ref{ass:model}(d), because each $\psi^{D}_j$ is an orthogonal score. With this choice, the GMM estimand is
\begin{equation}
\label{eq:latetheta0}
\theta_0=\frac{\tau^{D\prime}\Omega_0^{-1}\tau^{Y}}{\tau^{D\prime}\Omega_0^{-1}\tau^{D}}
=\sum_{j=1}^{d_z}\omega_j\,\theta^{(j)},
\qquad
\omega_j=\frac{\varpi_j^{-1}\big(\tau^{D}_j\big)^{2}}{\sum_{k=1}^{d_z}\varpi_k^{-1}\big(\tau^{D}_k\big)^{2}},
\qquad
\bar m_0=\tau^{Y}-\tau^{D}\theta_0 ,
\end{equation}
with $\bar m_{0,j}=\tau^{D}_j\big(\theta^{(j)}-\theta_0\big)$. Because $\Omega_0$ is diagonal and
positive,  the weights in \eqref{eq:latetheta0} are \emph{non-negative and sum to one}: the
pseudo-true value is a convex combination of the instrument-specific LATEs, an average treatment effect for a mixture of the complier groups, with instrument $j$ receiving weight proportional to its complier share when $\varpi_j \propto \tau^{D}_j$. \citet{chowkasahara2026} also characterize the estimand of over-identified IV-GMM as a combination of the instrument-specific Wald estimands whose weights sum to one, and show that the diagonality of the weighting matrix keeps every weight non-negative. For a general weighting matrix, a weight can be negative, which is why we do not consider the efficient weight or the
two-stage least squares weight; with an additively separable
specification in the covariates $X$, it need not be a convex combination of covariate-specific
effects either \citep{angristimbens1995,mtw2021,sloczynski2021,blandhol2022}.  In our setup, the scores \eqref{eq:latescores} are already nonparametric as the covariates $X$ enter only through the conditional expectations \eqref{eq:latebeta}, and can be estimated by machine learning with cross-fitting.\footnote{The diagonal of the two-stage least squares weight, $\Omega_0=\operatorname{diag}\big(\EP[\psi^{D}_j(W,\beta_0,\alpha_0)^{2}]\big)$ can be considered here. The weights $\omega_j$ remain positive; however, Assumption~\ref{ass:model}(d) fails and the correction of Remark~\ref{rem:nocorrection} is required. See Appendix \ref{app:general} for general results under efficient weights.}

Substituting \eqref{eq:lateab}--\eqref{eq:latezeta} into \eqref{eq:psirdmlclosed}, the misspecification-robust score is
\begin{equation}
\label{eq:latepsirdml}
\psi^{*}(w,\theta_0,\eta_0)
=-\,\tau^{D\prime}\Omega_0^{-1}\psi(w,\theta_0,\beta_0,\alpha_0)
-\big(\psi^{D}(w,\beta_0,\alpha_0)-\tau^{D}\big)'\Omega_0^{-1}\bar m_0 ,
\end{equation}
with $J_0=G_0'\Omega_0^{-1}G_0=\tau^{D\prime}\Omega_0^{-1}\tau^{D}$ and the Jacobian term in \eqref{eq:latepsirdml} remains whenever $\bar m_0\neq0$ with known weights. The standard Neyman orthogonal score of $\psi$ is
\[
G_0'\Omega_0^{-1}\psi(w,\theta_0,\beta_0,\alpha_0)
=-\,\tau^{D\prime}\Omega_0^{-1} \psi(w,\theta_0, \beta_0,\alpha_0)\in\R .
\]

The DML estimator of LATEs takes the following explicit form. For each
$k\in[K]$ and each instrument $j$, let $\widehat\ell_{j,k}$, $\widehat\nu_{j,k}$ and
$\widehat p_{j,k}$ be machine-learning or nonparametric estimators of $\ell_{0j}$, $\nu_{0j}$ and $p_{0j}$ based on $(W_i)_{i\in I_k^c}$, and for $i\in I_k$ form the cross-fitted scores
\[
\widehat\psi^{Y}_{i}:=\psi^{Y}\big(W_i,\widehat\beta_k,\widehat\alpha_k\big),
\qquad
\widehat\psi^{D}_{i}:=\psi^{D}\big(W_i,\widehat\beta_k,\widehat\alpha_k\big),
\qquad i\in I_k,
\]
with $\widehat\beta_k=(\widehat\ell_{j,k},\widehat\nu_{j,k})_j$ and
$\widehat\alpha_k=(\widehat p_{j,k})_j$. The cross-fitted Jacobian and moment of the DML estimator are the fold averages
\[
\widehat\tau^{Y}:=\frac1N\sum_{k=1}^{K}\sum_{i\in I_k}\widehat\psi^{Y}_{i},
\qquad
\widehat\tau^{D}:=\frac1N\sum_{k=1}^{K}\sum_{i\in I_k}\widehat\psi^{D}_{i},
\qquad
\widehat\Gamma=-\widehat\tau^{D},
\qquad
\widehat\Omega=\Omega_0 ,
\]
and we have 
\begin{equation}
\label{eq:lateest}
\check\theta
=\frac{\widehat\tau^{D\prime}\Omega_0^{-1}\widehat\tau^{Y}}{\widehat\tau^{D\prime}\Omega_0^{-1}\widehat\tau^{D}}
=\sum_{j=1}^{d_z}\widehat\omega_j\,\widehat\theta^{(j)},
\qquad
\widehat\theta^{(j)}=\frac{\widehat\tau^{Y}_j}{\widehat\tau^{D}_j},
\end{equation}
where $\widehat\omega_j$ is $\omega_j$ of \eqref{eq:latetheta0} with $\widehat\tau^{D}$ in place of
$\tau^{D}$. The estimated scores and Jacobian are
\begin{equation}
\label{eq:latepsihat}
\begin{gathered}
\widehat\psi^{*}_i=-\widehat\tau^{D\prime}\Omega_0^{-1}\big(\widehat\psi^{Y}_{i}-\check\theta\,\widehat\psi^{D}_{i}\big)
-\big(\widehat\psi^{D}_{i}-\widehat\tau^{D}\big)'\Omega_0^{-1}\widehat{\bar m},
\qquad i\in I_k,\\
\widehat J_0=\widehat\tau^{D\prime}\Omega_0^{-1}\widehat\tau^{D},
\qquad
\widehat\sigma^2=\widehat J_0^{-1}\,\frac1N\sum_{k=1}^{K}\sum_{i\in I_k}\big(\widehat\psi^{*}_i\big)^{2} \widehat J_0^{-1\prime},
\end{gathered}
\end{equation}
where  $\widehat{\bar m}:= \frac1N\sum_{k=1}^{K}\sum_{i\in I_k}(\widehat\psi^{Y}_{i} - \check\theta\widehat\psi^{D}_{i})$ is the cross-fitted moment. 

For each $j$, $\widehat\theta^{(j)}$ in \eqref{eq:lateest} is the single-instrument DML estimator of \citet[Theorem~5.2]{ccddhnr2018}, and the point estimate $\check\theta$ is a convex combination of these, with non-negative weights $\widehat\omega_j$.

\subsection{Asymptotic distribution}
\label{ssec:lateasym}

\begin{assumption}[Regularity conditions for the LATE]
\label{ass:late}
Let $q>2$, $\epsilon\in(0,1/2)$ and $0<c_0\le c_1$, $C<\infty$ be constants and
$\{\delta_N\},\{\Delta_N\}$ sequences of positive constants converging to zero with
$\delta_N\ge N^{-[(1-2/q)\wedge1/2]}$. For all
$P\in\mathcal P_N$, the following conditions hold.
\begin{enumerate}[label=(\alph*),leftmargin=2.4em]
\item $D\in\{0,1\}$, $Z\in\{0,1\}^{d_z}$, $\|Y\|_{P,q}\le C$, and
$\epsilon\le p_{0j}(X)\le1-\epsilon$ almost surely for every $j$.
\item $(\beta_0,\alpha_0)$ is given by \eqref{eq:latebeta} and
$\|\EP[Y^{2}\mid Z,X]\|_{P,\infty}\le C$.
\item  For every $j$,  $\tau^{D}_j$ and $\varpi_j$  lie in
$[c_0,c_1]$; and $\EP\big[\psi^{*}(W,\theta_0,\eta_0)^{2}\big]\ge c_0$.

\item $\theta_0$ in \eqref{eq:latetheta0} lies in the interior of $\Theta$, where $\Theta\subset\R$
is a compact interval.
\item With $P$-probability at least $1-\Delta_N$,  the cross-fitted estimators satisfy
$(\widehat\beta_k,\widehat\alpha_k)\in\mathcal B_N\times\mathcal A_N$ for all $k$, where all
$\beta=(\ell_j,\nu_j)_j\in\mathcal B_N$ and $\alpha=(p_j)_j\in\mathcal A_N$ satisfy, for all $j$ and
$z\in\{0,1\}$,
\[
\epsilon\le p_j(X)\le1-\epsilon
\quad\text{and}\quad
\nu_j(z,X)\in[0,1] \ \text{ a.s.},
\qquad
\|\ell_j(z,\cdot)-\ell_{0j}(z,\cdot)\|_{P,q}\le C,
\]
\[
\|\ell_j(z,\cdot)-\ell_{0j}(z,\cdot)\|_{P,2}
\vee\|\nu_j(z,\cdot)-\nu_{0j}(z,\cdot)\|_{P,2}
\vee\|p_j-p_{0j}\|_{P,2}\le\delta_N,
\]
and
\begin{equation}
\label{eq:laterates}
\|p_j-p_{0j}\|_{P,2}\Big(\textstyle\sum_{z\in\{0,1\}}\|\ell_j(z,\cdot)-\ell_{0j}(z,\cdot)\|_{P,2}
+\sum_{z\in\{0,1\}}\|\nu_j(z,\cdot)-\nu_{0j}(z,\cdot)\|_{P,2}\Big)\le\delta_N N^{-1/2}.
\end{equation}
\end{enumerate}
\end{assumption}

Assumption~\ref{ass:late} is \citet[Assumption~5.2]{ccddhnr2018} imposed on each instrument. With $d_z>1$, the system is over-identified and $\theta_0$ is the pseudo-true value \eqref{eq:latetheta0} rather than a common LATE. Accordingly, the non-degeneracy in Assumption~\ref{ass:late}(c) is imposed on the misspecification-robust score \eqref{eq:latepsirdml} rather than the conventional score $\psi$. The rate condition \eqref{eq:laterates} is the same as the product rate of
\citet[Assumption~5.2(g)]{ccddhnr2018}. The additional terms from the Jacobian and weight under misspecification are products of two different first-step errors, so the same product rate condition \eqref{eq:laterates} suffices.

\begin{theorem}[DML inference on a convex combination of LATEs]
\label{thm:late}
Suppose Assumption~\ref{ass:late} holds. Then the DML estimator $\check\theta$ of
\eqref{eq:lateest} satisfies
\[
\sqrt N\,\sigma^{-1}\big(\check\theta-\theta_0\big)\ \dto\ \mathcal N(0,1),
\qquad
\sigma^{2}=\frac{\EP\big[\psi^{*}(W,\theta_0,\eta_0)^{2}\big]}{\big(\tau^{D\prime}\Omega_0^{-1}\tau^{D}\big)^{2}} ,
\]
uniformly over $P\in\mathcal P_N$. Moreover $\widehat\sigma^{2}$ of \eqref{eq:latepsihat} satisfies
$\widehat\sigma^2=\sigma^2+o_P(1)$, and, for a nominal level $a\in(0,1)$, the confidence interval
$\mathrm{CI}_N=[\,\check\theta\pm\Phi^{-1}(1-a/2)\,\widehat\sigma/\sqrt N\,]$ is uniformly
asymptotically valid for $\theta_0$,
\[
\sup_{P\in\mathcal P_N}\Big|\Pr\nolimits_{P}\big(\theta_0\in\mathrm{CI}_N\big)-(1-a)\Big|\longrightarrow0 .
\]
\end{theorem}

Under the over-identifying restriction $\theta^{(1)}=\dots=\theta^{(d_z)}$, we have $\bar m_0=0$, the
second term of \eqref{eq:latepsirdml} drops out, and Theorem~\ref{thm:late} reduces to a pooled
version of \citet[Theorem~5.2]{ccddhnr2018}, where $\check\theta$ still estimates the
interpretable causal parameter \eqref{eq:latetheta0} at rate $\sqrt N$ with valid inference.

%==================================================================
\section{Simulations}
\label{sec:sim}

This section illustrates the main results and the DML inference in the partially linear IV model of Section~\ref{sec:pliv} using Monte Carlo simulations.  Our simulation designs closely follow the linear instrumental variables model of \citet{chs2015aer}, 
\begin{equation}
\label{eq:simmodel}
Y=\theta^{*}D+X'\theta_y+Z'\delta+\varepsilon,
\qquad
D=X'\theta_d+Z'\gamma+V,
\end{equation}
where $\theta^{*}$ is the structural coefficient on the endogenous regressor $D$, $X$ is the
vector of controls, $Z$ is the vector of instruments, and $\gamma$ is the first stage. Here, some instruments are allowed to have a direct effect $\delta$ on the outcome $Y$. The instruments are valid when $\delta=0$, in which case \eqref{eq:simmodel} is the partially linear IV model \eqref{eq:plivmodel2} with
$g_0(X)=X'\theta_y$.

The controls, instruments and errors are drawn as follows:
\[
\begin{gathered}
X\sim\mathcal N(0,\Sigma_x),\quad \Sigma_{x,kj}=0.5^{|j-k|},
\qquad
Z=(X_1,\dots,X_{d_z})'+\xi,\quad \xi\sim\mathcal N\big(0,0.25\,I_{d_z}\big),\\[4pt]
(\varepsilon,V)'\sim\mathcal N\!\left(0,\begin{pmatrix}1&0.6\\0.6&1\end{pmatrix}\right),
\end{gathered}
\]
with $X\in\R^{p_x}$, $p_x=200$, $d_z=5$, and $\xi$, $(\varepsilon,V)$ and $X$ mutually
independent. We set the structural coefficient as $\theta^{*}=1$,  control coefficients
$\theta_y=\theta_d=1/j^{2}$ for $j=1,\dots,p_x$,  and the
first stage as $\gamma_j=1/j$.

For Design A (invalid instruments) in \eqref{eq:simmodel}, the GMM estimand is $\theta_0=\theta^{*}+\gamma'\delta/\gamma'\gamma$. We report $\delta=0$, the correctly specified benchmark, and $\delta=(0,\,0.6,\,-0.6,\,0,\,0)'$, for which some instruments are invalid.  We also consider Design B (heterogeneous effects with valid instruments). This design keeps $\delta=0$ in \eqref{eq:simmodel}, so all instruments are exogenous and excluded. Instead, we let  $G\in\{A,B\}$ be a latent type with $\Pr(G=B)=1/2$, drawn independently of $(X,\xi,\varepsilon,V)$, and replace $\theta^{*}$ and $\gamma$ in \eqref{eq:simmodel} by $\theta_G$ and $\gamma_G$, with $\theta_A=\theta^{*}-\tau/2$, $\theta_B=\theta^{*}+\tau/2$ and $\gamma_A=0.6\,(1,1,1,0,0)'$, $\gamma_B=0.6\,(0,0,0,1,1)'$.   We report $\tau=0$, the correctly specified benchmark, which imposes a homogeneous treatment effect, $\theta_A=\theta_B$, and $\tau=2$, the heterogeneous case, for which $\theta_A=0$ and $\theta_B=2$.  Each instrument identifies the effect for the subpopulation it moves, a local average treatment effect, with $\theta^{(j)}=\theta_A$ for $j\le3$ and $\theta^{(j)}=\theta_B$ for $j\ge4$. The pseudo-true value is the convex combination $\theta_0=(3\theta_A+2\theta_B)/5=0.8$, and the over-identifying restrictions fail because those subpopulations differ, not because any instrument is invalid.

 \subsection{Results}
\label{ssec:simresults}
 
In both designs the DML2 estimator \eqref{eq:plivest} is the same cross-fitted two-stage least
squares estimator, with the cross-fitted residuals $(\widehat{\tilde Y}_i,\widehat{\tilde D}_i,\widehat{\tilde Z}_i)$. We report the two different standard errors $\widehat\sigma/\sqrt N$ with
$\widehat\sigma^2$ from \eqref{eq:plivpsihat}; the misspecification-robust standard error uses all three
terms of $\widehat\psi^{*}_i$, and the conventional one uses the first term alone. The nuisance parameters are the conditional expectations \eqref{eq:plivbeta} that are estimated by
the post-LASSO estimator of \citet{bellonichernozhukov2013} with the data-dependent penalty of \citet{bcch2012} and $K=5$
folds.  We report $2000$ replications with sample sizes $N\in\{500,1000,2000\}$ and nominal level $95\%$.\footnote{In an unreported simulation exercise, we also compare the parametric 2SLS estimator, either with a naive choice of controls among the $p_x=200$ candidates or without controls, with the 2SLS estimator with the first-step LASSO considered here. The parametric estimator is inconsistent for $\theta_0$, and its coverage is essentially zero even with the misspecification-robust standard errors.}

For each design and sample size, Table~\ref{tab:sim} reports the median absolute
deviation of $\check\theta$ from $\theta_0$, the standard deviation of $\check\theta$,
the average of each standard error, and the coverage of the corresponding $95\%$ confidence
interval for $\theta_0$.

\begin{table}[!ht]
\centering
\small
\caption{Monte Carlo results for the partially linear IV model.}
\label{tab:sim}
\begin{tabular}{lcccccc}
\toprule
 & & & \multicolumn{2}{c}{average s.e.} & \multicolumn{2}{c}{coverage} \\
\cmidrule(lr){4-5}\cmidrule(lr){6-7}
$N$ & MAD & SD($\check\theta$) & conventional & robust & conventional & robust \\
\midrule
\multicolumn{7}{l}{\textit{Design A, $\delta=0$ (correctly specified), $\theta_0=1$}}\\[2pt]
500 & 0.0517 & 0.0759 & 0.0744 & 0.0752 & 0.927 & 0.928 \\
1000 & 0.0366 & 0.0530 & 0.0525 & 0.0528 & 0.946 & 0.947 \\
2000 & 0.0263 & 0.0381 & 0.0370 & 0.0371 & 0.942 & 0.942 \\
\addlinespace
\multicolumn{7}{l}{\textit{Design A, $\delta\neq0$ (invalid instruments), $\theta_0=1.068$}}\\[2pt]
500 & 0.0656 & 0.0959 & 0.0786 & 0.0946 & 0.885 & 0.936 \\
1000 & 0.0457 & 0.0667 & 0.0551 & 0.0664 & 0.890 & 0.952 \\
2000 & 0.0323 & 0.0476 & 0.0389 & 0.0467 & 0.893 & 0.945 \\
\addlinespace
\multicolumn{7}{l}{\textit{Design B, $\tau=0$ (correctly specified), $\theta_0=1$}}\\[2pt]
500 & 0.0910 & 0.1342 & 0.1302 & 0.1382 & 0.924 & 0.933 \\
1000 & 0.0660 & 0.0949 & 0.0931 & 0.0959 & 0.938 & 0.948 \\
2000 & 0.0456 & 0.0667 & 0.0664 & 0.0674 & 0.949 & 0.953 \\
\addlinespace
\multicolumn{7}{l}{\textit{Design B, $\tau=2$ (heterogeneous effects), $\theta_0=0.8$}}\\[2pt]
500 & 0.2631 & 0.3847 & 0.3416 & 0.4033 & 0.938 & 0.962 \\
1000 & 0.1837 & 0.2742 & 0.2425 & 0.2812 & 0.921 & 0.957 \\
2000 & 0.1367 & 0.2021 & 0.1724 & 0.1980 & 0.909 & 0.949 \\
\bottomrule
\end{tabular}
\begin{flushleft}
\footnotesize
\textit{Notes.} For each design and sample size, we report the median of $|\check\theta-\theta_0|$ (MAD), and the standard deviation of the cross-fitted 2SLS estimator. We consider the two standard errors of \eqref{eq:plivpsihat}, and report average SE across simulations, and the coverage of the confidence intervals. $2000$ replications, $K=5$ folds, nominal level $95\%$. The first-step estimation of the nuisance parameters is based on the post-LASSO.
\end{flushleft}
\end{table}

First of all, the robust standard errors based on the misspecification-robust orthogonal scores work well even under correct specification. In both designs, the two standard errors and the coverage frequencies are almost the similar. Second, under misspecification, the robust standard error is about $15$--$20\%$ larger than the conventional one. However, the conventional standard error is much smaller than the standard deviation of the DML estimator $\check\theta$ while the robust one matches it very closely. Consequently, the conventional confidence interval undercovers at all sample sizes: in Design~A, at $0.885$, $0.890$ and $0.893$ against the robust one $0.936$, $0.952$ and $0.945$; in Design~B, at $0.938$, $0.921$ and $0.909$, while the robust coverage stays close to the nominal level, $0.962$,
$0.957$ and $0.949$.

\begin{figure}[!ht]
\centering
\includegraphics[width=\textwidth]{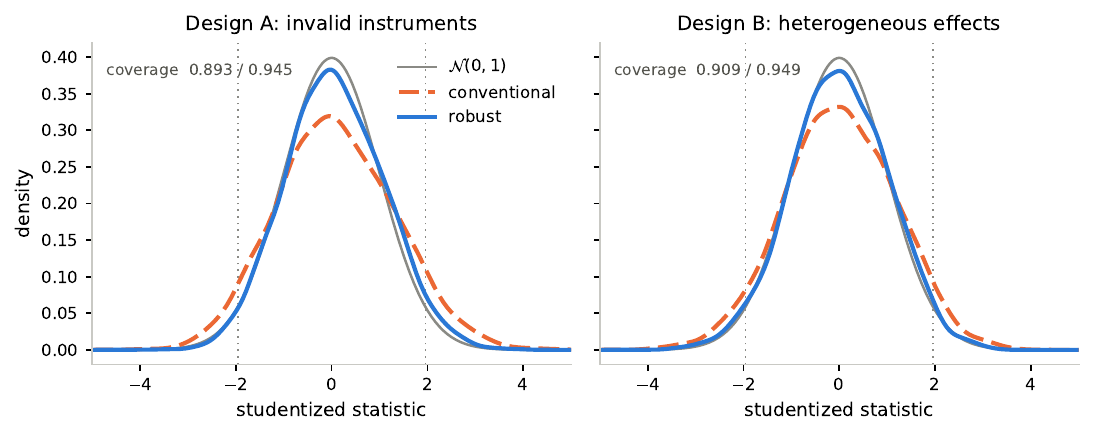}
\caption{Comparison of conventional and robust standard errors for the DML estimator. The density of the studentized statistic $(\check\theta-\theta_0)/\widehat{\mathrm{se}}$ at $N=2000$ under misspecification, over $2000$ replications. Left: Design~A, invalid instruments. Right: Design~B, heterogeneous effects.  The dashed  line: the conventional standard error;   solid line:  the misspecification-robust standard error of Theorem~\ref{thm:pliv}.}
\label{fig:sim}
\end{figure}

Figure~\ref{fig:sim} plots the density of the studentized statistic
$(\check\theta-\theta_0)/\widehat{\mathrm{se}}$ against the standard normal, separately
for the two standard errors (sample size $N=2000$). In both designs, the t-statistics with robust standard error closely follow the reference normal density, while the conventional one is more dispersed. This is in line with the coverage results in Table~\ref{tab:sim}.

%==================================================================
\section{Conclusion}
\label{sec:conc}

This paper develops Neyman orthogonal scores for GMM that remain valid when the moment conditions
fail to hold. Misspecification-robust orthogonal score has a closed form; the
conventional orthogonal score plus adjustment terms for the Jacobian and weighting
matrix, both proportional to the population moment at the GMM estimand and therefore negligible
under correct specification. The construction extends to semiparametric moment condition models,
where the moment functions depend on nonparametric or machine learning first steps, and the DML
estimator with cross-fitting based on the proposed score provides uniformly valid inference for the
GMM estimand. For linear scores with a one-step weighting
matrix, the point estimate is unchanged, and only the standard error has to be recomputed. In the
over-identified partially linear IV model, the estimator is cross-fitted two-stage least squares on
the residualized data, and the robust standard error uses only the same conditional expectations
and sample moments of the residuals; in the simulations the robust standard error is $15$--$20\%$ larger than the conventional one, and the conventional confidence interval undercovers. In the local average treatment effects model with several instruments, over-identification comes from
treatment-effect heterogeneity rather than from invalid instruments, a diagonal weighting matrix
makes the estimand a convex combination of the instrument-specific LATEs, and the robust variance
makes a confidence interval uniformly valid.

%==================================================================
\bibliographystyle{econometrica}

%==================================================================
\appendix
\section{Proofs}
\label{app:proofs}

\subsection{Proof of Lemma~\ref{lem:ccddhnr}}
\label{app:lem1}

\begin{proof}
The argument is the same as that of \citet[Lemma~2.3]{ccddhnr2018} and we provide it here for completeness. Since the nuisance parameters $\eta=(\beta',\vecop(\mu)')'$ are finite-dimensional, it suffices to show that the partial derivatives of
$\eta\mapsto\EP[\psi(W,\theta_0,\eta)]$ vanish at $\eta_0$. We note that $\EP[\psi(W,\theta_0,\beta,\mu_0)]=\mu_0\,\bar{m}(\gamma)$ with
$\gamma=(\theta_0',\beta')'$, so
\[
\partial_{\beta'}\EP[\psi(W,\theta_0,\eta_0)]=\mu_0G_{\beta,0}
=A'\Omega^{-1}G_{\beta,0}-A'\Omega^{-1}G_{\beta,0}\big(G_{\beta,0}'\Omega^{-1}G_{\beta,0}\big)^{-1}G_{\beta,0}'\Omega^{-1}G_{\beta,0}=0
\]
by \eqref{eq:mu0generic}.
Further, $\partial_{\vecop(\mu)'}\EP[\psi(W,\theta_0,\eta_0)]=\EP[m(W,\theta_0,\beta_0)]'\otimes
I_{d_\theta}=0$, and this gives \eqref{eq:lem1proof}. Since the derivatives are
exact zeros for arbitrary finite-dimensional deviations, orthogonality holds with the unrestricted realization set $T_N=T$.
\end{proof}

\subsection{Proof of Proposition~\ref{prop:ortho}}
\label{app:prop1}

\begin{proof}

The just-identified stacked system \eqref{eq:h}--\eqref{eq:hmoment} is a GMM problem of exactly the form considered in Section~\ref{ssec:nogmm} with a target $\theta$, a finite-dimensional nuisance $\lambda$, and a known moment function $h(w, \theta, \lambda)$ whose moment conditions hold at the (pseudo-)true parameter values. The stacked system is always just-identified rather than possibly over-identified for the original moment function $m$, which Lemma~\ref{lem:ccddhnr} also allows. 

First, $\EP[\psi^{*}(W,\theta_0,\eta_0)]=\mu_0\EP[h(W,\theta_0,\lambda_0)]=0$ by
\eqref{eq:hmoment}. The argument is similar to that of Lemma~\ref{lem:ccddhnr}, with the finite-dimensional nuisance $\eta=(\lambda',\vecop(\mu)')'$. For the $\lambda$-block, $\EP[\psi^{*}(W,\theta_0,\lambda,\mu_0)]=\mu_0\EP[h(W,\theta_0,\lambda)]$,
so $\partial_{\lambda'}\EP[\psi^{*}(W,\theta_0,\eta_0)]=\mu_0J_\lambda$, and by
\eqref{eq:mu0stack}
\[
\mu_0J_\lambda
=A_h'\Omega_h^{-1}J_\lambda-A_h'\Omega_h^{-1}J_\lambda\big(J_\lambda'\Omega_h^{-1}J_\lambda\big)^{-1}J_\lambda'\Omega_h^{-1}J_\lambda
=0 ,
\]
where $J_\lambda'\Omega_h^{-1}J_\lambda$ is nonsingular because $J_\lambda$ has full column
rank and $\Omega_h$ is positive definite. For the $\mu$-block, $\mu\mapsto\mu h$ is
linear, so $\partial_{\vecop(\mu)'}\EP[\psi^{*}(W,\theta_0,\eta_0)]
=\EP[h(W,\theta_0,\lambda_0)]'\otimes I_{d_\theta}=0$ by \eqref{eq:hmoment}. These give \eqref{eq:prop1proof} and complete the proof. 
\end{proof}

\subsection{Proof of Proposition~\ref{prop:closed}}
\label{app:prop2}

\begin{proof}
By using the identity $\vecop(ABC)=(C'\otimes A)\vecop(B)$, the first-order condition \eqref{eq:foc} has the following representations
\[
Q(\gamma) =\big(I_{d_\gamma}\otimes\bar m(\gamma)'\Omega(\gamma)^{-1}\big)\vecop G(\gamma)
=\big(\bar m(\gamma)'\otimes G(\gamma)'\big)\vecop\big(\Omega(\gamma)^{-1}\big).
\]
Then, using the chain rule and $\partial\vecop(\Omega^{-1})=-(\Omega^{-1}\otimes\Omega^{-1})\,\partial\vecop\Omega$ gives
\begin{equation}
\label{eq:Hexplicit}
H = \partial_{\gamma'}Q(\gamma)\big|_{\gamma=\gamma_0} = 
G_0'\Omega_0^{-1}G_0
+\big(I_{d_\gamma}\otimes\bar m_0'\Omega_0^{-1}\big)\dot G
-\big(\bar m_0'\Omega_0^{-1}\otimes G_0'\Omega_0^{-1}\big)\dot\Omega, 
\end{equation}
where $\bar m_0=\bar m(\gamma_0)$, $\dot G:=\partial_{\gamma'}\vecop G(\gamma)|_{\gamma_0}\in\R^{d_md_\gamma\times d_\gamma}$ and
$\dot\Omega:=\partial_{\gamma'}\vecop\Omega(\gamma)|_{\gamma_0}\in\R^{d_m^2\times d_\gamma}$.

Since $\EP[h(W,\theta,\lambda)]=\big(\vecop(\Gamma-G(\gamma))',\ \vecop(\Omega-\Omega(\gamma))',\
(\Gamma'\Omega^{-1}\bar m(\gamma))'\big)'$, the partial derivatives with respect to $\big(\gamma',\ \vecop(\Gamma)',\ \vecop(\Omega)'\big)'$ give
\begin{equation}
\label{eq:Jblocks}
J= \EP\big[\partial_{(\theta',\lambda')}h(W,\theta_0,\lambda_0)\big]= \big[\,J_\theta \ \ J_\lambda\,\big] = 
\begin{pmatrix}
-\dot G & I_{d_md_\gamma} & 0\\[2pt]
-\dot\Omega & 0 & I_{d_m^2}\\[2pt]
G_0'\Omega_0^{-1}G_0 & I_{d_\gamma}\otimes\bar m_0'\Omega_0^{-1} & -\big(\bar m_0'\Omega_0^{-1}\otimes G_0'\Omega_0^{-1}\big)
\end{pmatrix}.
\end{equation}

Next, we set \begin{equation}
\label{eq:mu0star}
M:=\Big[-\big(I_{d_\gamma}\otimes\bar m_0'\Omega_0^{-1}\big)\ \ \
\bar m_0'\Omega_0^{-1}\otimes G_0'\Omega_0^{-1}\ \ \ I_{d_\gamma}\Big],
\qquad
P:=\big[I_{d_\theta}\ \ -H_{\theta\beta}H_{\beta\beta}^{-1}\big],
\qquad
\mu_0^\star:=PM .
\end{equation}
By \eqref{eq:Hexplicit} and \eqref{eq:Jblocks}, $MJ=[\,H\ \ 0\ \ 0\,]$. Let $H = [H_\theta \quad H_\beta]$. Then, $\mu_0^\star J_\theta=PH_\theta=H_{\theta\theta}-H_{\theta\beta}H_{\beta\beta}^{-1}H_{\beta\theta}=J_0$ and  $\mu_0^\star J_\lambda=[\,PH_\beta\ \ 0\ \ 0\,]=0$, since $PH_\beta=H_{\theta\beta}-H_{\theta\beta}H_{\beta\beta}^{-1}H_{\beta\beta}=0$.

Now let $\mu_0$ be given by \eqref{eq:mu0stack}, so that
$\mu_0J_\lambda=0$. With $\mu_0J_\theta$ nonsingular, 
$(\mu_0J_\theta)^{-1}\mu_0J= (\mu_0J_\theta)^{-1}\mu_0 [J_{\theta} \ J_{\lambda}]= [\,I_{d_\theta}\ \ 0\,]=J_0^{-1}\mu_0^\star J$, and since $J$ is
nonsingular,
\begin{equation}
\label{eq:mu0rel}
(\mu_0J_\theta)^{-1}\mu_0=J_0^{-1}\mu_0^\star ,
\end{equation}
where both sides are the first $d_\theta$ rows of $J^{-1}$. Note that $\mu_0^\star$ is itself of
the form \eqref{eq:mu0stack} with $\Omega_h=I_{d_h}$ and $A_h=\mu_0^{\star\prime}$, because
$\mu_0^\star\big(I_{d_h}-J_\lambda(J_\lambda'J_\lambda)^{-1}J_\lambda'\big)=\mu_0^\star$ by
$\mu_0^\star J_\lambda=0$. For this choice $\mu_0=\mu_0^\star$ and $\mu_0J_\theta=J_0$, and the score is $s_\theta(w)-H_{\theta\beta}H_{\beta\beta}^{-1}s_\beta(w)$ itself.

By the definition of $M$ in \eqref{eq:mu0star}, and $h(w, \theta, \lambda)$ evaluated at  $(\theta_0,\lambda_0)$,  we have
\[
M\,h(w,\theta_0,\lambda_0)
=G_0'\Omega_0^{-1}m(w,\gamma_0)
+\big(\partial_{\gamma'}m(w,\gamma_0)-G_0\big)'\Omega_0^{-1}\bar m_0
-G_0'\Omega_0^{-1}\big(\zeta(w,\gamma_0)-\Omega_0\big)\Omega_0^{-1}\bar m_0
=s(w),
\]
so $\mu_0^\star h(w,\theta_0,\lambda_0)=P\,s(w)=s_\theta(w)-H_{\theta\beta}H_{\beta\beta}^{-1}s_\beta(w)$. Further,  $\EP[s(W)]=M\,\EP[h(W,\theta_0,\lambda_0)]=0$ by
\eqref{eq:hmoment}.
By \eqref{eq:mu0rel},
$\psi^{*}(w,\theta_0,\eta_0)=\mu_0h(w,\theta_0,\lambda_0)=(\mu_0J_\theta) J_0^{-1}\mu_0^\star h(w,\theta_0,\lambda_0)$,
which gives \eqref{eq:psiexplicit}. 

 The stacked estimator solves the sample just-identified system $N^{-1}\sum_{i=1}^Nh(W_i,\widehat\theta,\widehat\lambda)=0$
(Remark~\ref{rem:sample}), so under standard regularity conditions
\citep[Theorem~3.4]{neweymcfadden1994}
$\sqrt N((\widehat\theta{}',\widehat\lambda{}')'-(\theta_0',\lambda_0')')
=-J^{-1}N^{-1/2}\sum_{i=1}^Nh(W_i,\theta_0,\lambda_0)+o_P(1)$. By \eqref{eq:mu0rel}, its first
$d_\theta$ rows are
$-(\mu_0J_\theta)^{-1}N^{-1/2}\sum_{i=1}^N\psi^{*}(W_i,\theta_0,\eta_0)+o_P(1)
=-J_0^{-1}N^{-1/2}\sum_{i=1}^N \big(s_\theta(W_i)-H_{\theta\beta}H_{\beta\beta}^{-1}s_\beta(W_i)\big)+o_P(1)$, which is
\eqref{eq:if} and it does not depend on the choice of
$(A_h,\Omega_h)$. 
\end{proof}

\subsection{Proof of Theorem~\ref{thm:ortho}}
\label{app:thmortho}

\begin{proof}
The stacked equation \eqref{eq:hfun} is \eqref{eq:h} with $\gamma$ replaced by $\theta$, $m$ replaced by
$\psi(\cdot,\cdot,\beta,\alpha)$ and no finite-dimensional $\beta$ in $\lambda$ for any $\beta\in\mathcal B$ and $\alpha\in\mathcal A$. Hence the calculations in the proof of  Proposition~\ref{prop:closed} apply with $d_\gamma$ replaced by $d_\theta$, $P=I_{d_\theta}$, and
$\bar m$, $G$, $\Omega$ and $Q$ in \eqref{eq:focfun} and \eqref{eq:nuGfun}, where
$\EP[\psi(W,\theta,\beta_0,\alpha_0)]=\bar m(\theta)$ holds by Assumption \ref{ass:model}(c). In particular, $J$ is
given by \eqref{eq:Jblocks}, so that $J_\lambda$ has full column rank and $\mu_0$ of the form
\eqref{eq:mu0stack}  is well defined. $J_0=\partial_{\theta'}Q(\theta_0)$ is
similarly calculated as in \eqref{eq:Hexplicit}, which is \eqref{eq:J0fun}.  Further $\mu_0^\star=M$ of \eqref{eq:mu0star}
satisfies $MJ=[\,J_0\ \ 0\,]$; and, as in the proof of Proposition~\ref{prop:closed},
\eqref{eq:mu0rel} gives $\mu_0=\mu_0J_\theta J_0^{-1}M$ for any $\mu_0$ of the form
\eqref{eq:mu0stack} with $\mu_0J_\theta$ nonsingular. The normalization $\mu_0J_\theta=J_0$  yields $\mu_0=M$; $M$ itself is of the form \eqref{eq:mu0stack} with
$A_h=M'$ and $\Omega_h=I_{d_h}$, and more generally, since \eqref{eq:mu0stack} is linear in
$A_h'$, replacing $A_h'$ by $J_0(\mu_0J_\theta)^{-1}A_h'$ premultiplies $\mu_0$ by
$J_0(\mu_0J_\theta)^{-1}$. For any other choice of $(A_h,\Omega_h)$, $\psi^{*}$ is premultiplied by
the nonsingular matrix $\mu_0J_\theta J_0^{-1}$, which changes neither the solution of the
estimating equation nor the influence function $-(\mu_0J_\theta)^{-1}\psi^{*}$.
Moreover, a calculation similar to that in Appendix~\ref{app:prop2} gives, for all $\theta\in\Theta$, $\beta\in\mathcal B$ and $\alpha\in\mathcal A$,
\begin{equation}
\label{eq:muhfun}
\begin{aligned}
M\,h(w,\theta,\lambda_0,\beta,\alpha)
&=G_0'\Omega_0^{-1}\psi(w,\theta,\beta,\alpha)
+\big(\partial_{\theta'}\psi(w,\theta,\beta,\alpha)-G_0\big)'\Omega_0^{-1}\bar m_0\\
&\qquad-G_0'\Omega_0^{-1}\big(\zeta(w,\theta,\beta,\alpha)-\Omega_0\big)\Omega_0^{-1}\bar m_0 .
\end{aligned}
\end{equation}
Setting $(\beta,\alpha)=(\beta_0,\alpha_0)$ gives \eqref{eq:psirdmlclosed}, and
$\partial_{\theta'}\EP[\psi^{*}(W,\theta,\eta_0)]|_{\theta_0}=\mu_0J_\theta=J_0$. Further,
\eqref{eq:rdmlmoment} follows from
$\EP[\psi^{*}(W,\theta_0,\eta_0)]=\mu_0\EP[h(W,\theta_0,\lambda_0,\beta_0,\alpha_0)]=0$ by
\eqref{eq:hmomentfun}.

For orthogonality, the derivatives with respect to the finite-dimensional components
$(\lambda,\mu)$ are $[\,\mu_0J_\lambda,\ \EP[h(W,\theta_0,\lambda_0,\beta_0,\alpha_0)]'\otimes I_{d_\theta}\,]=0$
exactly as in the proof of Proposition~\ref{prop:ortho}, by \eqref{eq:mu0stack} and
\eqref{eq:hmomentfun}. For $(\beta,\alpha)$, write
$D_f:=\partial_{(\beta,\alpha)}\EP[f(W,\theta_0,\beta_0,\alpha_0)][\beta-\beta_0,\alpha-\alpha_0]$ for
$(\beta,\alpha)\in\mathcal B_N\times\mathcal A_N$, which exists by Assumption~\ref{ass:model}(b).
By \eqref{eq:muhfun},
\[
\partial_{(\beta,\alpha)}\EP\big[\psi^{*}(W,\theta_0,\lambda_0,\mu_0,\beta_0,\alpha_0)\big][\beta-\beta_0,\alpha-\alpha_0]
=G_0'\Omega_0^{-1}D_\psi+D_{\partial_{\theta'}\psi}'\,\Omega_0^{-1}\bar m_0
-G_0'\Omega_0^{-1}D_\zeta\,\Omega_0^{-1}\bar m_0 .
\]
Here $D_\psi=0$ by Assumption~\ref{ass:model}(c) at $\theta_0$, and $D_\zeta=0$ by Assumption~\ref{ass:model}(d).
For the Jacobian channel, the interchange of derivatives in Assumption~\ref{ass:model}(b) states
that, for all $(\beta,\alpha)\in\mathcal B_N\times\mathcal A_N$,
\begin{equation}
\label{eq:interchange}
\begin{aligned}
&\partial_{(\beta,\alpha)}\EP\big[\partial_{\theta'}\psi(W,\theta_0,\beta_0,\alpha_0)\big][\beta-\beta_0,\alpha-\alpha_0]\\
&\qquad=\partial_{\theta'}\Big(\partial_{(\beta,\alpha)}\EP\big[\psi(W,\theta,\beta_0,\alpha_0)\big][\beta-\beta_0,\alpha-\alpha_0]\Big)\Big|_{\theta=\theta_0} .
\end{aligned}
\end{equation}
Since \eqref{eq:unifortho} holds for every $\theta\in\Theta_0$ and $\theta_0$ is interior to
$\Theta_0$, the inner derivative on the right-hand side of \eqref{eq:interchange}
vanishes identically on a neighbourhood of $\theta_0$ and
\[
D_{\partial_{\theta'}\psi}
=\partial_{\theta'}\Big(\partial_{(\beta,\alpha)}\EP\big[\psi(W,\theta,\beta_0,\alpha_0)\big][\beta-\beta_0,\alpha-\alpha_0]\Big)\Big|_{\theta=\theta_0}=0 .
\]
This shows that derivatives vanish for every
$\eta\in T_N$, so $\psi^{*}$ is Neyman orthogonal at $(\theta_0,\eta_0)$ with respect to $T_N$.
\end{proof}

\subsection{Proof of Theorem~\ref{thm:dml}}
\label{app:thmdml}

\begin{proof}
Fix any sequence $\{P_N\}_{N\ge1}$ with $P_N\in\mathcal P_N$ for every $N$. Since the
sequence is arbitrary, it suffices to establish the conclusions under $P_N$, as in the proof of
\citet[Theorem~3.1]{ccddhnr2018}. We write $O_P$ for $O_{P_N}$ throughout, and all $O_P(\cdot)$ statements below hold uniformly over $P\in\mathcal P_N$, because the bounds
depend only on the constants in Assumptions~\ref{ass:model}--\ref{ass:rates}, which hold for
every $P\in\mathcal P_N$. Let $\mathcal E_N$ be the event, of probability at least $1-\Delta_N$, on which
$(\widehat\beta_k,\widehat\alpha_k)\in\mathcal B_N\times\mathcal A_N$ for all $k\in[K]$. Because
$\Delta_N\to0$, $\Pr_{P}(\mathcal E_N)\to1$, so every bound established below on $\mathcal E_N$
holds unconditionally up to a term that is $o_P(1)$. Throughout, $f$ denotes any one of the three functions $f(w,\beta,\alpha)\in\big\{\,\psi(w,\theta_0,\beta,\alpha),  \psi^{a}(w,\beta,\alpha), \zeta(w,\beta,\alpha)\,\big\},$
and we have that
\[
\EP\big[\psi(W,\theta_0,\beta_0,\alpha_0)\big]=\bar m_0,
\qquad
\EP\big[\psi^{a}(W,\beta_0,\alpha_0)\big]=G_0,
\qquad
\EP\big[\zeta(W,\beta_0,\alpha_0)\big]=\Omega_0.
\]
We also define the first-step bias and the centred cross-fitted average as
\[
\begin{aligned}
b_f(\beta,\alpha)&:=\EP\big[f(W,\beta,\alpha)\big]-\EP\big[f(W,\beta_0,\alpha_0)\big],\\
\Delta_f&:=\frac1N\sum_{k=1}^{K}\sum_{i\in I_k}f(W_i,\widehat\beta_k,\widehat\alpha_k)-\EP\big[f(W,\beta_0,\alpha_0)\big],
\end{aligned}
\]
and abbreviate $\Delta_\psi$, $\Delta_a$ and $\Delta_\zeta$ for the different choices of $f$, respectively. By Assumption~\ref{ass:linear},
$\partial_{\theta'}\psi(w,\theta,\beta,\alpha)=\psi^{a}(w,\beta,\alpha)$ does not depend on $\theta$
and $\psi(w,\theta,\beta,\alpha)=\psi(w,\theta_0,\beta,\alpha)+\psi^{a}(w,\beta,\alpha)(\theta-\theta_0)$,
so the cross-fitted quantities of Definition~\ref{def:dml2} satisfy
\[
\widehat\Gamma=G_0+\Delta_a,
\qquad
\widehat\Omega=\Omega_0+\Delta_\zeta,
\qquad
\widehat{\bar m}=\bar m_0+\Delta_\psi+\widehat\Gamma\big(\widetilde\theta-\theta_0\big).
\]
The sample means of the last two terms of \eqref{eq:psistarhat} vanish and
so the DML2 estimator $\check\theta$ solves \eqref{eq:dml2first} and, whenever
$\widehat\Gamma'\widehat\Omega^{-1}\widehat\Gamma$ is nonsingular,
\begin{equation}
\label{eq:thm2rep}
\check\theta-\theta_0=-\big(\widehat\Gamma'\widehat\Omega^{-1}\widehat\Gamma\big)^{-1}
\widehat\Gamma'\widehat\Omega^{-1}\big(\bar m_0+\Delta_\psi\big) .
\end{equation}

For each $f$  $\in \big\{\psi(w,\theta_0,\beta,\alpha),  \psi^{a}(w,\beta,\alpha), \zeta(w,\beta,\alpha)\big\}$,
\begin{equation}
\label{eq:thm2decomp}
\Delta_f=\frac1N\sum_{i=1}^{N}\Big(f(W_i,\beta_0,\alpha_0)-\EP\big[f(W,\beta_0,\alpha_0)\big]\Big)
+\bar b_f+O_P\big(\delta_NN^{-1/2}\big),
\qquad
\bar b_f:=\frac1K\sum_{k=1}^{K}b_f\big(\widehat\beta_k,\widehat\alpha_k\big).
\end{equation}
The first term is $O_P(N^{-1/2})$ by Chebyshev's inequality and Assumption~\ref{ass:rates}(b). The
remainder is 
$N^{-1}\sum_{i\in I_k}\{f(W_i,\widehat\beta_k,\widehat\alpha_k)-f(W_i,\beta_0,\alpha_0)-b_f(\widehat\beta_k,\widehat\alpha_k)\}$.
Because $(\widehat\beta_k,\widehat\alpha_k)$ is constructed from $(W_j)_{j\in I_k^c}$ alone, conditionally on $(W_j)_{j\in I_k^c}$, its
summands are independent with mean zero and, conditional on
$\mathcal E_N$,  second moments $\|f(W,\widehat\beta_k,\widehat\alpha_k)-f(W,\beta_0,\alpha_0)\|_{P,2}^{2}\le\delta_N^{2}$ by
Assumption~\ref{ass:rates}(b). Conditionally on $(W_j)_{j\in I_k^c}$, the remainder term in \eqref{eq:thm2decomp} is $O_P(\delta_NN^{-1/2})$ by Chebyshev's inequality, and unconditionally by Lemma~6.1 of \citet{ccddhnr2018}. 

Assumptions~\ref{ass:rates}(b) and (c) give, on $\mathcal E_N$,
\begin{equation}
\label{eq:thm2orders}
\begin{aligned}
&\|\Delta_a\|+\|\Delta_\zeta\|=O_P(\delta_N),
\qquad
\|\Delta_\psi\|+\|\bar m_0\|\big(\|\Delta_a\|+\|\Delta_\zeta\|\big)=O_P(N^{-1/2}),\\
&\big\|\bar b_{\psi}\big\|+\|\bar m_0\|\big(\|\bar b_{\psi^{a}}\|+\|\bar b_\zeta\|\big)\le\delta_NN^{-1/2},
\end{aligned}
\end{equation}
where $\|\bar m_0\|\le C$ by Assumption~\ref{ass:rates}(b). By \eqref{eq:thm2orders} and
Assumption~\ref{ass:rates}(d), $\widehat\Omega$ is nonsingular with probability approaching one and
$\widehat\Omega^{-1}=\Omega_0^{-1}-\Omega_0^{-1}\Delta_\zeta\Omega_0^{-1}+O_P(\|\Delta_\zeta\|^{2})$.
Expanding and using $G_0'\Omega_0^{-1}\bar m_0=0$ from \eqref{eq:focfun},
\begin{equation}
\label{eq:thm2lin}
\widehat\Gamma'\widehat\Omega^{-1}\big(\bar m_0+\Delta_\psi\big)
=G_0'\Omega_0^{-1}\Delta_\psi+\Delta_a'\Omega_0^{-1}\bar m_0-G_0'\Omega_0^{-1}\Delta_\zeta\Omega_0^{-1}\bar m_0+R,
\end{equation}
where the remainder term $R=O_P(\delta_N)O_P(N^{-1/2})$ by \eqref{eq:thm2orders}.

Substituting \eqref{eq:thm2decomp} into  \eqref{eq:thm2lin}, we have 
\[
\widehat\Gamma'\widehat\Omega^{-1}\big(\bar m_0+\Delta_\psi\big)
=\frac1N\sum_{i=1}^{N}\psi^{*}(W_i,\theta_0,\eta_0)+O_P\big(\delta_NN^{-1/2}\big), 
\]
where the bias parts are bounded by  \eqref{eq:thm2orders} together with 
Assumption~\ref{ass:rates}(d).  Also $\widehat\Gamma'\widehat\Omega^{-1}\widehat\Gamma=J_0+O_P(\delta_N)$, whose singular values are
bounded away from zero by Assumption~\ref{ass:rates}(d). Chebyshev's inequality and the second-moment bound established below give $N^{-1/2}\sum_i\psi^{*}(W_i,\theta_0,\eta_0)=O_P(1)$. Therefore \eqref{eq:thm2rep} gives
\[
\sqrt N\big(\check\theta-\theta_0\big)=-J_0^{-1}\frac{1}{\sqrt N}\sum_{i=1}^{N}\psi^{*}(W_i,\theta_0,\eta_0)+O_P(\delta_N).
\]
By \eqref{eq:psirdmlclosed}, $\psi^{*}(\cdot,\theta_0,\eta_0)$ is a linear combination of
$\psi(\cdot,\theta_0,\beta_0,\alpha_0)$, $\psi^{a}(\cdot,\beta_0,\alpha_0)$ and
$\zeta(\cdot,\beta_0,\alpha_0)$ with coefficients bounded by Assumption~\ref{ass:rates}(d), and thus  $\psi^{*}(\cdot,\theta_0,\eta_0)$ has
bounded $q$-th moment with $q>2$ by Assumption~\ref{ass:rates}(b), while the eigenvalues of
$\EP[\psi^{*}\psi^{*\prime}]$ are bounded below by Assumption~\ref{ass:rates}(d). Then the Lindeberg-Feller central limit theorem, whose condition is implied by Lyapunov's condition, together with the Cram\'er-Wold device, yields \eqref{eq:dmlclt} uniformly over $P\in\mathcal P_N$.

Now, we show the consistency of the variance estimator. First of all, $\widehat J_0=\widehat\Gamma'\widehat\Omega^{-1}\widehat\Gamma=J_0+O_P(\delta_N)$ and
$\widehat{\bar m}-\bar m_0=\Delta_\psi+\widehat\Gamma(\widetilde\theta-\theta_0)=O_P(\delta_N)$ by
$\|\widetilde\theta-\theta_0\|=O_P(\delta_N)$ and $\delta_N\ge N^{-1/2}$.  Let $\psi^{*}_{0,i}:=\psi^{*}(W_i,\theta_0,\eta_0)$ and $\widehat\psi^{*}_i:=\psi^{*}(W_i,\check\theta,\widehat\eta_k)$ for $i\in I_k$. By
\eqref{eq:psistarhat} and $\psi(W_i,\check\theta,\widehat\beta_k,\widehat\alpha_k)=\psi(W_i,\theta_0,\widehat\beta_k,\widehat\alpha_k)+\psi^{a}(W_i,\widehat\beta_k,\widehat\alpha_k)(\check\theta-\theta_0)$, each term of $\widehat\psi^{*}_i-\psi^{*}_{0,i}$ is either one of the differences $\widehat\Gamma-G_0$, $\widehat\Omega^{-1}-\Omega_0^{-1}$, $\widehat{\bar m}-\bar m_0$, $\check\theta-\theta_0$, all
$O_P(\delta_N)$, multiplied by $f(W_i,\widehat\beta_k,\widehat\alpha_k)$, or a
bounded matrix multiplied by $f(W_i,\widehat\beta_k,\widehat\alpha_k)-f(W_i,\beta_0,\alpha_0)$, for
$f$ in $\big\{\psi(w,\theta_0,\beta,\alpha),  \psi^{a}(w,\beta,\alpha), \zeta(w,\beta,\alpha)\big\}$. Conditional on $\mathcal E_N$, the second moments of
$f(W_i,\widehat\beta_k,\widehat\alpha_k)$ and of
$f(W_i,\widehat\beta_k,\widehat\alpha_k)-f(W_i,\beta_0,\alpha_0)$ are bounded by $C^{2}$ and by
$\delta_N^{2}$ by Assumption~\ref{ass:rates}(b), so the conditional Markov inequality and Lemma~6.1 of
\citet{ccddhnr2018} give
$N^{-1}\sum_{i=1}^{N}\|\widehat\psi^{*}_i-\psi^{*}_{0,i}\|^{2}=O_P(\delta_N^{2})$. Since
$N^{-1}\sum_{i=1}^{N}\|\psi^{*}_{0,i}\|^{2}=O_P(1)$, the Cauchy--Schwarz inequality gives
$N^{-1}\sum_{i=1}^{N}(\widehat\psi^{*}_i\widehat\psi^{*\prime}_i-\psi^{*}_{0,i}\psi^{*\prime}_{0,i})=O_P(\delta_N)$,
and 
\[
\frac1N\sum_{i=1}^{N}\psi^{*}_{0,i}\psi^{*\prime}_{0,i}-\EP\big[\psi^{*}\psi^{*\prime}\big]
=O_P\big(N^{-[(1-2/q)\wedge1/2]}\big)=O_P(\delta_N),
\]
by Chebyshev's inequality and the H\"older inequality with $q\ge4$, and by the von Bahr-Esseen
inequality with $p=q/2$ when $q\in(2,4)$, similarly as in the proof of
\citet[Theorem~3.2, equation~(A.25)]{ccddhnr2018}. Hence $\widehat\sigma^2=\sigma^2+O_P(\delta_N)$,
and uniform validity of the confidence intervals follows from \eqref{eq:dmlclt}. 
\end{proof}

\subsection{Proof of Theorem~\ref{thm:pliv}}
\label{app:pliv}

\begin{proof}
It suffices to verify Assumptions~\ref{ass:model}, \ref{ass:linear} and \ref{ass:rates}; the
conclusions then follow from Theorems~\ref{thm:ortho} and \ref{thm:dml}. First of all, Assumption~\ref{ass:linear} holds as in \eqref{eq:plivab}, and \eqref{eq:plivzeta} does not depend on $\theta$. For
$\beta=(\ell,r)\in\mathcal B_N$ and $\alpha=z\in\mathcal A_N$ write
\begin{equation}
\label{eq:plivdev}
\delta_\ell:=\ell-\ell_0,
\qquad
\delta_r:=r-r_0,
\qquad
\delta_z:=z-z_0,
\qquad
e(\theta):=\delta_\ell-\delta_r\theta
\end{equation}
for the deviations from the true values \eqref{eq:plivbeta}, and put $U(\theta):=\tilde Y-\tilde D\theta$
and $e:=e(\theta_0)$, so that $\tilde U=U(\theta_0)$. We have 
\begin{equation}
\label{eq:plivfun}
\begin{aligned}
\psi(w,\theta,\beta,\alpha)&=\big(Z-z(X)\big)\big(Y-\ell(X)-(D-r(X))\theta\big)
=(\tilde Z-\delta_z)\big(U(\theta)-e(\theta)\big),\\
\psi^{a}(w,\beta,\alpha)&=-\big(Z-z(X)\big)\big(D-r(X)\big)=-(\tilde Z-\delta_z)(\tilde D-\delta_r),\\
\zeta(w,\alpha)&=\big(Z-z(X)\big)\big(Z-z(X)\big)'=(\tilde Z-\delta_z)(\tilde Z-\delta_z)',
\end{aligned}
\end{equation}
which are equal at $(\beta_0,\alpha_0)$ to $\tilde ZU(\theta)$, $-\tilde Z\tilde D$ and $\tilde Z\tilde Z'$, respectively. 

For every $\theta\in\Theta$,
\begin{equation}
\label{eq:plivexact}
\begin{aligned}
&\EP\big[\psi(W,\theta,\beta,\alpha)\big]=\bar m(\theta)+\EP\big[\delta_z\,e(\theta)\big],
\qquad
\EP\big[\psi^{a}(W,\beta,\alpha)\big]=G_0-\EP\big[\delta_z\delta_r\big],\\
&\EP\big[\zeta(W,\alpha)\big]=\Omega_0+\EP\big[\delta_z\delta_z'\big],
\end{aligned}
\end{equation}
since  \eqref{eq:plivdev} are functions of $X$, $\EP[\tilde Z|X]=0$ and $\EP[U(\theta)|X]=0$ by \eqref{eq:plivbeta}, so every cross term in \eqref{eq:plivfun} has mean zero. 
 Assumption~\ref{ass:model}(b) follows from \eqref{eq:plivexact}, since the three
maps are (at most) quadratic polynomials in the deviations with second derivatives bounded under
Assumption~\ref{ass:pliv}(b)--(c), and its interchange condition holds because $\psi$ is linear in $\theta$.   Assumption~\ref{ass:model}(a) holds by Assumption~\ref{ass:pliv}(a), from the definition of $\bar m(\theta)$ and  $\Omega$, and by Assumption~\ref{ass:pliv}(d). Its dominance condition also holds by $\EP\|\tilde Z\tilde D\|<\infty$ under Assumption~\ref{ass:pliv}(b) and the Cauchy--Schwarz inequality.

For Assumption~\ref{ass:model}(c), \eqref{eq:psimean} holds  because $\bar m(\theta)=\EP[Z\tilde Y]-\EP[Z\tilde D]\theta=\EP[\tilde Z\tilde Y]-\EP[\tilde Z\tilde D]\theta$.  Along the segment from $(\beta_0,\alpha_0)$ to $(\beta,\alpha)$, indexed by
$s\in[0,1)$, the three  expansions read
$\bar m(\theta)+s^{2}\EP[\delta_ze(\theta)]$, $G_0-s^{2}\EP[\delta_r\delta_z]$ and $\Omega_0+s^{2}\EP[\delta_z\delta_z']$, whose
derivatives at $s=0$ all vanish. Since $\theta$ and the direction were arbitrary, these give \eqref{eq:unifortho} and \eqref{eq:zetaortho}. Thus, Assumption~\ref{ass:model}(c) and (d) hold. Assumption~\ref{ass:pliv}(e) implies Assumption~\ref{ass:rates}(a). For  Assumption~\ref{ass:rates}(b), the three functions in \eqref{eq:plivfun} are products of two factors with bounded $L^{q}(P)$ norms
by Assumption~\ref{ass:pliv}(b) and (e), hence have bounded $L^{q/2}(P)$ norms with $q/2>2$ by
H\"older's inequality.  

Expanding \eqref{eq:plivfun} at $\theta=\theta_0$, their deviations from the
values at $(\beta_0,\alpha_0)$ are
\[
-\delta_z\tilde U-\tilde Ze+\delta_ze,
\qquad
\delta_z\tilde D+\tilde Z\delta_r-\delta_z\delta_r,
\qquad
-\delta_z\tilde Z'-\tilde Z\delta_z'+\delta_z\delta_z', 
\]
respectively. Each linear term in the deviations $\delta$ has
bounded conditional second moments by Assumption~\ref{ass:pliv}(c). The bound for $\tilde U$
follows from $\EP[\tilde U^{2}|X]\le2\EP[\tilde Y^{2}|X]+2\theta_0^{2}\EP[\tilde D^{2}|X]$ and the
compactness of $\Theta$.  Each quadratic term is a product of two deviations $\delta$, bounded by Lyapunov's inequality,
$\|\delta\|_{P,4}\le\|\delta\|_{P,2}^{t}\|\delta\|_{P,q}^{1-t}$ with $t=(q-4)/(2q-4)\in(0,\tfrac12)$,
which gives $\|\delta_z\delta_r\|_{P,2}\le\|\delta_z\|_{P,4}\|\delta_r\|_{P,4}\le C\delta_N^{2t}$ and
likewise for $\delta_ze$ and $\delta_z\delta_z'$.  Assumption~\ref{ass:rates}(b) therefore holds with $q/2$ in place of $q$
and
\[
\bar\delta_N=C\,\delta_N^{2t}\vee N^{-[(1-4/q)\wedge1/2]}
\]
in place of $\delta_N$. Note that $\bar\delta_N\ge\delta_N$, and $\bar\delta_N \rightarrow 0$ because $q>4$, and so Assumption~\ref{ass:rates}(c)  and (d) continue to hold with $\bar\delta_N$.

Finally, \eqref{eq:plivexact} and the Cauchy--Schwarz inequality give
\[
\big\|\EP[\delta_ze]\big\|+\|\bar m_0\|\Big(\big\|\EP[\delta_z\delta_r]\big\|+\big\|\EP[\delta_z\delta_z']\big\|\Big)
\le C\Big(\|\delta_z\|_{P,2}\big(\|\delta_\ell\|_{P,2}+\|\delta_r\|_{P,2}\big)+\|\bar m_0\|\,\|\delta_z\|_{P,2}^{2}\Big),
\]
which is at most $C\delta_NN^{-1/2}$ by Assumption~\ref{ass:pliv}(e), while
$\|\EP[\delta_z\delta_r]\|+\|\EP[\delta_z\delta_z']\|\le2\delta_N^{2}\le\delta_N$.  These give Assumption~\ref{ass:rates}(c).  Assumption~\ref{ass:rates}(d) is Assumption~\ref{ass:pliv}(d), with $G_0=-\Pi$ and $\Omega_0=S$. 

Then, Theorem~\ref{thm:dml} gives the limit distribution for the DML estimator $\check \theta$, and \eqref{eq:plivpsihat}
uses $\widetilde\theta=\check\theta$, which satisfies $\|\widetilde\theta-\theta_0\|=O_P(N^{-1/2})$, so that $\widehat\sigma^2$ is consistent and the intervals are uniformly valid.
\end{proof}

\subsection{Proof of Theorem~\ref{thm:late}}
\label{app:late}

\begin{proof}

It suffices to verify Assumptions~\ref{ass:model}, \ref{ass:linear} and \ref{ass:rates}, and then the conclusions follow from
Theorems~\ref{thm:ortho} and \ref{thm:dml}. For $z\in\{0,1\}$, let
\[
\delta^{(z)}_{\ell j}:=\ell_j(z,\cdot)-\ell_{0j}(z,\cdot),
\qquad
\delta^{(z)}_{\nu j}:=\nu_j(z,\cdot)-\nu_{0j}(z,\cdot),
\qquad
\delta_{pj}:=p_j-p_{0j},
\]
denote the deviations of $\beta=(\ell_j,\nu_j)_j\in\mathcal B_N$ and
$\alpha=(p_j)_j\in\mathcal A_N$ from the true values \eqref{eq:latebeta}.

For the third term in the score $\psi^{Y}_j$ of \eqref{eq:latescores},
\[
\EP\Big[\frac{Z_j\big(Y-\ell_j(1,X)\big)}{p_j(X)}\,\Big|\,X\Big]
=\frac{p_{0j}}{p_j}\big(\ell_{0j}(1,X)-\ell_j(1,X)\big)
=-\frac{p_{0j}}{p_j}\,\delta^{(1)}_{\ell j},
\]
by definition of \eqref{eq:latebeta}, and $\EP[Z_j\mid X]=p_{0j}(X)$ and
$\EP[Z_jY\mid X]=p_{0j}(X)\ell_{0j}(1,X)$ by the law of iterated expectations. Similarly, the last term of $\psi^{Y}_j$, which involves $1-Z_j$, has conditional mean $\big((1-p_{0j})/(1-p_j)\big)\delta^{(0)}_{\ell j}$. Adding these to $\ell_j(1,X)-\ell_j(0,X)=\ell_{0j}(1,X)-\ell_{0j}(0,X)+\delta^{(1)}_{\ell j}-\delta^{(0)}_{\ell j}$, we have
\begin{equation}
\label{eq:lateexact}
\EP\big[\psi^{Y}_j(W,\beta,\alpha)\big]=\tau^{Y}_j+b^{Y}_j,
\qquad
b^{Y}_j:=\EP\Big[\frac{\delta^{(1)}_{\ell j}\,\delta_{pj}}{p_j}\Big]
+\EP\Big[\frac{\delta^{(0)}_{\ell j}\,\delta_{pj}}{1-p_j}\Big].
\end{equation}
The same calculation for the score  $\psi^{D}_j$ with $(Y,\ell_j,\ell_{0j})$ replaced by
$(D,\nu_j,\nu_{0j})$ gives $\EP[\psi^{D}_j(W,\beta,\alpha)]=\tau^{D}_j+b^{D}_j$, where $b^{D}_j$ is $b^{Y}_j$ of \eqref{eq:lateexact} with $\delta^{(z)}_{\nu j}$ in place of $\delta^{(z)}_{\ell j}$. Consequently, by \eqref{eq:latepsi} and \eqref{eq:latemean}, 
\begin{equation}
\label{eq:latepsibias}
\EP\big[\psi(W,\theta,\beta,\alpha)\big]=\bar m(\theta)+b^{Y}-\theta\,b^{D}
\qquad\text{for every }\theta\in\Theta
\end{equation} 
where $b^Y = (b_1^Y,\dots,b_{d_z}^{Y})'\in\R^{d_z}$ and $b^D = (b_1^D,\dots,b_{d_z}^{D})'\in\R^{d_z}$.

Assumption~\ref{ass:model}(a) holds by Assumption~\ref{ass:late}(d), and 
$\bar m (\theta)$ is linear in $\theta$, and $\Omega\equiv\Omega_0$ is constant. $\Omega_0$ and  $J_0=\tau^{D\prime}\Omega_0^{-1}\tau^{D}$ are positive definite by Assumption~\ref{ass:late}(c). Finally,  $\partial_{\theta'}\psi=\psi^{a}=-\psi^{D}$ is free of $\theta$, and each $\psi^{D}_j$ is bounded, because  $D$ and $Z_j$ are binary, $\nu_j(z,X)\in[0,1]$ and
$p_j\in[\epsilon,1-\epsilon]$ by Assumption~\ref{ass:late}(e). By \eqref{eq:lateexact} and \eqref{eq:latepsibias}, the maps $(\beta,\alpha)\mapsto\EP[\psi(W,\theta,\beta,\alpha)]$, $(\beta,\alpha)\mapsto\EP[\partial_{\theta'}\psi(W,\theta,\beta,\alpha)]$  differ from their values at  $(\beta_0,\alpha_0)$ only through $b^{Y}$ and $b^{D}$. Along the segment
$(\beta_0,\alpha_0)+s\big(\beta-\beta_0,\alpha-\alpha_0\big)$, $s\in[0,1)$, the numerators in
\eqref{eq:lateexact} scale with $s^{2}$, while the denominators $p_j$ and $1-p_j$ are bounded away from zero by Assumption~\ref{ass:late}(a) and (e). Hence, $b^{Y}_j, b^{D}_j$ and their first two derivatives are bounded by Assumption~\ref{ass:late}(e). By construction, $\EP[\zeta]$ is constant. Thus, all three maps are twice continuously Gateaux differentiable on $\mathcal B_N\times\mathcal A_N$, and Assumption~\ref{ass:model}(b) holds.

For Assumption~\ref{ass:model}(c), $b^{Y}$ and $b^{D}$ vanish at $(\beta_0,\alpha_0)$, so that
\eqref{eq:latepsibias} gives $\bar m(\theta)=\EP[m(W,\theta,\beta_0)]$ by \eqref{eq:latemean}. Furthermore, from the proof of part (b),  derivatives of $b^Y, b^D$ at $s=0$ vanish for every $\theta\in\Theta_0$ and every direction, which implies \eqref{eq:unifortho}.  Assumption~\ref{ass:model}(d) and Assumption~\ref{ass:linear} hold by construction. 

Assumption~\ref{ass:rates}(a) is imposed in Assumption~\ref{ass:late}(e).  For part (b), $\zeta$ is constant and $\psi^{a}=-\psi^{D}$ is bounded, while each term of $\psi^{Y}_j$ is a
bounded function of $(Z_j,X)$ multiplied by $Y$ or $\ell_j(z,X)$, whose $L^{q}(P)$ norms are
bounded by Assumption~\ref{ass:late}(a) and (e),  together with
$\|\ell_{0j}(z,\cdot)\|_{P,q}\le\epsilon^{-1/q}\|Y\|_{P,q}$, which follows from the conditional
Jensen inequality and Assumption~\ref{ass:late}(a). Thus, $\psi$, $\psi^{a}$ and $\zeta$ have bounded
$L^{q}(P)$ norms. 

For the $L^{2}(P)$ deviations, take the following terms of $\psi^{Y}_j$,
\[
\frac{Z_j\big(Y-\ell_j(1,X)\big)}{p_j}-\frac{Z_j\big(Y-\ell_{0j}(1,X)\big)}{p_{0j}}
=Z_j\big(Y-\ell_{0j}(1,X)\big)\frac{p_{0j}-p_j}{p_jp_{0j}}-\frac{Z_j\,\delta^{(1)}_{\ell j}}{p_j} .
\]
The $L^{2}(P)$ norm of the second term is bounded by $\epsilon^{-1}\|\delta^{(1)}_{\ell j}\|_{P,2}$, and the
first term has squared $L^{2}(P)$ norm bounded by 
$\epsilon^{-4}\EP\big[\delta_{pj}^{2}\,\EP[(Y-\ell_{0j}(1,X))^{2}\mid X]\big]$
$\le C\epsilon^{-4}\|\delta_{pj}\|_{P,2}^{2} \le \epsilon^{-4} C\delta_N^2$, since $\EP[(Y-\ell_{0j}(1,X))^{2}\mid X]\le2\EP[Y^{2}\mid X]+2\ell_{0j}(1,X)^{2}$ by Assumption~\ref{ass:late}(b) and (e). The remaining terms in $\psi^{Y}_j$ with $(1-Z_j)$ are handled identically, and so are those of $\psi^{D}_j$, with binary $D$ in place of $Y$. Therefore,   Assumption~\ref{ass:rates}(b) holds.

For  Assumption~\ref{ass:rates}(c), derivation \eqref{eq:lateexact} and 
Assumption~\ref{ass:late}(e), together with the Cauchy--Schwarz inequality give
\[
\big\|b^{Y}\big\|+\big\|b^{D}\big\|
\le C\sum_{j=1}^{d_z}\|\delta_{pj}\|_{P,2}\Big(\sum_{z\in\{0,1\}}\|\delta^{(z)}_{\ell j}\|_{P,2}
+\sum_{z\in\{0,1\}}\|\delta^{(z)}_{\nu j}\|_{P,2}\Big)\le C\delta_NN^{-1/2}
\]
by \eqref{eq:laterates}. Then, Assumption~\ref{ass:rates}(c) holds by the above bound,  Assumption~\ref{ass:late}(d) and
$\|\bar m_0\|=\|\EP[\psi(W,\theta_0,\beta_0,\alpha_0)]\| \le C$. Finally,  Assumption~\ref{ass:rates}(d) holds by Assumption~\ref{ass:late}(c). 

By Theorem~\ref{thm:dml}, the DML estimator has the stated limit distribution, and \eqref{eq:latepsihat}
uses $\widetilde\theta=\check\theta$, which satisfies $\|\widetilde\theta-\theta_0\|=O_P(N^{-1/2})$, so that $\widehat\sigma^2$ is consistent and the intervals are uniformly valid.
\end{proof}

%==================================================================
\section{Non-linear scores and general weighting}
\label{app:general}

In this appendix, we consider the construction of orthogonal scores and the properties of the DML for non-linear scores and general weightings by relaxing Assumption~\ref{ass:model}(d) (one-step weight) and Assumption~\ref{ass:linear} (linear scores) in Section~\ref{sec:theory}. Therefore, we can consider nonlinear moment conditions and the iterated GMM weights, where
$\zeta(w,\theta,\beta,\alpha)=\psi(w,\theta,\beta,\alpha)\psi(w,\theta,\beta,\alpha)'$ depends on
$\theta$ and on nuisance parameters $(\beta,\alpha)$. 

The orthogonal score construction in Theorem \ref{thm:ortho} allows general nonlinear scores, and we consider the orthogonal score $\psi^{*}$ of the following form with the general weight function correction term  $\kappa (\cdot)$
\begin{equation}
\label{eq:genpsi}
\psi^{*}(w,\theta,\eta):=\mu\,h(w,\theta,\lambda,\beta,\alpha)+\kappa(w,\theta,\eta),
\end{equation}
where the first part is the orthogonal score \eqref{eq:psirdml} of Section~\ref{ssec:funcscore} based on the stacked moment function with the nuisance parameter that is enlarged with an additional nuisance parameter $\chi$, with true value $\chi_0$, 
\begin{equation}
\label{eq:geneta}
\eta:=\big(\lambda',\vecop(\mu)',\beta,\alpha,\chi\big),
\qquad
\eta_0:=\big(\lambda_0',\vecop(\mu_0)',\beta_0,\alpha_0,\chi_0\big).
\end{equation}
 Evaluated at the true nuisance and with
$(A_h,\Omega_h)$ chosen so that $\mu_0J_\theta=J_0$, the first term has the closed form
\eqref{eq:psirdmlclosed}, so that for every $\theta\in\Theta$
\begin{equation}
\label{eq:genpsiclosed}
\begin{aligned}
\psi^{*}(w,\theta,\eta_0)
&=G_0'\Omega_0^{-1}\psi(w,\theta,\beta_0,\alpha_0)
+\big(\partial_{\theta'}\psi(w,\theta,\beta_0,\alpha_0)-G_0\big)'\Omega_0^{-1}\bar m_0\\
&\qquad-G_0'\Omega_0^{-1}\big(\zeta(w,\theta,\beta_0,\alpha_0)-\Omega_0\big)\Omega_0^{-1}\bar m_0
+\kappa(w,\theta,\eta_0),
\end{aligned}
\end{equation}
and the Jacobian is
\begin{equation}
\label{eq:genJ0}
J_0=G_0'\Omega_0^{-1}G_0
+\big(I_{d_\theta}\otimes\bar m_0'\Omega_0^{-1}\big)\,\partial_{\theta'}\vecop G(\theta_0)
-\big(\bar m_0'\Omega_0^{-1}\otimes G_0'\Omega_0^{-1}\big)\,\partial_{\theta'}\vecop\Omega(\theta_0).
\end{equation}

As noted in Remark \ref{rem:nocorrection}, the correction term
$\kappa(w,\theta,\eta)$ is required to relax Assumption~\ref{ass:model}(d) in Theorem  \ref{thm:ortho}. In Assumption~\ref{ass:kappa} below, we give the condition on $\kappa$, which exists under mild conditions and can be constructed using the Riesz representer or influence-function adjustment by \citet{ichimuranewey2022}. 

Now we define the DML estimator $\check\theta$ for general non-linear scores and the efficient weight. We make the whole nuisance parameters fold-specific, as in the original definition of \citet{ccddhnr2018}, whereas Definition~\ref{def:dml2} constructs $(\widehat\Gamma,\widehat\Omega,\widehat{\bar m})$ as averages over all folds for linear scores and a one-step weight, and exploits the resulting closed form of $\check\theta$ in Theorem~\ref{thm:dml}.

\begin{definition}[DML estimator]
\label{def:dml2gen}
\begin{enumerate}[label=(\alph*),leftmargin=2.4em]
\item Take a $K$-fold random partition $(I_k)_{k=1}^K$ of $[N]$ with $|I_k|=n=N/K$.
\item For each $k\in[K]$, construct an ML estimator $\widehat\eta_k:=(\widehat\lambda_k',\vecop(\widehat\mu_k)',\widehat\beta_k,\widehat\alpha_k,\widehat\chi_k)$ of $\eta_0$ from $(W_i)_{i\in I_k^c}$ alone. To be specific,  construct $\widehat\beta_k$, $\widehat\alpha_k$ and a preliminary estimator $\widetilde\theta_k$ of $\theta_0$ from $(W_i)_{i\in I_k^c}$, and evaluate the cross-fitted Jacobian, weight matrix and moment $\widehat\Gamma_k$, $\widehat\Omega_k$, $\widehat{\bar m}_k$, and $\widehat\chi_k$ at $\widetilde\theta_k$, again using $(W_i)_{i\in I_k^c}$ only. Then set $\widehat\lambda_k:=(\vecop(\widehat\Gamma_k)',\vecop(\widehat\Omega_k)')'$ and $\widehat\mu_k=\Big[-\big(I_{d_\theta}\otimes\widehat{\bar m}_k'\widehat\Omega_k^{-1}\big)\ \ \
\widehat{\bar m}_k'\widehat\Omega_k^{-1}\otimes\widehat\Gamma_k'\widehat\Omega_k^{-1}\ \ \ I_{d_\theta}\Big]$ as in  Definition~\ref{def:dml2}. Then, by \eqref{eq:genpsi},
\[
\begin{aligned}
\psi^{*}(W_i,\theta,\widehat\eta_k)
&=\widehat\Gamma_k'\widehat\Omega_k^{-1}\psi(W_i,\theta,\widehat\beta_k,\widehat\alpha_k)
+\big(\partial_{\theta'}\psi(W_i,\theta,\widehat\beta_k,\widehat\alpha_k)-\widehat\Gamma_k\big)'\widehat\Omega_k^{-1}\widehat{\bar m}_k\\
&\qquad-\widehat\Gamma_k'\widehat\Omega_k^{-1}\big(\zeta(W_i,\theta,\widehat\beta_k,\widehat\alpha_k)-\widehat\Omega_k\big)\widehat\Omega_k^{-1}\widehat{\bar m}_k
+\kappa(W_i,\theta,\widehat\eta_k) .
\end{aligned}
\]
\item Construct the DML estimator $\check\theta$ in either of the two ways below:

(DML1) For each $k\in[K]$, let $\check\theta_k$ be an approximate solution of
\begin{equation}
\label{eq:dml1gen}
\Big\|\frac1n\sum_{i\in I_k}\psi^{*}\big(W_i,\check\theta_k,\widehat\eta_k\big)\Big\|
\le\inf_{\theta\in\Theta}\Big\|\frac1n\sum_{i\in I_k}\psi^{*}\big(W_i,\theta,\widehat\eta_k\big)\Big\|+\varepsilon_N ,
\end{equation}
where $\varepsilon_N=o(\delta_N N^{-1/2})$, and $\delta_N$ is some sequence of positive constants converging to zero. Then, aggregate over the folds, $\check\theta:=K^{-1}\sum_{k=1}^{K}\check\theta_k$.

(DML2) Construct $\check\theta$ as an approximate solution of
\begin{equation}
\label{eq:dml2gen}
\Big\|\frac1N\sum_{k=1}^{K}\sum_{i\in I_k}\psi^{*}\big(W_i,\check\theta,\widehat\eta_k\big)\Big\|
\le\inf_{\theta\in\Theta}\Big\|\frac1N\sum_{k=1}^{K}\sum_{i\in I_k}\psi^{*}\big(W_i,\theta,\widehat\eta_k\big)\Big\|+\varepsilon_N .
\end{equation}
\end{enumerate}
\end{definition}

\subsection{Asymptotic distribution of the DML for non-linear scores and general weighting}
\label{ssec:genasym}

\begin{assumption}[Weight correction]
\label{ass:kappa}
For all $P\in\mathcal P_N$, the correction term $\kappa(w,\theta,\eta)$ in \eqref{eq:genpsi} satisfies the following conditions: $\EP[\kappa(W,\theta_0,\eta_0)]=0$ and $\partial_{\theta'}\EP[\kappa(W,\theta,\eta_0)]|_{\theta=\theta_0}=0$. The Gateaux derivatives of $\EP[\kappa(W,\theta_0,\eta_0)]$ in the directions of $\lambda$, $\vecop(\mu)$ and $\chi$ vanish. For all $(\beta,\alpha)\in\mathcal B_N\times\mathcal A_N$,
\begin{equation}
\label{eq:genkappa}
\begin{aligned}
&\partial_{(\beta,\alpha)}\EP\big[\kappa(W,\theta_0,\eta_0)\big][\beta-\beta_0,\alpha-\alpha_0]\\
&\qquad=G_0'\Omega_0^{-1}\,\partial_{(\beta,\alpha)}\EP\big[\zeta(W,\theta_0,\beta_0,\alpha_0)\big][\beta-\beta_0,\alpha-\alpha_0]\,\Omega_0^{-1}\bar m_0 .
\end{aligned}
\end{equation}

\end{assumption}

Assumption~\ref{ass:kappa} replaces Assumption~\ref{ass:model}(d) and provides the conditions on $\kappa$.  The correction term $\kappa$ takes into account the derivative of the weight function with respect to the first step, and the assumption is used only for the orthogonality of $\psi^{*}$ in Theorem~\ref{thm:general} below. Assumption~\ref{ass:kappa} keeps the moment condition and leaves the Jacobian unchanged. 

\begin{assumption}[Regularity conditions for non-linear scores]
\label{ass:general}
Let $0<c_0\le c_1$,  $c_2>1$, $q>2$, and $v>0$ be constants and $\{\delta_N\}$, $\{\Delta_N\}$,
$\{\tau_N\}$ sequences of positive constants converging to zero, with $\delta_N\ge N^{-1/2+1/q}\log N$ and $N^{-1/2}\log N\le\tau_N\le\delta_N$.  For all $N\ge3$ and $P\in\mathcal P_N$, the following conditions
hold.
\begin{enumerate}[label=(\alph*),leftmargin=2.4em]

\item $\Theta$ is bounded and contains a ball of radius $c_1N^{-1/2}\log N$ centred at $\theta_0$
of \eqref{eq:focfun}.
\item The map $(\theta,\eta)\mapsto\EP[\psi^{*}(W,\theta,\eta)]$ is twice continuously
Gateaux-differentiable on $\Theta\times T$, where $T:=\Lambda\times\R^{d_\theta d_h}\times\mathcal B\times\mathcal A\times\mathcal X$ is convex. For every $k\in[K]$,  $\widehat\eta_k$ belongs to the nuisance realization set $T_N$ with probability at least $1-\Delta_N$, where $T_N:=\Lambda_N\times\mathcal M_N\times\mathcal B_N\times\mathcal A_N\times\mathcal X_N\subset T$  contains $\eta_0$ and is constrained by conditions given below, and $\Lambda_N$, $\mathcal M_N$ and $\mathcal X_N$ are neighbourhoods of $\lambda_0$, $\mu_0$ and $\chi_0$.

\item For all $\theta\in\Theta$,
$2\|\EP[\psi^{*}(W,\theta,\eta_0)]\|\ge\|J_0(\theta-\theta_0)\|\wedge c_0$, where $J_0$ of
\eqref{eq:genJ0} has singular values between $c_0$ and $c_1$.  The variance of the score is non-degenerate. All eigenvalues of
$\EP[\psi^{*}(W,\theta_0,\eta_0)\psi^{*}(W,\theta_0,\eta_0)']$ are bounded below by $c_0$.

\item  For each $\eta\in T_N$, the function class
\[
\mathcal F_\eta:=\big\{\psi^{*}_j(\cdot,\theta,\eta):j=1,\dots,d_\theta,\ \theta\in\Theta\big\}
\]
is suitably measurable with a measurable envelope $F_\eta$ satisfying $\|F_\eta\|_{P,q}\le c_1$,
and its uniform covering entropy obeys
\[
\sup_Q\ \log N\big(t\,\|F_\eta\|_{Q,2},\ \mathcal F_\eta,\ \|\cdot\|_{Q,2}\big)\ \le\ v\log(c_2/t)
\qquad\text{for all }0<t\le1,
\]
where $N(s,\mathcal F,\|\cdot\|)$ denotes the $s$-covering number of $\mathcal F$ with respect to
$\|\cdot\|$ and the supremum runs over all finitely discrete probability measures $Q$ on the sample space.
\item The following conditions on the statistical rates $r_N, r_N'$, and $ \lambda_N'$ hold:
\[
\begin{aligned}
r_N&:=\sup_{\eta\in T_N,\ \theta\in\Theta}
\big\|\EP[\psi^{*}(W,\theta,\eta)]-\EP[\psi^{*}(W,\theta,\eta_0)]\big\| 
\le\delta_N\tau_N,\\
r_N'&:=\sup_{\eta\in T_N,\ \|\theta-\theta_0\|\le\tau_N}
\big(\EP\big\|\psi^{*}(W,\theta,\eta)-\psi^{*}(W,\theta_0,\eta_0)\big\|^{2}\big)^{1/2}, \textnormal{ and } r_N'\log^{1/2}(1/r_N')\le\delta_N,\\ 
\lambda_N'&:=\sup_{r\in(0,1),\ \eta\in T_N,\ \|\theta-\theta_0\|\le\tau_N}
\big\|\partial_r^{2}\EP\big[\psi^{*}\big(W,\theta_0+r(\theta-\theta_0),\eta_0+r(\eta-\eta_0)\big)\big]\big\|\le\delta_NN^{-1/2}.
\end{aligned}
\]
\end{enumerate}
\end{assumption}

Assumption~\ref{ass:general} is similar to \citet[Assumptions~3.3 and 3.4]{ccddhnr2018} imposed on the misspecification-robust score
$\psi^{*}(W,\theta,\eta)$ and the enlarged nuisance parameter $\eta=\big(\lambda',\vecop(\mu)',\beta,\alpha,\chi\big)$.  As discussed in \citet{ccddhnr2018}, these are mild regularity conditions and standard in moment condition models.  Under correct specification ($\bar m_0=0$),  $\psi^{*}=G_0'\Omega_0^{-1}\psi$ and the conditions in Assumption~\ref{ass:general}
reduce to those of  \citet{ccddhnr2018}; under misspecification ($\bar m_0 \neq 0$), we use conditions on a different Jacobian $J_0$ with additional smoothness conditions (Assumption~\ref{ass:model}(a)), and the identification condition at the GMM estimand \eqref{eq:focfun}, instead of the true value.

\begin{remark}[Partially linear IV model (cont.)]
\label{rem:plivit}In the partially linear IV model with the efficient weights $\zeta(w,\theta,\beta,\alpha)=\psi(w,\theta,\beta,\alpha)\psi(w,\theta,\beta,\alpha)'$,
the target becomes the iterated GMM estimand, i.e., the fixed point of \eqref{eq:focfun} with
$\Omega(\theta)=\EP[\tilde Z\tilde Z'U(\theta)^2]$, where $U(\theta)=\tilde Y-\tilde D\theta$, so that
$\Omega_0=\Omega(\theta_0)=\EP[\tilde Z\tilde Z'\tilde U^2]$ with $\tilde U=U(\theta_0)$. The weight
function now depends on the first step and violates \eqref{eq:zetaortho}: its derivative in the direction $(\ell-\ell_0,r-r_0,z-z_0)$ does not vanish in general. The correction term is
\begin{equation}
\label{eq:plivkappa}
\kappa(w,\theta,\eta)=-\chi_1(X)\big(Z-z(X)\big)-2\,\chi_2(X)\big(Y-\ell(X)-(D-r(X))\theta\big),
\end{equation}
with the additional nuisance parameter $\chi=(\chi_1,\chi_2)$, whose true value
$\chi_0=(\chi_{10},\chi_{20})$ is

\begin{align*}
\chi_{10}(X)&:=\EP\big[\tilde U^2\big(( \bar m_0'\Omega_0^{-1}\tilde Z)\Pi'\Omega_0^{-1}+(\Pi'\Omega_0^{-1}\tilde Z)\bar m_0'\Omega_0^{-1}\big)\,\big|\,X\big]\in\R^{1\times d_z},\\
\chi_{20}(X)&:=\EP\big[(\Pi'\Omega_0^{-1}\tilde Z)(\bar m_0'\Omega_0^{-1}\tilde Z)\tilde U\,\big|\,X\big]\in\R.
\end{align*}
Both are conditional expectations, which can be estimated by machine learning, and both are proportional to $\bar m_0$, so that the correction vanishes under correct specification. It is straightforward to verify that $\kappa$ in \eqref{eq:plivkappa} satisfies Assumption~\ref{ass:kappa}. Then, the
orthogonal score is the analogue of \eqref{eq:plivpsirdml},
\[
\begin{aligned}
\psi^{*}(w,\theta_0,\eta_0)
&=-\,\Pi'\Omega_0^{-1}\tilde Z\tilde U
-\big(\tilde Z\tilde D-\Pi\big)'\Omega_0^{-1}\bar m_0\\
&\qquad+\Pi'\Omega_0^{-1}\big(\tilde Z\tilde Z'\tilde U^{2}-\Omega_0\big)\Omega_0^{-1}\bar m_0
-\chi_{10}(X)\tilde Z-2\,\chi_{20}(X)\tilde U,
\end{aligned}
\]
with the Jacobian \[
J_0=\Pi'\Omega_0^{-1}\Pi-2\,\Pi'\Omega_0^{-1}\EP\big[\tilde Z\tilde Z'\tilde U\tilde D\big]\Omega_0^{-1}\bar m_0,
\]
in place of $J_0=\Pi'S^{-1}\Pi$ in Section \ref{sec:pliv}.
\end{remark}

\begin{theorem}[Orthogonal scores and DML inference for non-linear scores and general weights]
\label{thm:general}
Let $\psi^{*}$ be given by \eqref{eq:genpsi} with $(A_h,\Omega_h)$ chosen so that $\mu_0J_\theta=J_0$.
\begin{enumerate}[label=(\roman*),leftmargin=2.4em]
\item Under Assumption~\ref{ass:model}(a)--(c) and Assumption~\ref{ass:kappa},
$\EP[\psi^{*}(W,\theta_0,\eta_0)]=0$, and the score $\psi^{*}$ is Neyman orthogonal at
$(\theta_0,\eta_0)$.
\item If, in addition, Assumption~\ref{ass:general} holds, the estimator $\check\theta$ of
Definition~\ref{def:dml2gen}, in either its DML1 or its DML2 form, satisfies
\[
\sqrt N\,\sigma^{-1}\big(\check\theta-\theta_0\big)\ \dto\ \mathcal N\big(0,I_{d_\theta}\big),
\qquad
\sigma^{2}=J_0^{-1}\,\EP\big[\psi^{*}(W,\theta_0,\eta_0)\psi^{*}(W,\theta_0,\eta_0)'\big]\,J_0^{-1\prime},
\]
uniformly over $P\in\mathcal P_N$. Moreover, $\sigma^{2}$ can be replaced by a consistent estimator
$\widehat\sigma^{2}$ with $\widehat\sigma^{2}=\sigma^{2}+o_P(1)$ uniformly over
$P\in\mathcal P_N$, and the confidence intervals \eqref{eq:dmlci} built from $\check\theta$ and
such a $\widehat\sigma^{2}$ are uniformly asymptotically valid.
\end{enumerate}
\end{theorem}

Theorem~\ref{thm:general} covers the general non-linear scores and weightings, which the main results in Section~\ref{sec:theory} do not cover.  Above all, we allow the efficient weight  $\zeta=\psi\psi'$ of
Remarks~\ref{rem:nocorrection} and \ref{rem:plivit}, where $\zeta$ depends on both $\theta$ and the first step $(\beta,\alpha)$, and thus $\kappa\not\equiv0$. If $\psi$ and $\zeta$ do not depend on the nonparametric first step $(\beta,\alpha)$, then $\kappa\equiv0$, and
Theorem~\ref{thm:general} covers standard non-linear GMM with efficient weighting: $\theta_0$ is
the iterated GMM fixed point, $J_0$ is \eqref{eq:genJ0}, and $\sigma^{2}$ is the misspecification-robust variance (Proposition~\ref{prop:closed}) as in \citet{hallinoue2003} and \citet{hansenlee2021}. Furthermore, Theorem~\ref{thm:dml} is not a special case of Theorem~\ref{thm:general}, because the two use different estimators: Definition~\ref{def:dml2gen} makes $(\widehat\Gamma,\widehat\Omega,\widehat{\bar m})$ fold-specific as in \citet{ccddhnr2018}, whereas Definition~\ref{def:dml2} pools
 them over folds and exploits the closed form \eqref{eq:thm2rep} for linear scores, which requires neither the entropy condition (Assumption~\ref{ass:general}(d)) nor the global identification condition (Assumption~\ref{ass:general}(c)).

\subsection{Proof of Theorem~\ref{thm:general}}
\label{app:genproof}

\begin{proof}
(i) Note that Assumption~\ref{ass:linear} is not used in the proof of Theorem~\ref{thm:ortho}, and Assumption~\ref{ass:model}(d) implies $D_\zeta=\partial_{(\beta,\alpha)}\EP[\zeta(W,\theta_0,\beta_0,\alpha_0)][\beta-\beta_0,\alpha-\alpha_0] = 0$ for
$(\beta,\alpha)\in\mathcal B_N\times\mathcal A_N$. Therefore, similar arguments carry over from the proof of Theorem~\ref{thm:ortho}, and it remains to account for $\kappa$ and for $D_\zeta\neq0$.

The moment condition
 $\EP[\mu_0h(W,\theta_0,\lambda_0,\beta_0,\alpha_0)]+\EP[\kappa(W,\theta_0,\eta_0)]=0$ holds by \eqref{eq:rdmlmoment}, as in the proof of Theorem~\ref{thm:ortho}, with the enlarged nuisance parameter $\eta=\big(\lambda',\vecop(\mu)',\beta,\alpha,\chi\big)$, and  by Assumption~\ref{ass:kappa}. The derivatives of $\EP[\mu_0h]$ in $\lambda$ and $\vecop(\mu)$ vanish as in the proof of Theorem~\ref{thm:ortho}, and  those of $\EP[\kappa]$ in $\lambda$, $\vecop(\mu)$ and $\chi$ vanish by Assumption~\ref{ass:kappa}.
 
For $(\beta,\alpha)\in\mathcal B_N\times\mathcal A_N$, similar calculations in
the proof of Theorem~\ref{thm:ortho} give
\[
\partial_{(\beta,\alpha)}\EP\big[\psi^{*}(W,\theta_0,\eta_0)\big][\beta-\beta_0,\alpha-\alpha_0]
=G_0'\Omega_0^{-1}D_\psi+D_{\partial_{\theta'}\psi}'\,\Omega_0^{-1}\bar m_0
-G_0'\Omega_0^{-1}D_\zeta\,\Omega_0^{-1}\bar m_0+D_\kappa = 0,
\]
where the last two terms cancel because $D_\kappa=G_0'\Omega_0^{-1}D_\zeta\Omega_0^{-1}\bar m_0$ by \eqref{eq:genkappa}.  As in the proof of Theorem~\ref{thm:ortho},  $D_\psi=0$ and $D_{\partial_{\theta'}\psi}=0$ by Assumption~\ref{ass:model}(c), and $\partial_{\theta'}\EP[\kappa(W,\theta,\eta_0)]|_{\theta=\theta_0}=0$ by Assumption~\ref{ass:kappa}. The closed form \eqref{eq:genpsiclosed} can be similarly constructed as 
\eqref{eq:muhfun} at $(\beta_0,\alpha_0)$ plus $\kappa$, and the Jacobian
remains unchanged $\mu_0J_\theta=J_0$ of \eqref{eq:genJ0} by Assumption~\ref{ass:kappa}.

(ii)  We give the argument for the DML1 form of Definition~\ref{def:dml2gen} and note that the DML2 form is treated in the same way as done in \citet{ccddhnr2018}. We closely follow the proof of Theorem~3.3 of \citet{ccddhnr2018}, and their Lemma~6.3, and show that the score $\psi^{*}$ of \eqref{eq:genpsi}, the nuisance parameter $\eta$ of \eqref{eq:geneta} and the GMM estimand $\theta_0$ of \eqref{eq:focfun} satisfy Assumptions~3.3 and 3.4 of \citet{ccddhnr2018}. 

Fix a sequence $\{P_N\}_{N\ge1}$ with $P_N\in\mathcal P_N$ for
every $N$; since the sequence is arbitrary it suffices to establish the conclusion under $P_N$, and we write $O_P$ for $O_{P_N}$. Definition~\ref{def:dml2gen} makes the whole of $\widehat\eta_k$ fold-specific, and its DML1 and
DML2 forms are their Definitions~3.1 and 3.2 for the score $\psi^{*}$ with
$\varepsilon_N=o(\delta_NN^{-1/2})$. Note that  $\EP[\psi^{*}(W,\theta_0,\eta_0)]=0$ and the Neyman orthogonality holds by part (i), and thus Assumption 3.3 of \citet{ccddhnr2018} holds by Assumption~\ref{ass:general}(a)--(c), where $\partial_{\theta'}\EP[\psi^{*}(W,\theta,\eta_0)]|_{\theta=\theta_0}=J_0$ of \eqref{eq:genJ0} by the proof of part~(i). Similarly, Assumption~\ref{ass:general} also implies their Assumption~3.4 with the score $\psi^{*}$. The growth conditions $\delta_N\ge N^{-1/2+1/q}\log N$ and $N^{-1/2}\log N\le\tau_N\le\delta_N$ of their Theorem~3.3 are imposed in Assumption~\ref{ass:general}.

Fix $k\in[K]$. Conditional on $(W_i)_{i\in I_k^c}$, the nuisance $\widehat\eta_k$  lies in $T_N$ with probability at least $1-\Delta_N$. We condition on the event in which all $K$ first steps lie in $T_N$ with probability at least $1-K\Delta_N=1-o(1)$. Hence, Lemma~6.3 of \citet{ccddhnr2018}, which establishes approximate linearity of the subsample DML estimators, applies to the fold-$k$
estimator \eqref{eq:dml1gen} and gives, uniformly over $P\in\mathcal P_N$,
\[
\sqrt n\,\sigma^{-1}\big(\check\theta_k-\theta_0\big)
=-\frac{1}{\sqrt n}\sum_{i\in I_k}\sigma^{-1}J_0^{-1}\psi^{*}(W_i,\theta_0,\eta_0)+O_P(\delta_N),
\]
where $n = N/K$. Hence their Theorem~3.3 applies to the DML1 form of $\check\theta$ (averaging over $k \in [K]$) and gives,
uniformly over $P\in\mathcal P_N$,
\[
\sqrt N\,\sigma^{-1}\big(\check\theta-\theta_0\big)
=-\frac{1}{\sqrt N}\sum_{i=1}^{N}\sigma^{-1}J_0^{-1}\psi^{*}(W_i,\theta_0,\eta_0)+O_P(\rho_N)
\ \dto\ \mathcal N\big(0,I_{d_\theta}\big),
\]
where $\rho_N=N^{-1/2+1/q}\log N+r_N'\log^{1/2}(1/r_N')+N^{1/2}\lambda_N+N^{1/2}\lambda_N'
\le C\delta_N$ by the condition on $\delta_N$, and 
Assumption~\ref{ass:general}(e). The uniform validity of the confidence interval holds by their Theorem~3.3 together with their Corollary~3.1. 
\end{proof}

\end{document}